\documentclass{jfm}
\usepackage{graphicx}
\usepackage{epstopdf, epsfig}
\usepackage{amsfonts,amssymb,amsmath,mathbbol}   
\usepackage{caption}
\usepackage{subcaption}

\usepackage{xcolor} 

\def\sin{{\rm{sin}}}
\def\cos{{\rm{cos}}}

\def\v{{\rm{\bf v}}}
\def\e{{\rm{\bf e}}}

\def\n{{\rm \bf n}}
\def\t{{\rm \bf t}}

\def\I{{\rm \bf I}}

\usepackage{color} 

\shorttitle{Asymmetric freezing of sliding droplet} 
\shortauthor{S, Kavuri, G. Karapetsas, C. S. Sharma and K. C. Sahu } 
\title{Asymmetric freezing of a sliding droplet on an inclined surface}

\author
 {
 Sivanandan Kavuri,\aff{1}
 George Karapetsas,\aff{2}
 Chander Shekhar Sharma\aff{3}
  \and
  Kirti Chandra Sahu\aff{1}\corresp{\email{gkarapetsas@auth.gr; chander.sharma@iitrpr.ac.in; ksahu@che.iith.ac.in}}
  }

\affiliation
{
\aff{1}
Department of Chemical Engineering, Indian Institute of Technology Hyderabad, Sangareddy 502 284, Telangana, India
\aff{2}
Department of Chemical Engineering, Aristotle University of Thessaloniki, Thessaloniki 54124, Greece
\aff{3}
Department of Mechanical Engineering, Indian Institute of Technology Ropar, 140001 Rupnagar, India
}

\begin{document}

\maketitle

\begin{abstract}
We investigate the asymmetric freezing of a liquid droplet sliding on an inclined cold surface using numerical simulations based on the lubrication approximation. The combined effects of gravity, capillarity, and solidification kinetics on droplet motion, interfacial deformation, and the resulting frozen morphology are examined through systematic variations in substrate inclination, wettability, effective Bond number, and Stefan number. Our results show that sliding prior to and during the early stages of freezing plays a dominant role in governing the asymmetry of the frozen droplet. A tilted ice cusp forms at the droplet tip due to the competition between gravitational forces and capillary resistance, with its orientation and magnitude strongly dependent on substrate wettability and inclination. Greater inclination and increased wettability enhance asymmetry in droplet morphology. Further, highly wetting substrates favor capillary-driven retraction and induce transient liquid motion opposite to gravity during freezing. The evolution of contact-angle hysteresis at both the solid surface and the liquid–ice interface underscores the importance of early-time dynamics, when the unfrozen liquid remains mobile and gravitational effects are most pronounced. Decomposition of the liquid motion into capillary- and gravity-driven contributions provides physical insight into contact-line pinning, receding-edge thinning, and the development of asymmetric liquid–ice contact angles. Increasing the Stefan number accelerates freezing, limits sliding-induced deformation, and reduces both the cusp angle and the post-freezing contact-angle contrast. Overall, this study establishes a physical framework for understanding the morphology of frozen droplets on inclined substrates.
\end{abstract}


\section{Introduction} \label{sec:intro}
Freezing of liquid droplets on a cold surface is a fundamental multiphysics phenomenon relevant to a wide range of natural and engineering processes, including cloud microphysics \citep{pruppacher1980microphysics,hanna2008cloud}, freezing rain formation \citep{zerr1997freezing,mensah2015review}, cryopreservation \citep{mazur1970cryobiology}, food processing and storage \citep{james2015review,bonat2016influence,zhao2017freezing,zhu2021biomimetic}, freeze-drying technologies \citep{franks1998freeze,assegehegn2019importance}, thermal management \citep{sharma2022numerical}, controlled fabrication of porous materials and microstructures \citep{qian2011controlled,li2012freeze}, and icing phenomena on aircraft and civil infrastructure \citep{cao2015aircraft,myers1998modeling}, to name a few.

The solidification of a droplet is driven by a complex interplay of latent heat release, thermally induced phase change, interfacial wetting and capillary forces, gravitational and viscous effects, and substrate interactions. The nonlinear interaction of these processes makes droplet freezing a rich and challenging problem in interfacial fluid mechanics. Over the past decades, extensive research has yielded a detailed understanding of the freezing behaviour of sessile, stationary droplets on flat surfaces \citep{jung2012mechanism,jung2012frost,wang2022new,kavuri2023freezing,kavuri2025evaporation}. These studies have established that freezing proceeds in two distinct stages. The first is the recalescence stage, during which rapid ice nucleation and crystal growth occur as supercooling drives solidification, releasing latent heat and triggering a transient temperature rise \citep{feuillebois1995freezing,hindmarsh2003experimental,hu2010icing,jung2012frost}. This temperature rise has been attributed to the rapid restructuring and percolation of the hydrogen-bond network in water \citep{tavakoli2015freezing}. Once the droplet reaches its melting temperature, the process transitions to a slower, typically conduction-controlled freezing stage in which the solid–liquid interface gradually advances from the substrate toward the apex of the droplet \citep{zadravzil2006droplet,jung2012frost,kavuri2023freezing}. Concurrently, vapor condensation on the cold substrate surrounding the droplet leads to the formation of frost halos \citep{jung2012frost,kavuri2023freezing}.

A striking characteristic of droplet freezing is the development of a pointed ice tip during the final stage of solidification, where volume expansion arising from the density difference between liquid water and ice produces an inflection at the apex and a distinct cusp-like morphology \citep{sanz1987influence, anderson1996case, zadravzil2006droplet}. This phenomenon, governed by interfacial thermodynamics and curvature constraints and strongly influenced by environmental and material properties such as humidity, salinity, and substrate thermal properties, has led to extensive experimental research and sustained interdisciplinary interest in freezing of sessile droplets \citep{anderson1996case,hu2010icing,jung2012frost,zhang2016freezing,marin2014universality,jung2012mechanism,zhang2017freezing,zhang2019shape}. Further studies have revealed a strong sensitivity of freezing behaviour to ambient conditions, with \citet{kavuri2023freezing} and \citet{sebilleau2021air,kavuri2025evaporation} demonstrating that relative humidity and temperature play a key role in controlling frost halo formation and freezing duration. Using a two-dimensional coupled level-set and volume-of-fluid model together with an enthalpy-porosity approach, \citet{wan2023freezing} reported that increasing salinity slows the freezing process, resulting in longer solidification times for seawater droplets. Recently, \citet{lyu2023liquid} demonstrated that an environmental medium with high thermal conductivity enhances external cooling, alters the freezing-front dynamics, promotes earlier crust formation, and ultimately modifies the final frozen morphology of the droplet. The influence of surface roughness has also been explored \citep{suzuki2007hydrophobicity,hao2014freezing,fuller2024analysis}. Their studies showed that textured or hydrophobic surfaces introduce greater thermal resistance at the droplet–substrate interface, increase the freezing temperature, and prolong the overall freezing time.

The impact and solidification of water droplets on cold substrates have also been extensively investigated. \citet{thievenaz2019solidification,thievenaz2020retraction,thievenaz2020freezing} examined the coupled impact and freezing dynamics, elucidating contact-line pinning, dewetting and retraction behaviour, as well as the influence of interfacial ice growth on droplet spreading. Furthermore, \citet{knight1967contact,demmenie2023growth,demmenie2025partial,thievenaz2020retraction,sarlin2025macroscopic} demonstrated the partial wetting of water on ice, reporting a finite equilibrium contact angle of approximately $12^\circ$ near the melting point, along with its strong temperature dependence due to contact-line pinning induced by crystallization at lower temperatures. This is consistent with the thermodynamic interpretation of \citet{macdowell2026key}, which shows that finite contact angles can coexist with a nanometric premelting film. In addition, \citet{huerre2025freezing} provide a comprehensive review of the coupling between capillary flow and solidification, emphasizing the key mechanisms and open questions in freezing dynamics.

Despite significant advances in understanding the freezing of stationary droplets on horizontal substrates, a major gap remains between laboratory studies and real-world scenarios, where droplets rarely remain stationary either before or during freezing. In many practical situations, droplets slide, deform, and accelerate under gravity on inclined natural or engineered surfaces, such as aircraft wings, wind turbine blades, marine vessels, power transmission lines, heat exchangers, and fluid transport pipes \citep{geer1939analysis, gent2000aircraft, cao2015aircraft, kraj2010phases, makkonen1998modeling, chuvilin2001factors}. Under such conditions, droplet shapes are asymmetric, and gravity-driven motion induces persistent differences between advancing and receding sides, modifies internal pressure and curvature distributions, and alters heat transfer. While the hydrodynamics of sliding droplets have been studied extensively \citep{podgorski2001corners,kim2002sliding,le2005shape,savva2013droplet,thampi2013liquid,puthenveettil2013motion,karapetsas2013effect,benilov2015thin,park2017droplet}, virtually nothing is known about how transient motion, geometric asymmetry, and dynamic contact-line behaviour influence the subsequent freezing process and the final frozen morphology. Consequently, existing literature, which overwhelmingly assumes droplets are fixed or immobilized, leaves open fundamental questions about whether the well-established morphologies and universal freezing trajectories observed in sessile droplets persist when motion precedes or accompanies freezing, highlighting the critical role of surface inclination in influencing freezing dynamics.

The shape of a frozen droplet, including the asymmetry of the cusp-like morphology at its tip, is strongly influenced by its initial equilibrium shape \citep{jin2016impacta,zeng2022influence}, highlighting the importance of understanding the initial droplet morphology, particularly on inclined or more complex non-horizontal substrates. Numerous studies have investigated the dynamics of droplets sliding on inclined substrates, focusing on contact line behaviour, velocity, and shape evolution, albeit in the absence of freezing and evaporation \citep{podgorski2001corners,kim2002sliding,thiele2002sliding,le2005shape,savva2013droplet,thampi2013liquid,puthenveettil2013motion,karapetsas2013effect,benilov2015thin,park2017droplet}. \citet{podgorski2001corners} demonstrated that droplets of silicone oil and water on fluoro-polymer-coated surfaces exhibit rounded, cornered, or pearling shapes depending on the balance between viscous and surface tension forces, which also governs the dynamic contact angle at the trailing edge. On rough surfaces, droplets begin to slide when the substrate is tilted beyond a critical angle. \citet{kim2002sliding} showed that ethylene glycol, glycerine, and glycerine–water mixture droplets on polycarbonate surfaces exhibit sliding velocities consistent with a linear balance between gravitational and viscous–capillary forces. \citet{le2005shape} reported that, with increasing Bond number (the ratio between the gravitational and surface tension forces), a millimetre-sized drop on a partially wetting inclined plane elongates and undergoes morphological transitions from rounded to cornered, cusped, and ultimately pearling shapes. In addition, \citet{puthenveettil2013motion} found that at high Reynolds numbers, droplet velocity scales with the Bond number through a balance of inertial, viscous, and capillary forces, leading mercury droplets to form cornered shapes and water droplets to develop rivulets.

In parallel with these experimental investigations, several theoretical and numerical studies \citep{thiele2002sliding, karapetsas2013effect, savva2013droplet, thampi2013liquid, benilov2015thin, park2017droplet} have examined droplet dynamics on inclined substrates using long-wave, lubrication, and diffuse-interface models. \citet{thiele2002sliding} identified universal and non-universal regimes of sliding drops governed by substrate inclination, while \citet{karapetsas2013effect} showed that thermo-capillarity and temperature-dependent wettability significantly influence contact line motion and spreading. \citet{thampi2013liquid} distinguished rolling and sliding modes based on droplet geometry, viscosity contrast, and slip length. \citet{benilov2015thin} derived asymptotic expressions for thin drop velocities using a Navier-slip lubrication framework and validated their model with experiments. \citet{savva2013droplet} investigated two-dimensional droplets on chemically heterogeneous substrates using a long-wave evolution equation with slip, revealing stick–slip dynamics, substrate-induced hysteresis, and uphill motion under strong chemical gradients. \citet{park2017droplet} studied droplets on substrates with topographical defects using a thin-film evolution equation with a precursor layer, illustrating that pinning–depinning and residual droplet formation are controlled by interactions of the receding contact line with defect geometry.

The freezing of droplets on inclined surfaces has also attracted increasing attention in recent years due to its relevance to several of the applications discussed above. \citet{starostin2022universality,starostin2023effects} experimentally examined freezing on horizontal, inclined, inverted, and silicone-oil-lubricated surfaces, showing that the tip angles of frozen droplets remain nearly constant at $143^\circ \pm 11^\circ$, and analyzed how asymmetric cooling governs tip orientation. \citet{dang2022modelling} modeled the freezing of sessile droplets on inclined substrates and demonstrated that smaller droplets are more sensitive to inclination, whereas larger droplets exhibit only modest changes, with freezing occurring more readily on steeper surfaces, an effect that is particularly pronounced for small droplets. More recently, \citet{kumar2025understanding} conducted an experimental study on the sequential freezing and melting of impact-deposited and sessile droplets, investigating contact-line arrest, tip singularity, melting modes, and interfacial dynamics over a broad range of temperatures, impact velocities, and substrate inclination angles. Furthermore, in the studies of \citet{dang2022modelling, starostin2022universality, starostin2023effects, kumar2025understanding}, the contact line becomes arrested after the initial spreading or deformation stage, before the onset of solidification. Consequently, the freezing process proceeds from an essentially stationary droplet configuration.

In the present work, we theoretically investigate the freezing dynamics of a droplet sliding on a smooth substrate, where the droplet continues to move until contact-line arrest occurs due to solidification. Although droplets typically do not slide on smooth surfaces unless the inclination exceeds a critical angle or the droplet is sufficiently large for gravity to overcome capillary adhesion \citep{quere1998drops}, we assume that contact-angle hysteresis is sufficiently small to permit sliding over the range of inclinations considered. Under these conditions, asymmetric freezing can substantially alter post-solidification contact angles, shift the cusp singularity away from the surface normal, and is strongly influenced by surface wettability, substrate inclination, freezing rate, and the interplay between gravitational and capillary forces. To address these unresolved issues, we develop a unified theoretical–numerical framework that couples lubrication-based hydrodynamics, dynamic contact-line motion, and conduction-controlled solidification. In contrast to earlier studies that assume a frozen geometry or prescribe the motion of the freezing front, our model captures the fully self-consistent evolution of the liquid shape, freezing front, contact angles, and wetting state from the onset of solidification to complete freezing. By systematically varying substrate inclination and surface wettability, we uncover a rich hierarchy of freezing behaviours that are not observed in stationary droplets. Our results reveal that substrate inclination induces strong and persistent asymmetry in both droplet morphology and freezing dynamics. During the early stages, the advancing and receding contact angles diverge, resulting in an asymmetric liquid layer above the freezing front and a shift of the final cusp relative to the contact line normal, with the degree of asymmetry increasing monotonically with inclination angle of the substrate. Remarkably, even brief droplet motion prior to contact-line arrest leaves a lasting imprint on the frozen structure. As freezing progresses, a dynamical crossover emerges in which gravity dominates the early behaviour, while the diminishing liquid volume enhances capillary effects that gradually restore symmetry and drive the system toward a universal tip angle. These findings demonstrate that the freezing of moving droplets is not a trivial extension of freezing of sessile droplets, but a qualitatively distinct regime in which transient hydrodynamics, geometry, and phase change remain strongly coupled, providing the first mechanistic insight into asymmetric freezing governed by substrate tilt, surface wettability, and motion-induced geometric memory.

The remainder of this paper is structured as follows. In \S \ref{sec:model}, we present the problem formulation, governing equations, scaling considerations, and the numerical approach utilized within the framework of the lubrication approximation. In \S \ref{sec:dis}, we discuss the results, focusing on the mechanism governing the freezing of a sliding drop, supported by an extensive parametric study. Finally, concluding remarks are provided in \S \ref{sec:Conc}.

\section{Problem formulation} \label{sec:model}

\begin{figure}
\centering
\includegraphics[width=0.75\textwidth]{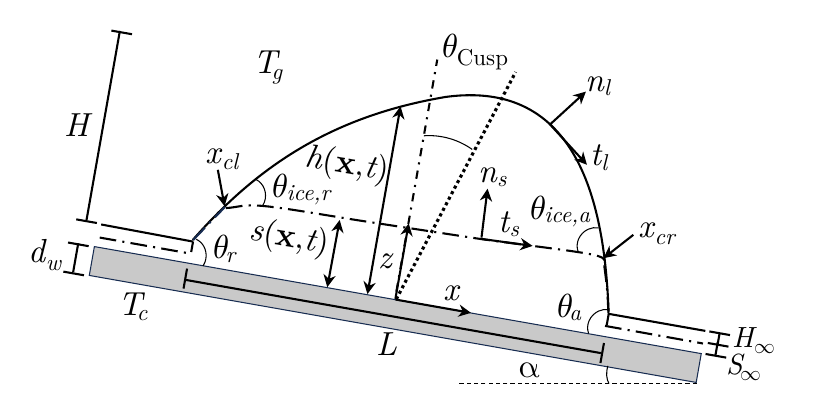}
\caption{Schematic of a sessile droplet freezing on a solid substrate inclined at an angle $\alpha$. The droplet length and maximum height are denoted by $L$ and $H$, respectively. The quantities $S_\infty$ and $H_\infty$ represent the initial thickness of the microscopic ice layer and the precursor film. The advancing and receding contact angles of the droplet are $\theta_a$ and $\theta_r$. The advancing and receding contact angles at the liquid–ice interface are $\theta_{ice,a}$ and $\theta_{ice,r}$. The left and right contact lines of the unfrozen droplet are denoted by $x_{cl}$ and $x_{cr}$. The ambient and substrate temperatures are $T_g$ and $T_c$, respectively. The cusp angle of the droplet, measured with respect to the $z$-axis, is denoted by $\theta_{\text{Cusp}}$.}
\label{fig:geom}
\end{figure}

We examine the freezing dynamics of a sessile droplet on an inclined substrate using numerical simulations based on the lubrication approximation. A schematic of the two-dimensional (2D) droplet at a given time, $t$, during the freezing process, illustrating the physical parameters used in the model, is shown in figure~\ref{fig:geom}. A Cartesian coordinate system $(x,z)$ is employed in the theoretical modeling, whose origin is located at the center of the droplet on the substrate. The substrate, inclined at an angle $\alpha$, has a thickness $d_w$, thermal conductivity $\lambda_w$, and specific heat capacity $C_{pw}$, with its bottom surface maintained at a constant temperature $T_c$. Initially (at $t = 0$), when freezing begins, the droplet on the inclined substrate is considered to have reached its equilibrium shape and steady sliding velocity, with the initial recalescence phase completed and the liquid at the melting temperature ($T_m$). Since our focus is on a droplet that is already sliding, this assumption ensures that the droplet is in a steady state before freezing starts. Prior to the propagation of the freezing front, a very rapid recalescence phase occurs~\citep{hu2010icing,jung2012frost,wang2022new,hindmarsh2003experimental,feuillebois1995freezing}, lasting only a few milliseconds. During this phase, the release of latent heat forms an ice crystal scaffold (or mushy phase) and induces a transient rise in temperature. As the recalescence timescale is much shorter than the total freezing time, it is reasonable to assume that this phase has already concluded and that the liquid is at $T_m$ when the freezing front begins to propagate. This initial recalescence explains why the unfrozen liquid does not exhibit a contact angle of $12^\circ$ at the liquid–ice interface immediately after freezing onset, as noted in previous studies~\citep{huerre2025freezing,knight1967contact,sarlin2025macroscopic,thievenaz2020retraction,demmenie2023growth,demmenie2025partial}.

At $t = 0$, a uniformly thin ice layer of thickness $S_\infty$ has already formed along the substrate, indicating the onset of the slower solidification stage governed by heat conduction. The droplet length, maximum height, initial ice layer thickness, precursor film thickness, and the advancing and receding contact angles at the droplet–substrate and liquid–ice interfaces are denoted by $L$, $H$, $S_\infty$, $H_\infty$, $\theta_a$, $\theta_r$, $\theta_{ice,a}$, and $\theta_{ice,r}$, respectively. The angle formed by the singularity peak of the droplet with the vertical ($z$-axis) at the droplet center is denoted by $\theta_{\text{Cusp}}$. The liquid is treated as an incompressible, Newtonian fluid with constant density $\rho_l$, dynamic viscosity $\mu_l$, specific heat capacity $C_{pl}$, and thermal conductivity $\lambda_l$. The liquid–ice and liquid–gas interfaces are represented by $z = s(x,t)$ and $z = h(x,t)$, respectively. The surface tension at the liquid–gas interface is denoted by $\gamma_{lg}$. The density $\rho_s$, specific heat capacity $C_{ps}$, and thermal conductivity $\lambda_s$ of the solid (ice) phase are also assumed to be constant.

Within the lubrication approximation, the droplet is very thin, with its maximum height $H$ much smaller than its length $L$ ($H \ll L$). Although this approximation strictly applies only to thin droplets, previous studies show that lubrication-based models provide reasonably accurate predictions for contact angles up to $60^\circ$~\citep{charitatos2020thin,Tembely2019}. Consequently, the small aspect ratio $\epsilon = 2H/L \ll 1$ justifies using lubrication theory to derive the evolution equations describing the freezing dynamics of the sessile droplet, including the liquid, ice, and gas phases. In the present study, we choose the equilibrium contact angle range of $12^\circ$–$30^\circ$ to ensure that this thin-droplet assumption remains valid. The classical no-slip condition at a moving contact line leads to a stress singularity, which can be alleviated by introducing a thin precursor film. This approach has been widely used for droplet motion \citep{Schwartz1998, espin2015droplet, park2017droplet}, droplet evaporation and imbibition \citep{pham2019imbibition, wang2021dynamics, Karapetsas2016, williams2021}, and droplet freezing \citep{zadravzil2006droplet, Tembely2019, kavuri2023freezing, kavuri2025evaporation}. This approach successfully captures key features such as cusp formation, freezing front propagation, and volume expansion, while avoiding numerical difficulties at the contact line. The dimensional governing equations describing the freezing process of a sessile droplet on an inclined substrate are presented below.

\subsection{Dimensional governing equations}\label{sec:gov_eqn}

\subsubsection{Liquid phase}

The liquid-phase dynamics is governed by the conservation equations for mass, momentum, and energy, which are expressed as
\begin{equation}\label{eq:mom}
\rho_l \left( \frac{\partial \v}{\partial t} + \v \cdot \nabla \v \right) = - \nabla p + \mu_l \nabla^2 \v + \rho_l g ~\sin \alpha ~e_x-\rho_l g ~ \cos\alpha ~{e_z},
\end{equation}
\begin{equation}\label{eq:cont}
\nabla \cdot \v = 0,
\end{equation}
\begin{equation}\label{eq:energy_liquid}
\rho_l C_{pl} \left( \frac{\partial T_l}{\partial t} + \v \cdot \nabla T_l \right) = \lambda_l \nabla^2 T_l,
\end{equation}
where $\v$, $p$, and $T_l$ denote the velocity, pressure, and temperature in the liquid phase, respectively; $\nabla$ is the gradient operator; $g$ represents the acceleration due to gravity; and $e_x$ and $e_z$ denote the components of the unit vectors along the $x$- and $z$-directions, respectively.

At the free surface ($z=h(x,t)$), the outward unit normal to the liquid-gas interface ($\n_l$) is defined as
\begin{equation}
\n_l = (-h_{x}e_x + e_z)/\sqrt{(h_{x}^2+1)},
\end{equation}
 The tangential components of the velocity are identical in both phases and are given by $\v_{\tau} = \v - (\v \cdot \n_l) \n_l = \v_{lg} - (\v_{lg} \cdot \n_l) \n_l$. At $z = h(x,t)$, the velocity field satisfies the local mass, force, and energy balance between the liquid and gas phases \citep{Karapetsas2016}, which are given by
\begin{equation}
\rho_l (\v -\v_{lg}) \cdot \n_l = \rho_g (\v_g -\v_{lg}) \cdot \n_l,
\end{equation}
\begin{equation}
 \n_l \cdot \left[-p \I + \mu_l \left( \nabla \v + \nabla \v ^T \right) \right]  = -p_g \n_l + \gamma_{lg} \kappa_{lg} \n_l + \Pi \n_l ,
\end{equation}
\begin{equation}
 \lambda_l \nabla T_l \cdot \n_l - \lambda_g \nabla T_g \cdot \n_l -h_{tc}(T_{g}-T_{lg})= 0,
\label{energy1}
\end{equation}
where $\rho_g$, $\lambda_g$, $\v_g$, $T_g$, $\v_{lg}$ and $h_{tc}$ denote the density, thermal conductivity, velocity field, and temperature of the gas phase, respectively; $v_{lg}$ denotes the velocity of the liquid–gas interface, and $h_{tc}$ denotes the convective heat transfer coefficient. Here, $\I$ represents the identity tensor, $T_{lg}$ denotes the temperature at the liquid–gas interface, $\kappa_{lg} = - \nabla_{s,l} \cdot \n_l$ is the mean curvature of the free surface, and $\nabla_{s,l} = (I - \n_l \n_l) \cdot \nabla$ is the surface gradient operator. The disjoining pressure ($\Pi$), which accounts for the van der Waals interactions, is given by
\begin{equation} \label{eq:disj_press}
\Pi = A \left[ \left( \frac{B}{h - s} \right)^{n} - \left( \frac{B}{h - s} \right)^{m} \right],
\end{equation}
where $A = A_{\text{Ham}} / B^{n} \geq 0$ quantifies the magnitude of the intermolecular interaction energy between the liquid–gas and liquid–ice interfaces, and $B$ is the precursor layer thickness. Here, $n > m > 1$, and $A_{\text{Ham}}$ represents the dimensional Hamaker constant.

At the liquid–ice interface $(z = s(x,t))$, the velocity field satisfies the kinematic boundary condition:
\begin{equation} \label{eq:mass_bc_solid_ND}
\v = \v_{ls} - \left( \frac{J_s}{\rho_l} \right) \n_s,
\end{equation}
where $J_s$ denotes the freezing mass flux, and $\v_{ls}$ is the velocity of the liquid–ice interface. The outward unit normal to the liquid–ice interface ($\n_s$) is given by
\begin{equation}
\n_s = {\left ( {-s_x e_x + e_z} \right) / {\sqrt{s_x^2 + 1}}}.
\end{equation}

A no-slip condition is imposed at the liquid–ice interface, requiring the tangential component of the velocity to satisfy
\begin{equation} \label{eq:ice_no_slip}
\v \cdot \t_s = 0,
\end{equation}
where $\t_s$ is the outward unit tangential vector along the liquid–ice interface, defined as
\begin{equation} \label{eq:tangent}
\t_s = {\left (e_x + s_x e_z \right) / \sqrt{s_x^2 + 1}}.
\end{equation}

\subsubsection{Solid (ice) phase} \label{solid_ice_phase}

The temperature evolution in the solid (ice) phase is governed by
\begin{equation}\label{eq:energy_ice}
\rho_s C_{ps} \frac{\partial T_s}{\partial t} = \lambda_s \nabla^2 T_s,
\end{equation}
where $T_s$ represents the temperature in the solid phase.

At the solid substrate ($z=0$), the continuity of temperature is imposed, which is given by
\begin{equation} \label{eq2p12}
T_s = T_w,
\end{equation}
where $T_w$ denotes the substrate temperature at $z=0$.

At the freezing front ($z=s(x,t)$), the temperature satisfies
\begin{equation} \label{eq:temp_bc_liquid_solid}
T_s = T_l = T_f,
\end{equation}
where the equilibrium temperature at the freezing front, $T_f$, is assumed to be equal to the melting temperature, $T_m$.

The conservation of mass and energy at the liquid–solid interface (at $z = s(x,t)$) yields
\begin{equation} \label{eq:mass_bc_solid}
J_s = \rho_l (\v_{ls} - \v)\cdot \n_s = \rho_s (\v_{ls} - \v_s)\cdot \n_s,
\end{equation}
\begin{equation} \label{eq:energy_bc_solid}
\rho_s (\v_{ls} - \v_s)\cdot \n_s H_s - \lambda_s \nabla T_s \cdot \n_s
= \rho_l (\v_{ls} - \v)\cdot \n_s H_l - \lambda_l \nabla T_l \cdot \n_s,
\end{equation}
where $\v_s$ is the velocity of the ice phase, which is stationary in our study ($\v_s = 0$); $H_s$ and $H_l$ denote the enthalpy of the solid and liquid phases, respectively.

Combining eqs.~(\ref{eq:mass_bc_solid}) and (\ref{eq:energy_bc_solid}), and employing $\v_s = 0$, we obtain
\begin{equation} \label{eq:energy_bc_solid_final}
J_s \Delta H_{sl} - \lambda_s \nabla T_s \cdot \n_s + \lambda_l \nabla T_l \cdot \n_s = 0,
\end{equation}
where $\Delta H_{sl} = H_s - H_l$ represents the enthalpy jump across the liquid–ice interface. Then, substituting
$H_s = C_{ps}(T_f - T_m) + L_f(T_m)$ and $H_l = C_{pl}(T_f - T_m)$, we get
\begin{equation} \label{eq:DHsl}
\Delta H_{sl} = (C_{ps} - C_{pl})(T_f - T_m) + L_f,
\end{equation}
where $L_f$ is the latent heat of fusion defined with $T_m$ as the reference temperature. In the following, eq.~(\ref{eq:energy_bc_solid_final}) is used to determine the evolution of the freezing front, $s(x,t)$.

\begin{table}
\centering
\begin{tabular}{ccc}
Property   &     Notation       & Value  \\  \hline
Density of the liquid phase   & $\rho_{l}$ &  1000 Kg m$^{-3}$ \\
Density of the frozen solid phase   & $\rho_{s}$ & 900 Kg m$^{-3}$  \\
Density of the gas phase   & $\rho_{g}$ &  $5 \times 10^{-3}$ Kg m$^{-3}$ \\
Viscosity of the liquid phase   & $\mu_{l}$ &  $10^{-3}$ Pa$\cdot$s\\
Viscosity of the gas phase   & $\mu_{g}$ &  $1.81 \times 10^{-5}$ Pa$\cdot$s \\
Melting temperature  & $T_m$ &  273 K \\
Ambient temperature	 & $T_g$ &  273 K \\
Temperature at the bottom of the substrate   & $T_c$ & 243 K  \\ 
Thickness of the substrate   & $d_w$ & $1 \times 10^{-3}$ m \\  
Thermal conductivity of the substrate   & $\lambda_w$ &  384.75 W m$^{-1}$ K$^{-1}$\\ 
Thermal conductivity of the liquid phase   & $\lambda_l$ &  0.57 W m$^{-1}$ K$^{-1}$ \\ 
Thermal conductivity of the frozen solid phase   & $\lambda_s$ &  2.21 W m$^{-1}$ K$^{-1}$ \\
Thermal conductivity of the gas phase   & $\lambda_g$ &  0.02 W m$^{-1}$ K$^{-1}$ \\ 
Specific heat capacity of the substrate   & $C_{pw}$ & 300 - 3000 J Kg$^{-1}$K$^{-1}$ \\ 
Specific heat capacity of the liquid phase   & $C_{pl}$ &  4220 J Kg$^{-1}$K$^{-1}$\\
Specific heat capacity of the frozen solid phase   & $C_{ps}$ &  2050 J Kg$^{-1}$K$^{-1}$ \\
Specific heat capacity of the gas phase   & $C_{pg}$ & 4220 J Kg$^{-1}$K$^{-1}$  \\
Surface tension of the liquid-gas interface at $T_m$  & $\gamma_{lg}$ &  0.07 N m$^{-1}$ \\
Surface tension of the liquid-ice interface    & $\gamma_{ls}$ &  0.02 N m$^{-1}$ \\
Latent  heat  of  fusion   & $L_{f}$ &  $3.35\times10^5$ J Kg$^{-1}$\\
Accommodation  coefficient & $\alpha$ &  $\approx 1$ \\
\end{tabular}
\vspace{2mm}
\caption{Representative values of the physical parameters used in our simulations. The listed properties correspond to the water–air system and copper substrate.}
\label{T:wecr} 
\end{table}

\subsubsection{Solid substrate}

The energy equation governing heat transfer in the solid substrate is
\begin{equation}\label{eq:energy_solid}
\rho_w C_{pw} \frac{\partial T_w}{\partial t} = \lambda_w \nabla^2 T_w,
\end{equation}
where $\rho_w$ denotes the density of the substrate. This equation is supplemented by the continuity of heat flux at the ice–substrate interface ($z = 0$):
\begin{equation}
\lambda_s \frac{\partial T_s}{\partial z} = \lambda_w \frac{\partial T_w}{\partial z},
\end{equation}
and the temperature condition at the bottom boundary of the substrate ($z = -d_w$):
\begin{equation}
T_w = T_c.
\end{equation}

The fluid properties and ranges of physical parameters considered in this study are summarized in Table \ref{T:wecr}. All simulations are performed for water droplets on a highly conductive copper substrate, with the bottom surface maintained at approximately $-30^\circ$C and the ambient temperature close to the melting point of water. Under these conditions, evaporation is minimal and is therefore neglected in the present model. In the following, we nondimensionalize the governing equations and boundary conditions.

\subsection{Scaling} \label{sec:scaling}

The above governing equations and boundary conditions are nondimensionalized using the following scalings, where a tilde denotes the dimensionless variable:
\begin{equation}\label{all_scales}
\begin{gathered}
( x, z, h, D_{w} ) = l ( \tilde{x}, \epsilon \tilde{z}, \epsilon \tilde{h} , \epsilon \tilde{D_{w}}), ~ t = \frac{l}{U} \tilde{t}, ~ ( u, w ) = U ( \tilde{u}, \epsilon \tilde{w} ),\\ (u_g, w_g ) = U ( \tilde{u_{g}}, \epsilon \tilde{w_{g}} ), ~ (p,\Pi) = \frac{\mu_l U l}{H^2} (\tilde{p},\tilde{\Pi}), ~ T_i = \Delta T \; \tilde{T}_i  + T_c \; (i=l,s,w), \\ 
 ~ J_s = \epsilon \rho_s U \; \tilde{J}_s, \nabla = \frac{1}{l} \tilde{\nabla}, ~ \nabla_{s,i} = \frac{1}{l} \tilde{\nabla}_{s,i} \; (i=l,s),
\end{gathered}
\end{equation}
where $\Delta T = T_m - T_c$, $l = L/2$ (half-length of the droplet), $\epsilon = H/l$, $\widetilde{\nabla} = \e_x \widetilde{\partial}_x + \e_z \epsilon^{-1} \widetilde{\partial}z$, and $\widetilde{\nabla}{s,i} = (\mathbf{I} - \n \n) \cdot \widetilde{\nabla}$ $(i = l, s)$. The velocity scale is chosen as $U = \epsilon^3 \gamma_{lg} / \mu_l$, such that $Ca/\epsilon^2 = 1$, wherein $Ca = \mu_l U / (\epsilon \gamma_{lg})$ is the capillary number. Hereafter, the tilde notation is suppressed, and the partial derivatives ${\partial / \partial x}$, ${\partial / \partial z}$, and ${\partial / \partial t}$ are represented by the subscripts $x$, $z$, and $t$, respectively. By applying these scalings and invoking the lubrication approximation ($\epsilon \ll 1$), we obtain the dimensionless governing equations and boundary conditions for the liquid, ice, and gas phases.

\subsubsection{Liquid phase}
The governing equations for the liquid phase, written in dimensionless form, are as follows:
\begin{equation} \label{eq:xmom_scaled}
\partial^2_z u + \frac{Bo~\sin\alpha}{\epsilon} = \partial_x p,
\end{equation}
\begin{equation} \label{eq:zmom_scaled}
\partial_z p + Bo ~ \cos\alpha = 0,
\end{equation}
\begin{equation} \label{eq:cont_scaled}
\partial_x u + \partial_z w= 0,
\end{equation}
\begin{equation} \label{eq:Tl_scaled}
\partial^2_z T_l = 0.
\end{equation}
The boundary conditions at the liquid–gas interface ($z = h(x,t)$) are
\begin{equation} \label{eq:ph_scaled}
p = - \kappa_{lg} - \Pi,
\end{equation}
\begin{equation} \label{eq:tstress_scaled}
\partial_z u = 0,
\end{equation}
\begin{equation} \label{eq:ebc_scaled}
\partial_z T_l = Bi(T_v - T_l),
\end{equation}
\begin{equation} \label{eq:kin_scaled}
\partial_t h + u\partial_x h - w = 0.
\end{equation}
Here, $Bo = \rho_l g l^{2}/\gamma_{lg}$ is the Bond number, representing the ratio of gravitational to surface tension forces, and $Bi = h_{tc} H/\lambda_l$ is the Biot number, which compares the internal thermal resistance due to conduction to the external resistance associated with convective heat transfer.

The boundary conditions at the liquid–ice interface (at $z = s(x,t)$) are
\begin{equation} \label{eq:noslip_ls_scaled}
u = 0, \quad T_l = T_f,
\end{equation}
where $T_f = T_m = 1$.

The dimensionless disjoining pressure can be expressed as
\begin{equation}
\Pi = A_{n} \epsilon^{-2} \left[ \left( \frac{\beta}{h-s} \right)^{n} - \left( \frac{\beta}{h-s} \right)^{m} \right],
\label{eq:dimensionless_disj_press}
\end{equation}
where $\beta$ is of the same order as the equilibrium precursor film thickness \citep{pham2019imbibition}, and $A_{n} = H A / \gamma_{lg}$ is the dimensionless Hamaker constant. The competition between the attractive and repulsive contributions in eq.~(\ref{eq:dimensionless_disj_press}) determines the equilibrium contact angle, $\theta_{eq}$ \citep{Schwartz1998, zadravzil2006droplet, espin2015droplet, pham2019imbibition, Tembely2019}, which may be approximated by \citep{pham2019imbibition}
\begin{equation} \label{eq:CA}
\theta_{eq} \approx \sqrt{\beta A_{n}}.
\end{equation}

The mean curvatures of the liquid–gas and liquid–ice interfaces, retaining higher-order terms ($\epsilon^{2} h_x^{2}$ and $\epsilon^{2} s_x^{2}$), are given by
\begin{equation}
\kappa_{lg} = \frac{h_{xx}}{(1+\epsilon^{2} h_x^{2})^{3/2}},
\quad {\rm and} \quad
\kappa_{sl} = \frac{s_{xx}}{(1+\epsilon^{2} s_x^{2})^{3/2}},
\end{equation}
respectively.

\subsubsection{Solid (ice) phase}

The dimensionless energy conservation equation in the ice phase is given by
\begin{equation} \label{T_diff}
\partial^2_z T_s = 0.
\end{equation}
The corresponding energy balance at the liquid–ice interface ($z = s(x,t)$) is
\begin{equation} \label{Stefan_scaled}
Ste \left( \Lambda_s \partial_z T_s - \partial_z T_l \right) = J_s,
\end{equation}
where $\Lambda_s = \lambda_s / \lambda_l$ is the thermal conductivity ratio, and $Ste = \lambda_l \Delta T / (\epsilon \rho_s U L_f H)$ is the Stefan number, which represents the ratio of sensible (conductive) heat transport in the liquid to the latent heat released at the solidification front.

The nondimesionalization of the mass conservation at $z = s(x,t)$ (eq.~\ref{eq:mass_bc_solid_ND}) yields
\begin{equation} \label{eq:Js_scaled}
\partial_t s - w = D_s J_s
\end{equation}
where $J_s = \partial_t s$, $D_s = \rho_s / \rho_l$ denotes the density ratio of the solid to the liquid phase. 

At the ice–substrate interface ($z = 0$), the temperature satisfies
\begin{equation}
T_s = T_w.
\end{equation}

\subsubsection{Solid substrate}

The dimensionless energy equation for the solid substrate is given by
\begin{equation}
\partial^2_z T_w = 0.
\end{equation}
At the solid–ice interface ($z = 0$), the temperature field satisfies
\begin{equation}
\Lambda_s \partial_z T_s = \Lambda_w \partial_z T_w,
\end{equation}
where $\Lambda_w = \lambda_w / \lambda_l$ is the thermal conductivity ratio for the substrate.

At the bottom surface of the substrate $z = -D_w$, where $D_w = d_w/H$ is the scaled wall thickness, the temperature is prescribed as
\begin{equation}
T_w = 0.
\end{equation}

\subsection{Evolution equations}

Integrating eqs. (\ref{eq:xmom_scaled}) and (\ref{eq:zmom_scaled}) with respect to $z$, and applying the boundary conditions in eqs. (\ref{eq:tstress_scaled}), (\ref{eq:noslip_ls_scaled}), and (\ref{eq:ph_scaled}), yields
\begin{equation}
u = \left( \partial_x p - \frac{Bo ~ \sin\alpha}{\epsilon} \right)
\left( \frac{z^2 - s^2}{2} - h(z - s) \right),
\end{equation}
\begin{equation}
p = Bo (h - z)\cos\alpha - \kappa_{lg} - \Pi.
\end{equation}

Integrating the continuity equation (\ref{eq:cont_scaled}) and applying the kinematic condition (\ref{eq:kin_scaled}) along with the mass conservation equation (\ref{eq:Js_scaled}) yields the evolution equation
\begin{equation}
\partial_t h - \partial_t s = -\partial_x q_l - D_s J_s,
\end{equation}
where the volumetric flux is given by
\begin{equation}
q_l = \left( \partial_x p - \frac{Bo ~ \sin\alpha}{\epsilon} \right)
\left( -\frac{h^3}{3} + s h^2 - s^2 h + \frac{s^3}{3} \right).
\end{equation}

Similarly, integrating eq. (\ref{eq:Tl_scaled}) and applying the thermal boundary conditions (eqs.~\ref{eq:ebc_scaled} and \ref{eq:noslip_ls_scaled}) gives
\begin{equation}
T_l = \left[ Bi \left( T_v - T_l\big|_{h} \right) \right] (z - s) + T_f.
\label{eq262}
\end{equation}

The temperature field in the ice phase is given by
\begin{equation}
T_s = \frac{T_f}{D_w + s \Lambda_w/\Lambda_s}
\left( D_w + z  \frac{\Lambda_w}{\Lambda_s} \right).
\end{equation}

Substituting this expression, along with eq. (\ref{eq:Js_scaled}), in eq. (\ref{Stefan_scaled}) yields 
\begin{equation}\label{eq:Freezing_rate}
\partial_t s = Ste \left(  \frac{\Lambda_w T_f}{D_w+s \Lambda_w/\Lambda_s} -  Bi\left(T_{v}-T_{l}\big|_{h}\right)\right). 
\end{equation}

Finally, the temperature profile in the solid substrate is given by
\begin{equation}
T_w = \frac{T_f}{D_w + s \Lambda_w/\Lambda_s} (z + D_w).
\end{equation}

\subsection{Initial and boundary conditions}\label{Ics_Bcs}

In the theoretical model incorporated in the present study based on lubrication approximation, the droplet is placed on a thin precursor film that lies above a thin ice layer. By suitably selecting the dimensionless Hamaker constant $A_{n}$ (eqs. \ref{eq:dimensionless_disj_press} and \ref{eq:CA}), we ensure that the droplet attains the prescribed equilibrium contact angle with the substrate. This choice effectively balances the attractive and repulsive contributions of the disjoining pressure, thereby setting the equilibrium contact angle \citep{Schwartz1998,zadravzil2006droplet,espin2015droplet,pham2019imbibition,Tembely2019}. The initial conditions imposed on the domain are
\begin{eqnarray}
h(x,t=0) = \max \left(h_{\infty}+s_{\infty}-x^{2}+2xx_{0}+1-x_{0}^{2}, ~h_{\infty}+s_{\infty}\right).
\end{eqnarray}

The precursor layer thickness far from the droplet is defined as $h_{\infty} = (h - s_{\infty}) = H_{\infty}/H$, and is assumed to be uniform, flat, and of zero mean curvature. The parameter $x_{0}$ denotes the initial position of the contact line of the droplet. In all simulations, $\beta = 0.01$ and the initial thickness of the ice layer is taken as $s_{\infty} = 10^{-3}$. We confirm that reducing the value of the latter parameter by one or two orders of magnitude does not affect the results.

To prevent unphysical solidification within the precursor film, we introduce a thickness-dependent term that suppresses freezing in this region. This is essential, as solidification of the precursor layer would reintroduce the stress singularity, defeating its purpose. Following \citet{zadravzil2006droplet,Tembely2019,kavuri2023freezing,kavuri2025evaporation}, we incorporate a thickness-dependent Stefan number, $Ste(x)$, which inhibits freezing within the precursor film and is defined as
\begin{equation}
Ste(x) = \frac{1}{2}\left(1 + \tanh \left[4 \times 10^{3}\big((h - s) - 1.4\beta\big)\right]\right) Ste.
\label{pre_model1}
\end{equation}

The boundary conditions applied at the edges of the computational domain are
\begin{equation}
h_{x}(0,t) = h_{xxx}(0,t) = h_{x}(x_{\pm\infty},t) = h_{xxx}(x_{\pm\infty},t) = 0,
\end{equation}
\begin{equation}
s_{x}(0,t) = s_{xxx}(0,t) = s_{x}(x_{\pm\infty},t) = s_{xxx}(x_{\pm\infty},t) = 0,
\end{equation}
where $\pm x_{\infty}$ denotes the far-field boundary of the domain.

\subsection{Numerical method and validation}\label{method}

\begin{table}
\centering
\setlength{\tabcolsep}{1pt}
\renewcommand{\arraystretch}{1.2}
\begin{tabular}{p{4cm} p{7.0cm} p{2.3cm}}
\textbf{Dimensionless group} & \textbf{Description} & \textbf{Value} \\ 
$\epsilon = H/L$                                          & Droplet aspect ratio                          & $0.2$ \\
$D_w = d_w/H$                                             & Scaled wall thickness                         & $2.94$ \\
$T_v = (T_g - T_c)/(T_m - T_c)$                           & Scaled gas temperature                        & $1$ \\
$D_s = \rho_s/\rho_l$                                     & Density ratio (solid--liquid)                 & $0.9$ \\
$\Lambda_s = \lambda_s/\lambda_l$                         & Thermal conductivity ratio (ice--liquid)     & $3.89$ \\
$\Lambda_w = \lambda_w/\lambda_l$                         & Thermal conductivity ratio (wall--liquid)    & $675$ \\
$Bi = h_{tc}H/\lambda_l$                                  & Biot number                                   & $0.16$ \\
$A_n = H A/\gamma_{lg}$                                   & Scaled Hamaker constant                       & $10$ \\
$Ste = \lambda_l \Delta T / (\epsilon \rho_s U L_f H)$    & Stefan number                                 & $1.49\times10^{-3}$ \\
$Bo = \rho_l g l^{2}/\gamma_{lg}$                         & Bond number                                   & $0.4$ \\
\end{tabular}
\vspace{2mm}
\caption{Dimensionless groups and their typical values considered in the present study.}
\label{T:dim_groups}
\end{table}

\begin{figure}
\centering
\includegraphics[width=0.95\textwidth]{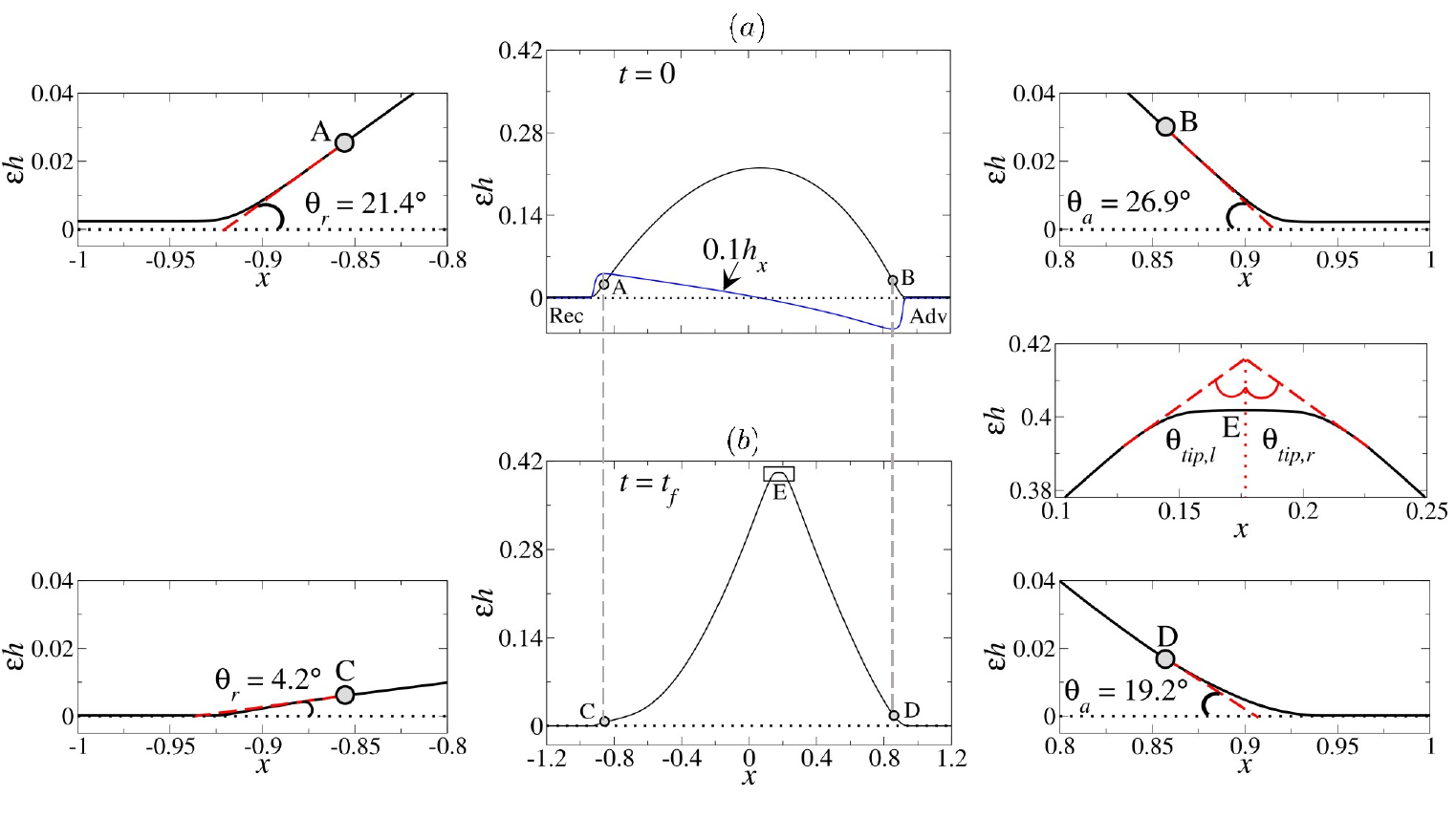}
\caption{Illustration of the procedure used to determine contact angles during (a) the initial and (b) final stages of droplet freezing. Insets at points A–D highlight the measurement locations, while the tip angle after complete freezing (point E) is shown in the right inset. The advancing ($\theta_a$) and receding ($\theta_r$) contact angles at $t = 0$ (points A and B) are obtained by drawing tangents at the locations of maximum slope in the profile of $h_x = \partial h/\partial x$, following \citet{espin2015droplet,Schwartz1998}. The blue solid line in panel (a) represents $0.1 h_x$. At $t = t_f$, tangents at the same locations (points C and D shown in in panel b) yield the final contact angles. The tip angle near the cusp (point E) is approximately $137^\circ$. Here, $\theta_{eq} = 26.5^\circ$, $\alpha = 75^\circ$, and the remaining parameters are listed in Table~\ref{T:dim_groups}.}
\label{fig:demo}
\end{figure}

\begin{figure}
\centering
\includegraphics[width=0.95\textwidth]{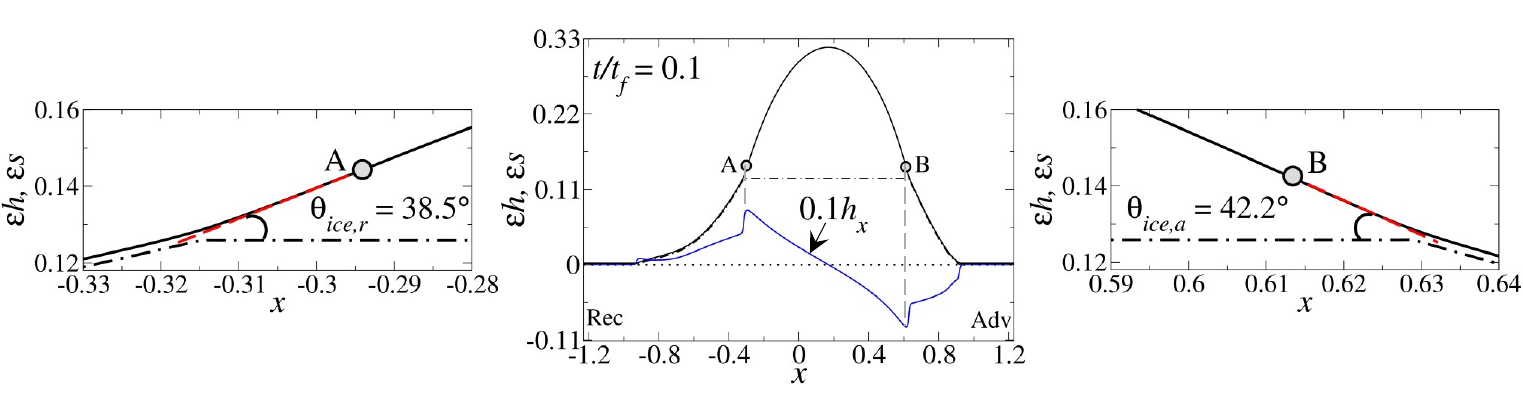}
\caption{Illustration of the procedure used to determine the liquid–ice contact angle ($\theta_{ice}$) at an early stage of freezing ($t/t_f = 0.1$). The receding ($\theta_{ice,r}$) and advancing ($\theta_{ice,a}$) contact angles (points A and B) are identified by drawing tangents at the locations of maximum slope in the profile of $h_x = \partial h/\partial x$, following \citet{espin2015droplet,Schwartz1998}. The blue solid line represents $0.1 h_x$. Here, $\theta_{eq} = 26.5^\circ$, $\alpha = 75^\circ$, and the remaining parameters are provided in Table~\ref{T:dim_groups}.}
\label{fig:demo_l_i}
\end{figure}

The dimensionless governing equations are discretized using the Galerkin finite element method, with weak forms formulated for each equation. The nonlinear system is solved iteratively using the Newton–Raphson method, and time integration is performed with an implicit Euler scheme featuring adaptive time stepping. The adaptive time step is controlled by the maximum residual from the preceding iteration, consistent with standard adaptive implicit Euler procedures. Linear algebra operations are carried out using the LAPACK library, and the complete solver is implemented in FORTRAN. Our model is validated for both droplet sliding on a smooth inclined substrate and droplet freezing on a horizontal substrate. We first simulated a non-freezing droplet sliding down an inclined surface, a configuration previously investigated by \citet{park2017droplet}. As shown in Appendix \ref{A1}, figure \ref{fig:Slid_val_A1}, the final equilibrium droplet profile closely matches the corresponding result reported in figure 2(c) of \citet{park2017droplet}. We further validated the freezing component of the model by reproducing the benchmark case of \citet{zadravzil2006droplet}. Appendix \ref{A2}, figure \ref{fig:Zad_val_A2}, illustrates the temporal evolution of the droplet interface, $h$ (solid line), and the freezing front, $s$ (dot–dashed line), showing good agreement with figure 15 of \citet{zadravzil2006droplet}. All simulations are performed on a one-dimensional mesh along the $x$-direction, spanning a domain of thirty-two dimensionless units, discretized using 25601 grid points. Extensive domain-independence and grid-convergence tests were conducted to optimize the domain size and grid resolution, taking into account computational time. Appendix \ref{A3} reports the grid-convergence results (figure \ref{fig:Grid_study}), indicating that grids with 12001, 25601, and 30001 points produce indistinguishable solutions. Additional details can be found in \cite{kavuri2023freezing, wang2024role, williams2021}. Below, we describe the procedure used to determine the contact angles at both the substrate and the liquid–ice interface.

\subsection{Calculation of contact angles} \label{sec:CA}

The apparent advancing ($\theta_a$) and receding ($\theta_r$) contact angles are obtained by locating the points of maximum slope on the droplet profile and drawing tangents from these points to the substrate before freezing begins ($t/t_f = 0$), where $t_f$ denotes the total freezing time. As shown in figure \ref{fig:demo}(a), the point of maximum slope on the receding side is marked as A, and that on the advancing side as B. The enlarged insets next to panel (a) highlight these points and illustrate how the corresponding contact angles are measured. After freezing is complete ($t/t_f = 1$), the horizontal positions of points A and B are retained and relabeled as C and D, representing the receding and advancing sides, respectively. Since the locations of maximum slope after freezing are no longer close to the contact line, the final contact angles are evaluated at the same horizontal positions identified prior to freezing. The insets in figure~\ref{fig:demo}(b) illustrate the procedure used to determine these angles. As noted earlier, we employ a penalty function (eq. \ref{pre_model1}) to preserve the precursor layer thickness far from the droplet and to maintain accuracy in predicting the freezing process. This method, however, limits the freezing rate when the liquid layer becomes comparable to the precursor thickness. As a result, the model becomes less accurate near the end of solidification and is unable to resolve the formation of the sharp tip. Recognizing this limitation and noting that the tip forms only at the very final freezing stage, we estimate the tip angle using the local geometry of the ice–gas interface near the tip following \citet{kavuri2023freezing}. As shown in figure \ref{fig:demo}(b), tangents are drawn near the tip (labeled as E) and the corresponding inset illustrates this construction. For the set of parameters considered in figure \ref{fig:demo}, the resulting tip angle is approximately $137^{\circ}$, which closely matches the experimental value reported by \citet{marin2014universality}.

The contact angle at the liquid–ice interface ($\theta_{ice}$) is calculated using a procedure analogous to that described above. The apparent advancing ($\theta_{ice,a}$) and receding ($\theta_{ice,r}$) angles are obtained by identifying the points of maximum slope within the liquid layer above the ice and drawing tangents from these points to the ice surface. As shown in the central panel of figure \ref{fig:demo_l_i}, the points of maximum slope on the receding and advancing sides are labeled A and B, respectively. The corresponding magnified insets highlight these locations and illustrate the procedure used to determine the receding and advancing contact angles. Additionally, to clearly illustrate the advancing and receding contact angles, both the contact length (diameter) and the drop height are presented using consistent scaling in figures 2 and 3. Accordingly, we plot $\epsilon h$ versus $x$ rather than $h$ versus $x$, since the contact length and drop height are scaled differently, $(x, z, h) = l (\tilde{x}, \epsilon \tilde{z}, \epsilon \tilde{h})$ (see, eq. \ref{all_scales}).

\section{Results and discussion} \label{sec:dis}

We begin by examining the asymmetric shape of the droplet and analyzing the contact angles at the two interfaces, where $\theta$ denotes the droplet–substrate contact angle and $\theta_{\text{ice}}$ represents the liquid–ice interface contact angle for a droplet placed on an inclined surface undergoing freezing. For comparison, the freezing dynamics of a droplet on a horizontal substrate are also presented. In all simulations, the droplet is first allowed to attain its equilibrium shape prior to the onset of freezing. This is achieved by suppressing freezing by setting the Stefan number, $Ste = 0$. Once the droplet reaches a steady shape while moving on the substrate, indicating equilibrium, its geometry and the pressure distribution along the liquid–gas interface are extracted and used as initial conditions for the subsequent simulations in which freezing is activated. The dimensionless parameters used in the freezing simulations are $Ste = 1.49 \times 10^{-3}$, $Bo = 0.4$, $A_{n} = 10$, $D_{s} = 0.9$, $\Lambda_{S} = 3.89$, $\Lambda_{W} = 675$, $Bi = 0.16$, $D_{w} = 2.94$, and $\epsilon = 0.2$, with their definitions summarized in Table~\ref{T:dim_groups}. This set of parameters corresponds to a water droplet exhibiting a contact angle of $17.5^\circ$ (eq.~\ref{eq:CA}) on a horizontal cold substrate maintained at $T_c=-30^\circ$. It is important to note that, although \citet{huerre2025freezing,knight1967contact,sarlin2025macroscopic,thievenaz2020retraction,demmenie2023growth,demmenie2025partial} report that the contact angle of water on ice near the melting temperature is approximately $12^\circ$, the liquid layer above the ice does not necessarily exhibit this equilibrium value once the recalescence phase has occurred. While the freezing front itself remains at the melting temperature, the unfrozen liquid layer can display larger apparent contact angles than $12^\circ$, as observed experimentally by \citet{marin2014universality,zhang2016freezing,zhang2019shape,jung2012frost,wang2022new}. In the discussion that follows, we will investigate a wide range of contact angles and examine its effect on the freezing process. For completeness, we present droplet shapes for $\theta_{eq} = 12^\circ$ and compare them with $\theta_{eq} = 17.5^\circ$ in figure~\ref{fig:Theta_eq_var_profile} of Appendix~\ref{sec:the_eq_alpha_75}.

\begin{figure}
\centering
\includegraphics[width=0.95\textwidth]{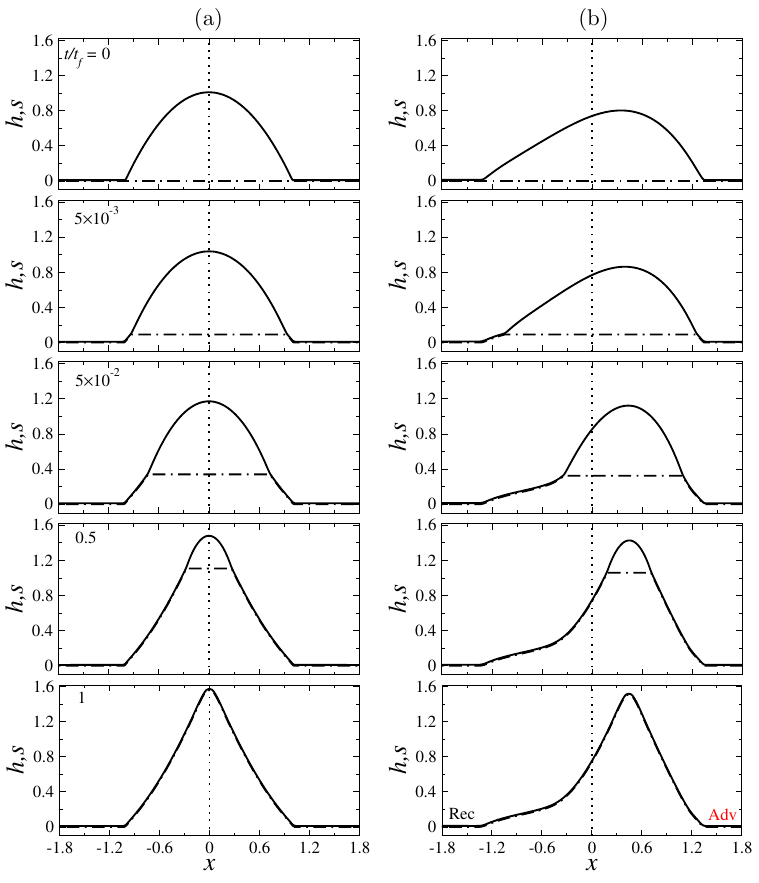} 
\caption{Evolution of the droplet shape, $h$ (solid lines), and the freezing front, $s$ (dot-dashed lines), on a substrate with (a) $\alpha = 0^\circ$ and (b) $\alpha = 75^\circ$. The normalized time, $t/t_f$, corresponding to each row is indicated in panel (a). Here, the equilibrium contact angle is fixed at $\theta_{eq} = 17.5^\circ$, with all other parameters listed in Table~\ref{T:dim_groups}. The corresponding total freezing time ($t_f$) is reported in Table~\ref{tab:tf_theta}.}
\label{fig:alpha_profile}
\end{figure}

\begin{figure}
\centering
\hspace{0.5cm}{\large (a)} \hspace{5.5cm}  {\large (b)} \\
\includegraphics[width=0.45\textwidth]{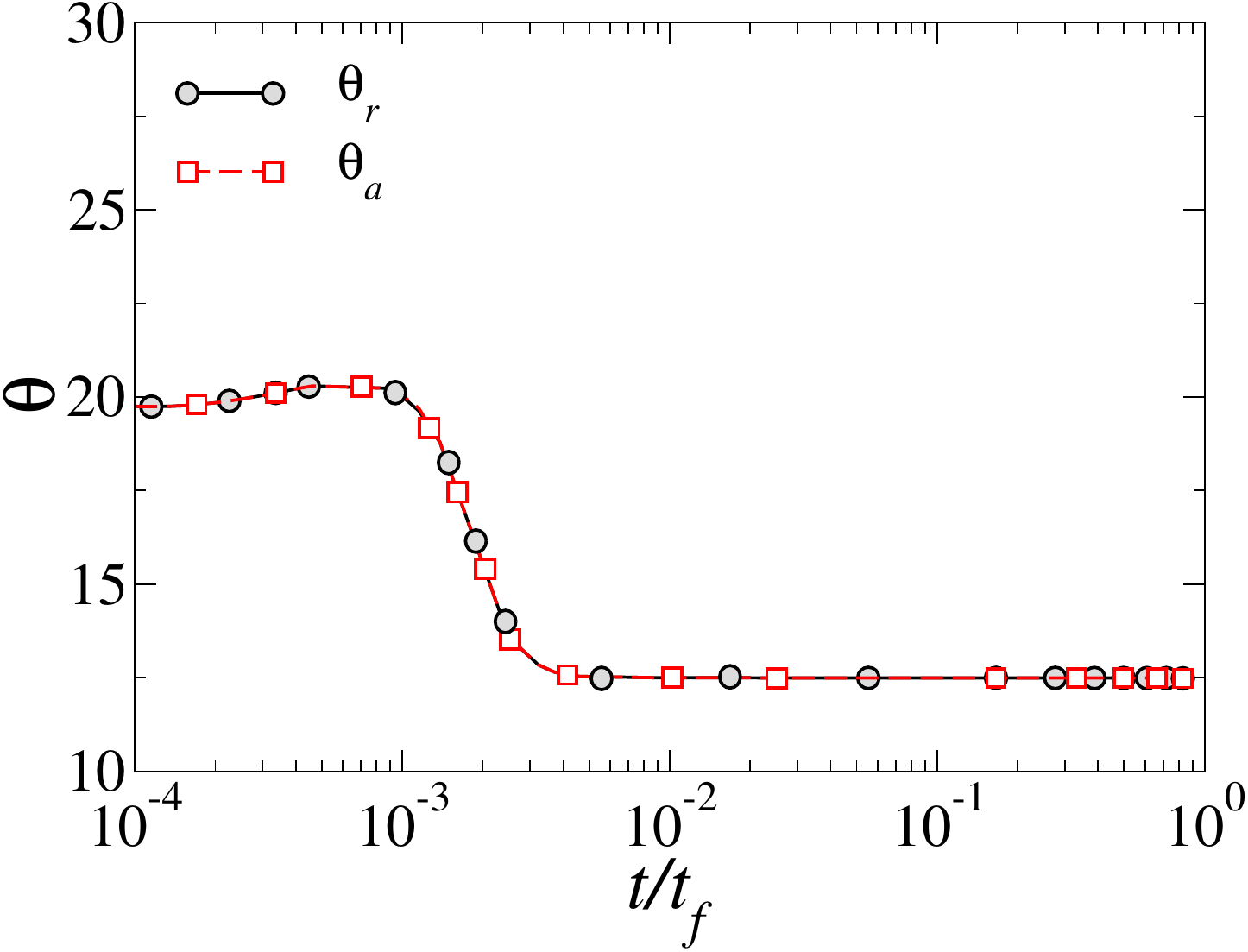} 
\includegraphics[width=0.45\textwidth]{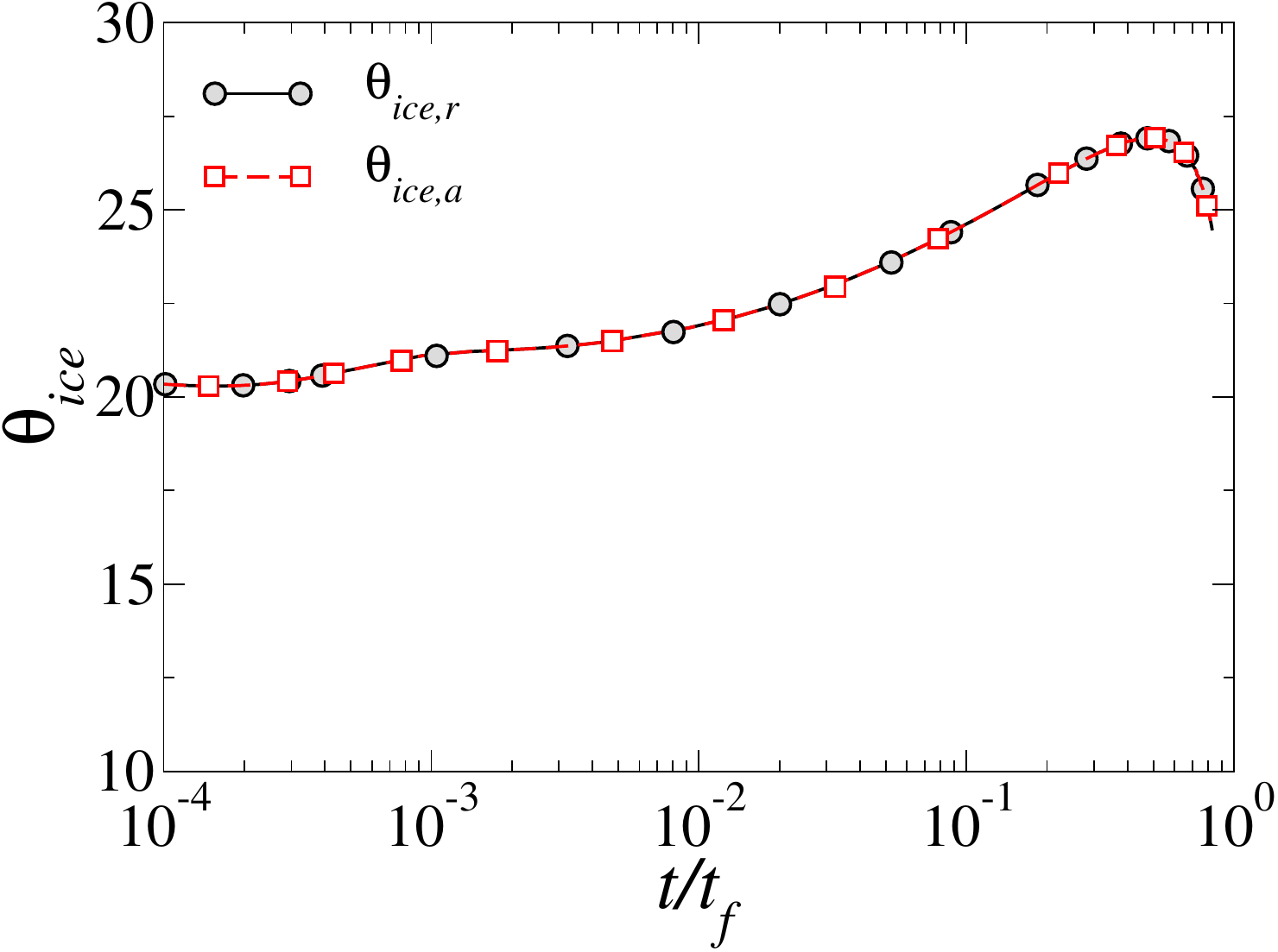} \\
\hspace{0.5cm}{\large (c)}   \hspace{5.5cm}  {\large (d)} \\
\includegraphics[width=0.45\textwidth]{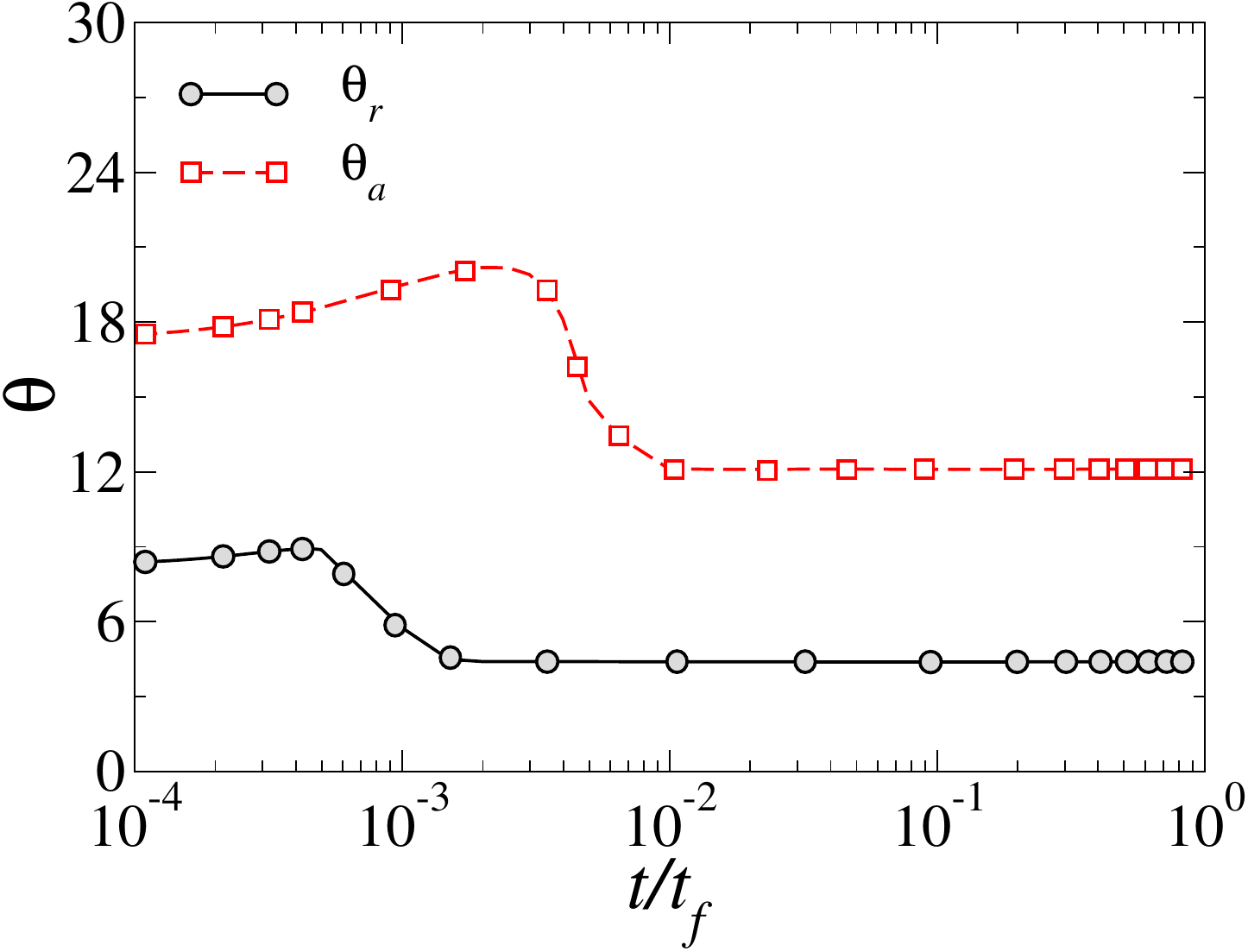}
\includegraphics[width=0.46\textwidth]{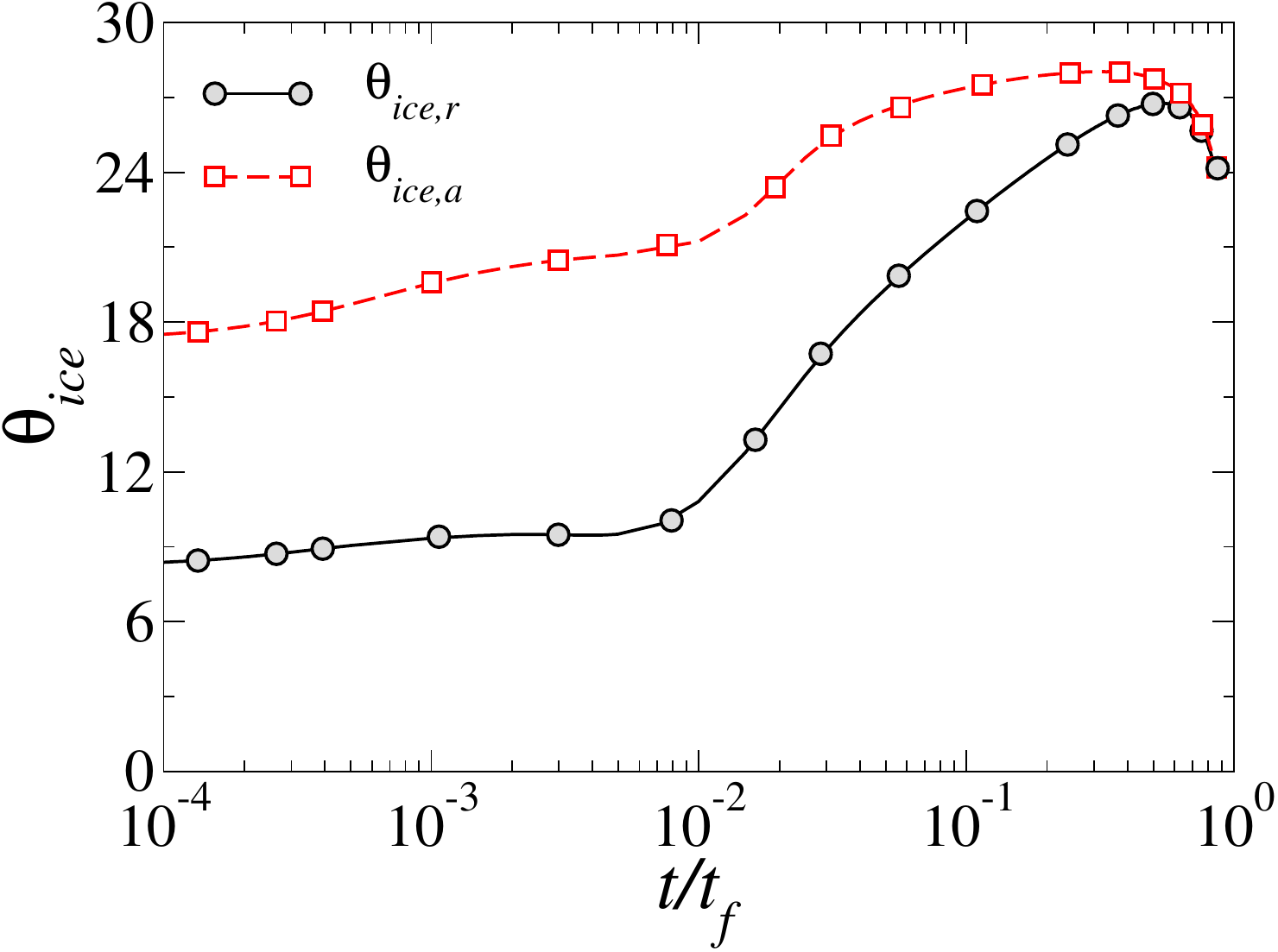}\\
\caption{Temporal evolution of (a, c) the droplet–substrate contact angle ($\theta$) and (b,d) the liquid–ice interface contact angle ($\theta_{ice}$). Panels (a,b) correspond to a horizontal substrate ($\alpha = 0^\circ$), while panels (c,d) correspond to an inclined substrate ($\alpha = 75^\circ$). In each panel, the contact angles on the advancing ($\theta_a$) and receding ($\theta_r$) sides are depicted. The equilibrium contact angle is fixed at $\theta_{eq} = 17.5^\circ$, with all other dimensionless parameters listed in Table~\ref{T:dim_groups}. The corresponding total freezing time ($t_f$) is reported in Table~\ref{tab:tf_theta}.}
\label{fig:theta_ice_profile}
\end{figure}

The temporal evolution of the droplet shape ($h$) and the freezing front ($s$) for substrates with $\alpha = 0^\circ$ and $\alpha = 75^\circ$ is shown in figure~\ref{fig:alpha_profile}(a) and \ref{fig:alpha_profile}(b), respectively. It can be seen that on a horizontal substrate (figure~\ref{fig:alpha_profile}a), the droplet solidifies symmetrically with respect to the normal (dotted line) passing through its centre. In contrast, on an inclined substrate with $\alpha = 75^\circ$ (figure~\ref{fig:alpha_profile}b), the initial droplet shape at $t/t_f = 0$ is asymmetric. This asymmetry persists during freezing, resulting in an asymmetrically solidified droplet, with the cusp formed at complete solidification ($t/t_f = 1$) oriented toward the advancing side. Moreover, the final frozen morphology also deviates significantly from the initial droplet shape. 

Figures~\ref{fig:theta_ice_profile}(a,b) and \ref{fig:theta_ice_profile}(c,d) show the temporal evolution of the droplet–substrate contact angle ($\theta$) and the liquid–ice interface contact angle ($\theta_{\text{ice}}$) for the same set of parameters used in figure~\ref{fig:alpha_profile}. For a droplet freezing on a horizontal substrate, the substrate contact angle ($\theta$) remains nearly constant until the contact line freezes at $t/t_f = 10^{-3}$. Beyond this point, $\theta$ decreases and then remains constant until solidification is complete, as shown in figure~\ref{fig:theta_ice_profile}(a). The liquid–ice interface angle ($\theta_{ice}$), however, continues to increase until $t/t_{f} = 0.65$, by which time the freezing front has already approached the droplet apex (figure~\ref{fig:theta_ice_profile}b). It can be seen that on a horizontal substrate, both $\theta$ and $\theta_{\text{ice}}$ are symmetric on either side of the droplet and overlap, as evident in figures~\ref{fig:theta_ice_profile}(a) and \ref{fig:theta_ice_profile}(b). In contrast, figure~\ref{fig:theta_ice_profile}(c,d) illustrates that for the inclined substrate, both the advancing ($\theta_a$) and receding ($\theta_r$) contact angles initially increase slightly before dropping sharply once the contact line freezes. Throughout the freezing process, the advancing side consistently maintains a higher contact angle than the receding side (figure~\ref{fig:theta_ice_profile}c). The earlier decrease in $\theta_r$ compared to $\theta_a$ arises from the difference in height between the points of maximum slope on the advancing and receding sides of the droplet, where the contact angles are determined by drawing a tangent at the point of maximum slope, as described in \S\ref{sec:CA}. Initially, the contact angles at the liquid–ice interface on both sides ($\theta_{ice,a}$ and $\theta_{ice,r}$) match their respective substrate contact angles and increase as freezing progresses, converging near the end of the process (figure~\ref{fig:theta_ice_profile}d). This convergence indicates a loss of asymmetry between the advancing and receding sides, which explains why, as reported by \citet{starostin2022universality,starostin2023effects}, the tip angle of frozen droplets remains nearly the same, around $143^\circ \pm 11^\circ$, on both horizontal and inclined substrates. This behaviour is expected since the effective Bond number ($Bo_m = Bo ~\sin\alpha$) for the remaining liquid towards the end of freezing becomes small, thus diminishing the effect of gravity on the liquid profile near the tip just before completion of freezing. With gravity playing a negligible role, the contact angles at the liquid–ice interface converge, leaving the tip angle largely independent of the inclination angle ($\alpha$). 

The equilibrium contact angle ($\theta_{eq}$) estimated from eq.~(2.34) differs slightly from the apparent contact angles obtained by drawing a tangent at the point of maximum slope, as described in \S~2.6 of the revised manuscript. This is illustrated in figure~\ref{fig:theta_ice_profile}(a) and figure \ref{fig:theta_ice_profile}(c), where eq.~(2.34) gives $\theta_{eq} = 17.5^\circ$, while the contact angle on the horizontal substrate, determined using the procedure in \S~2.6, is approximately $20^\circ$. For the inclined substrate ($\alpha = 75^\circ$), the average of the advancing and receding angles, $(\theta_a + \theta_r)/2$, computed using the same method, is about $13.5^\circ$. The discrepancy arises from the absence of a sharply defined contact line. The apparent contact angle is therefore inferred rather than directly measured, introducing inherent ambiguity. Throughout this study, we consistently use the method in \S~2.6 to estimate contact angles on both the substrate and the liquid–ice interface. In the following, we present a mechanistic framework to explain the observed asymmetry during droplet freezing by performing a large set of simulations in which the dimensionless parameters listed in Table~\ref{T:dim_groups} are systematically varied. For convenience, Tables~\ref{tab:tf_theta}-\ref{tab:Ste_tf} summarize the total dimensionless freezing time ($t_f$) for different values of the inclination angles $(\alpha)$, equilibrium contact angles $(\theta_{eq})$, Bond number $(Bo)$ and Stefan number $(Ste)$.

\begin{table}
\centering
\caption{Total dimensionless freezing time ($t_f$) for different values of the inclination angle $(\alpha)$ and equilibrium contact angle $(\theta_{eq})$. Here, $Bo = 0.4$ and the remaining dimensionless parameters are listed in Table~\ref{T:dim_groups}.}
\label{tab:tf_theta}
\setlength{\tabcolsep}{8pt}
\renewcommand{\arraystretch}{1.2}
\begin{tabular}{c c c c c c c c}
$\theta_{eq}$ (deg) & $\alpha=0^\circ$ & $\alpha=15^\circ$ & $\alpha=30^\circ$ & $\alpha=45^\circ$ & $\alpha=60^\circ$ & $\alpha=75^\circ$ \\
17.5 & 217 & 202 & 203 & 204 & 203 & 201 \\
21.2 & 258 & 256 & 258 & 261 & 263 & 265 \\
26.5 & 337 & 344 & 347 & 349 & 352 & 355 \\
30.6 & 412 & 422 & 424 & 427 & 430 & 433 \\
\end{tabular}
\end{table}

\begin{table}
\centering
\caption{Total dimensionless freezing time ($t_f$) for different values of the inclination angle $(\alpha)$ and Bond number $(Bo)$ at $\theta_{eq} = 26.5^\circ$. The remaining dimensionless parameters are listed in Table~\ref{T:dim_groups}.}
\label{tab:Bo_tf}
\setlength{\tabcolsep}{8pt}
\renewcommand{\arraystretch}{1.2}
\begin{tabular}{c c c c c c}
$\alpha$ (deg) & $Bo=0.2$ & $Bo=0.5$ & $Bo=0.95$ & $Bo=1$ & $Bo=1.25$ \\
15 & 346 & 343 & 340 & 339 & 338 \\
45 & 349 & 350 & 356 & 356 & 357 \\
75 & 350 & 356 & 366 & 365 & 352 
\end{tabular}
\end{table}

\begin{table}
\centering
\caption{Total dimensionless freezing time ($t_f$) for different values of the equilibrium contact angle $(\theta_{eq})$ at $\alpha=60^\circ$. The remaining dimensionless parameters are listed in Table~\ref{T:dim_groups}.}
\label{tab:tf_alpha60_compact}
\setlength{\tabcolsep}{8pt}
\renewcommand{\arraystretch}{1.2}
\begin{tabular}{c c c c c c c c}
$\theta_{eq}$ (deg) & $15.3^\circ$ & $17.5^\circ$ & $21.2^\circ$ & $24.1^\circ$ & $26.5^\circ$ & $28.7^\circ$ & $30.6^\circ$ \\
$t_f$ & 136 & 203 & 263 & 310 & 352 & 392 & 430 \\
\end{tabular}
\end{table}

\begin{table}
\centering
\caption{Total dimensionless freezing time ($t_f$) for different values of the the inclination angle $(\alpha)$ and Stefan number $(Ste)$ at $\theta_{eq}=26.5^\circ$. The remaining dimensionless parameters are listed in Table~\ref{T:dim_groups}.}
\label{tab:Ste_tf}
\setlength{\tabcolsep}{10pt}
\renewcommand{\arraystretch}{1.2}
\begin{tabular}{c c c c}
$\alpha$ (deg) & $Ste = 1.49\times10^{-3}$ & $Ste = 0.01$ & $Ste = 0.1$ \\
15 & 344 & 52 & 5.0 \\
45 & 349 & 52 & 5.0 \\
75 & 355 & 53 & 5.0 \\
\end{tabular}
\end{table}

\subsection{Effect of the sliding force}\label{Bo_var}

\begin{figure}
\centering
\hspace{0.5cm}{\large (a)}   \hspace{5.5cm}  {\large (b)} \\
\includegraphics[width=0.45\textwidth]{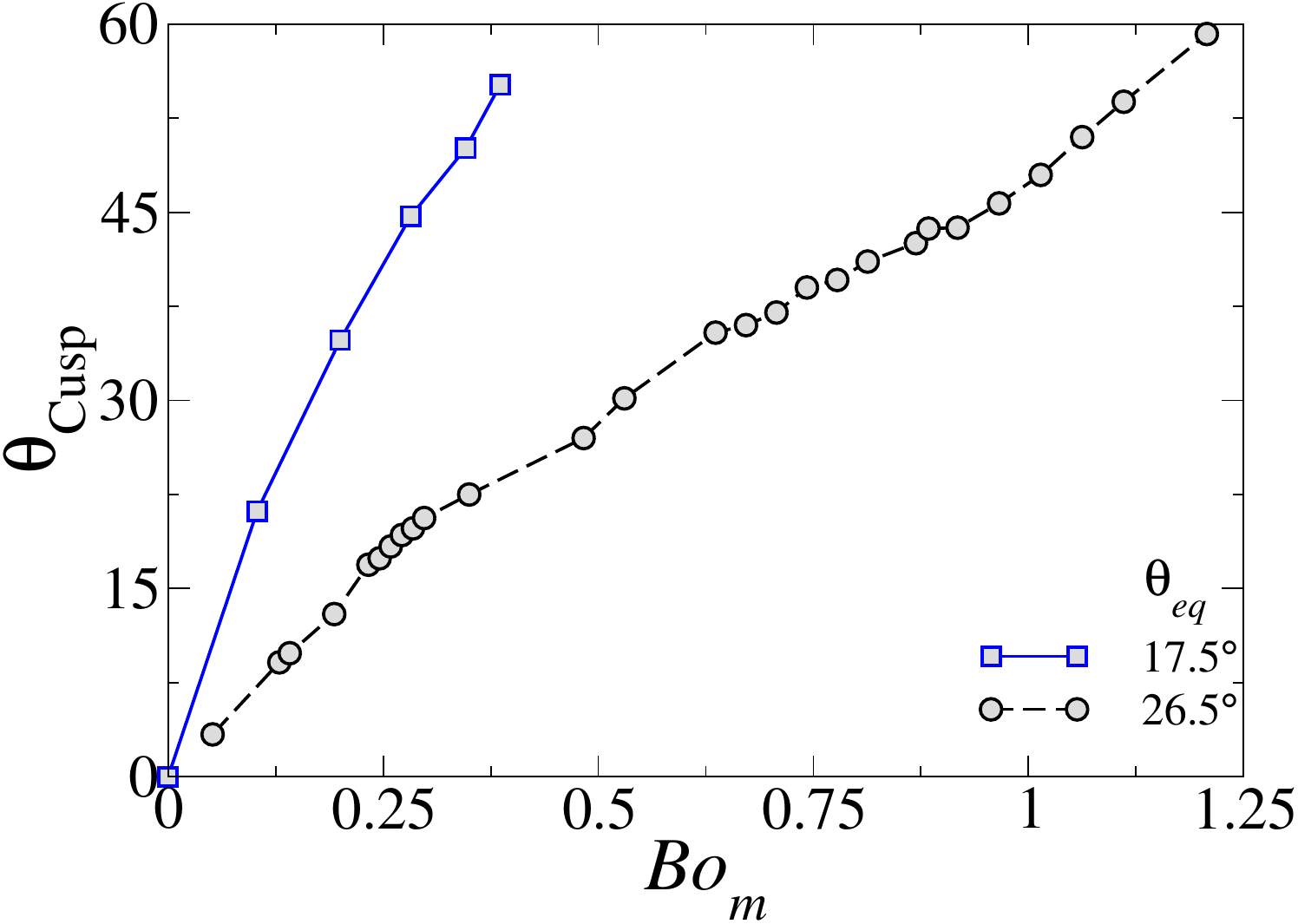} \includegraphics[width=0.45\textwidth]{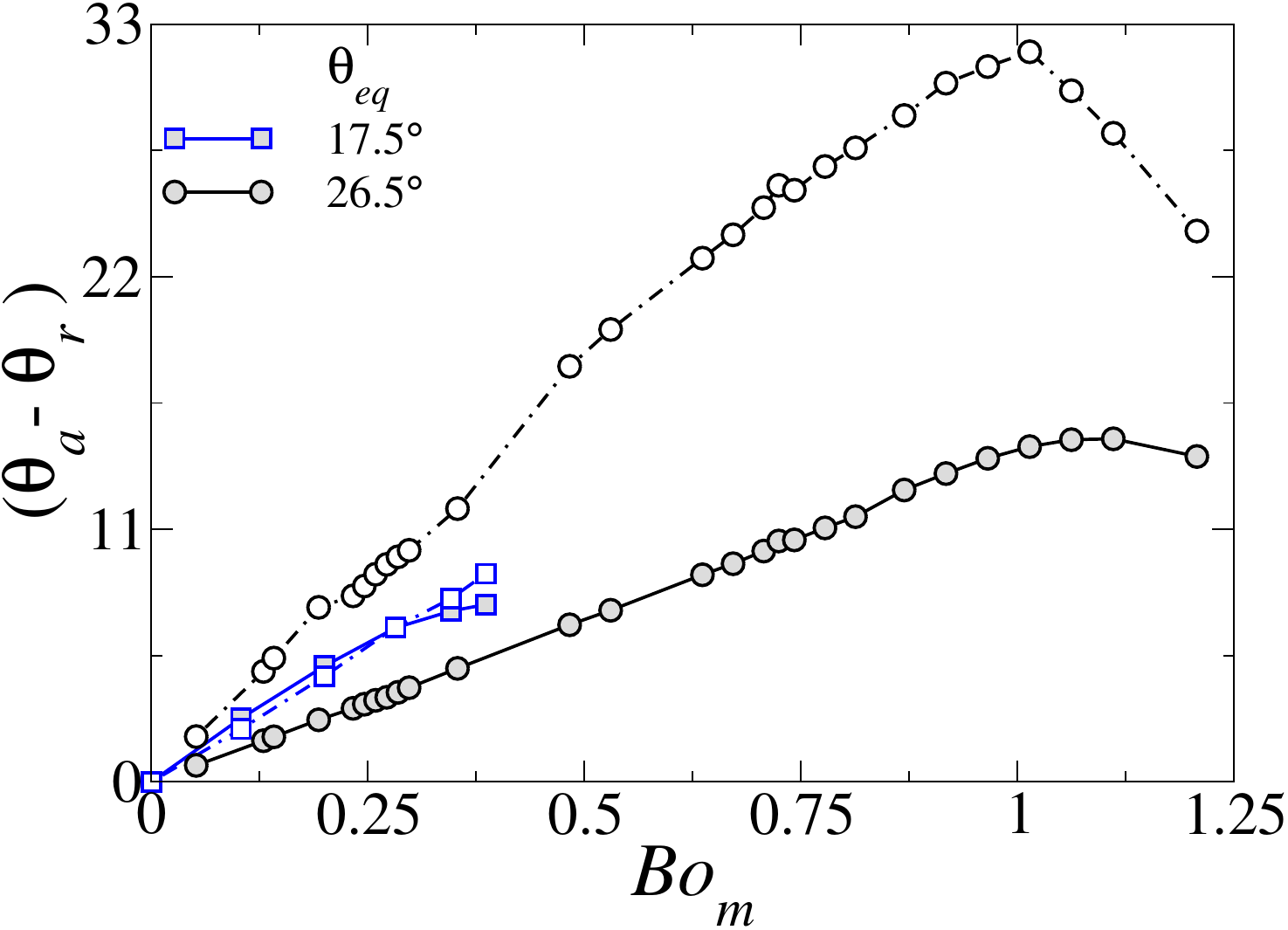}\\
\caption{Effect of the effective Bond number ($Bo_m = Bo ~\sin\alpha$) on (a) the cusp angle $\theta_{\rm Cusp}$ and (b) the droplet–substrate contact angle hysteresis $(\theta_a-\theta_r)$ at $t/t_f = 0$ (solid lines) and $t/t_f = 1$ (dot-dashed lines) for different equilibrium contact angles. The remaining dimensionless parameters are listed in Table~\ref{T:dim_groups}. The total freezing times ($t_f$) for different values of $\alpha$ and $Bo$ are provided in Tables~\ref{tab:tf_theta} and \ref{tab:Bo_tf}.}
\label{fig:Del_theta_BoSina_H}
\end{figure}

The freezing behaviour of a sliding droplet is strongly influenced by both the Bond number ($Bo$) and the inclination angle ($\alpha$), as they together determine the component of gravity acting tangentially along the inclined surface. To capture this combined effect, we employ the effective Bond number ($Bo_m = Bo ~ \sin\alpha$). This parameter represents the tangential gravitational contribution that drives droplet motion and provides a resultant measure of the coupled influence of gravity and inclination of the substrate on the freezing dynamics. The effect of $Bo_m$ on the cusp angle, $\theta_{\rm Cusp}$ (defined as the angle between the frozen tip formed at the end of solidification and the $z$-axis as shown in figure~\ref{fig:geom}) and the contact angle hysteresis ($\theta_a-\theta_r$) measured on the droplet–substrate interface at $t/t_f = 0$ and $t/t_f = 1$ for different values of the equilibrium contact angle $(\theta_{eq})$ are illustrated in figure~\ref{fig:Del_theta_BoSina_H}(a,b). 

\begin{figure}
\centering
\includegraphics[width=0.99\textwidth]{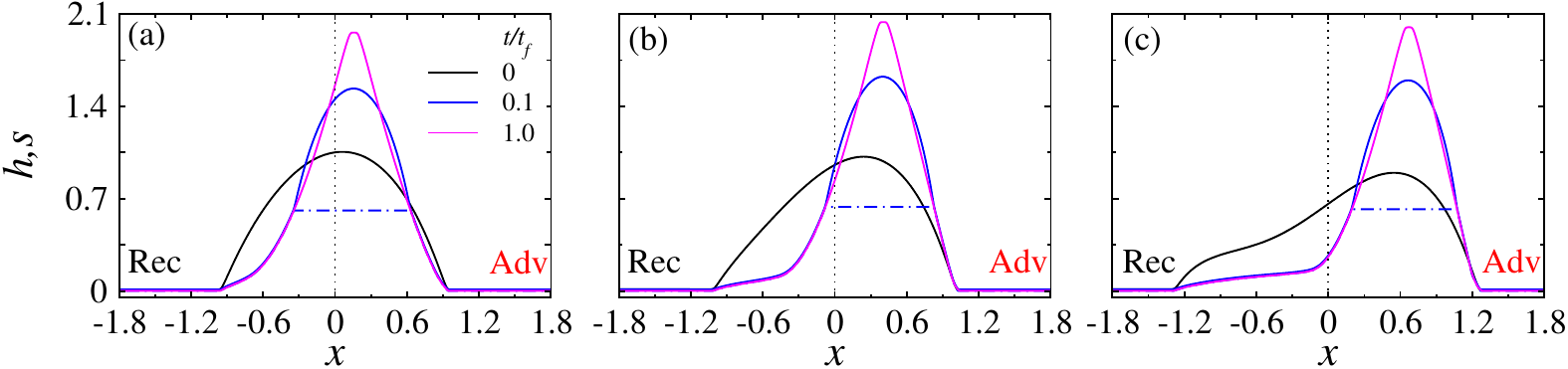}
\caption{Droplet shapes at different normalized times for various effective Bond numbers $Bo_m ~(= Bo ~\sin\alpha)$: (a) $Bo_m = 0.19$, (b) $Bo_m = 0.92$, and (c) $Bo_m = 1.2$. At $t/t_f = 1.0$, the freezing front coincides with the droplet surface. The equilibrium contact angle is $\theta_{eq} = 26.5^\circ$. The remaining dimensionless parameters and the corresponding total freezing time ($t_f$) are provided in Tables~\ref{T:dim_groups} and~\ref{tab:Bo_tf}, respectively.} 
\label{fig:Del_theta_Bo_H}
\end{figure}

The substrate wettability is characterized by $\theta_{eq}$, with higher $\theta_{eq}$ corresponding to lower wettability. It can be seen from figure~\ref{fig:Del_theta_BoSina_H}(a) that the cusp angle of the droplet increases with $Bo_m$ for both values of $\theta_{eq}$ considered. This behaviour can be attributed to the continuous sliding of the liquid droplet and the increasing asymmetry at higher inclination angles, which leads to a larger deviation of the frozen tip from the normal (i.e., a higher $\theta_{\rm Cusp}$). This trend is observed for both $\theta_{eq}=17.5^\circ$ and $\theta_{eq}=26.5^\circ$. For $\theta_{eq}=26.5^\circ$, the droplet–substrate contact angle hysteresis ($\theta_a-\theta_r$), measured before freezing ($t/t_f=0$) and after complete solidification ($t/t_f=1$), exhibits a non-monotonic dependence on $Bo_m$, as shown in figure~\ref{fig:Del_theta_BoSina_H}(b). This non-monotonic dependence is attributed to the long tail of the droplet that forms and eventually freezes when the sliding force is high. Figure~\ref{fig:Del_theta_Bo_H}(a–c) shows the evolution of droplet shapes for effective Bond numbers $Bo_m = 0.19$, 0.92, and 1.2, respectively. For each $Bo_m$, the droplet profiles are presented at the onset of freezing ($t/t_f = 0$), at an intermediate stage ($t/t_f = 0.1$), and at the completion of freezing ($t/t_f = 1$). It can be seen that as $Bo_m$ increases from 0.92 to 1.2, the initial droplet shape becomes increasingly elongated, with pronounced swelling on the advancing side and tapering on the receding side. The stronger sliding force at higher $Bo_m$ enhances droplet elongation, leading to a reduction in the droplet–substrate contact angle hysteresis ($\theta_a-\theta_r$) at both the initial and final stages of freezing.

\begin{figure}
\centering
\hspace{0.5cm}{\large (a)}\\
\includegraphics[width=0.45\textwidth]{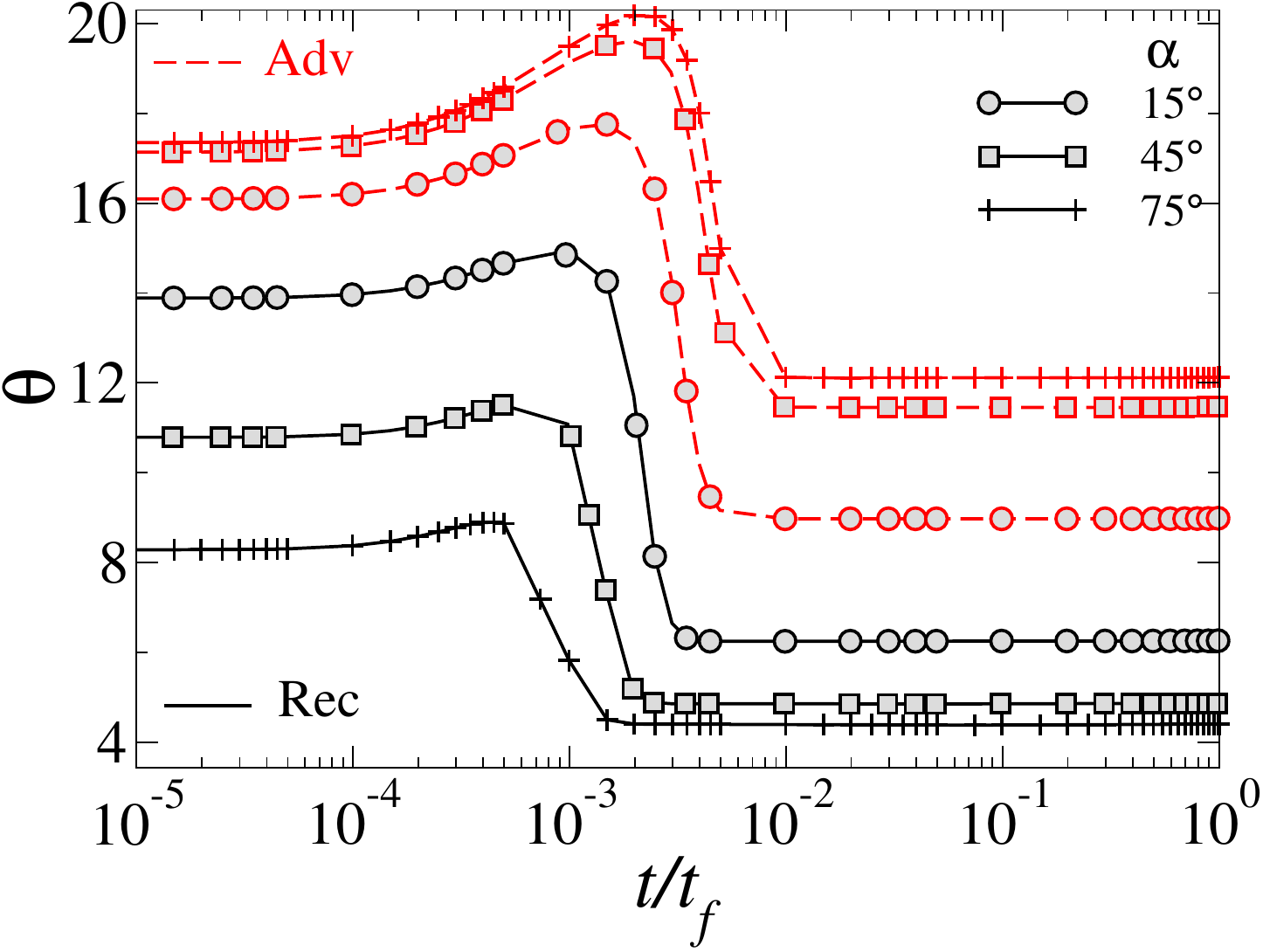} \\
\hspace{0.75cm}{\large (b)} \hspace{5.5cm}  {\large (c)} \\
\includegraphics[width=0.45\textwidth]{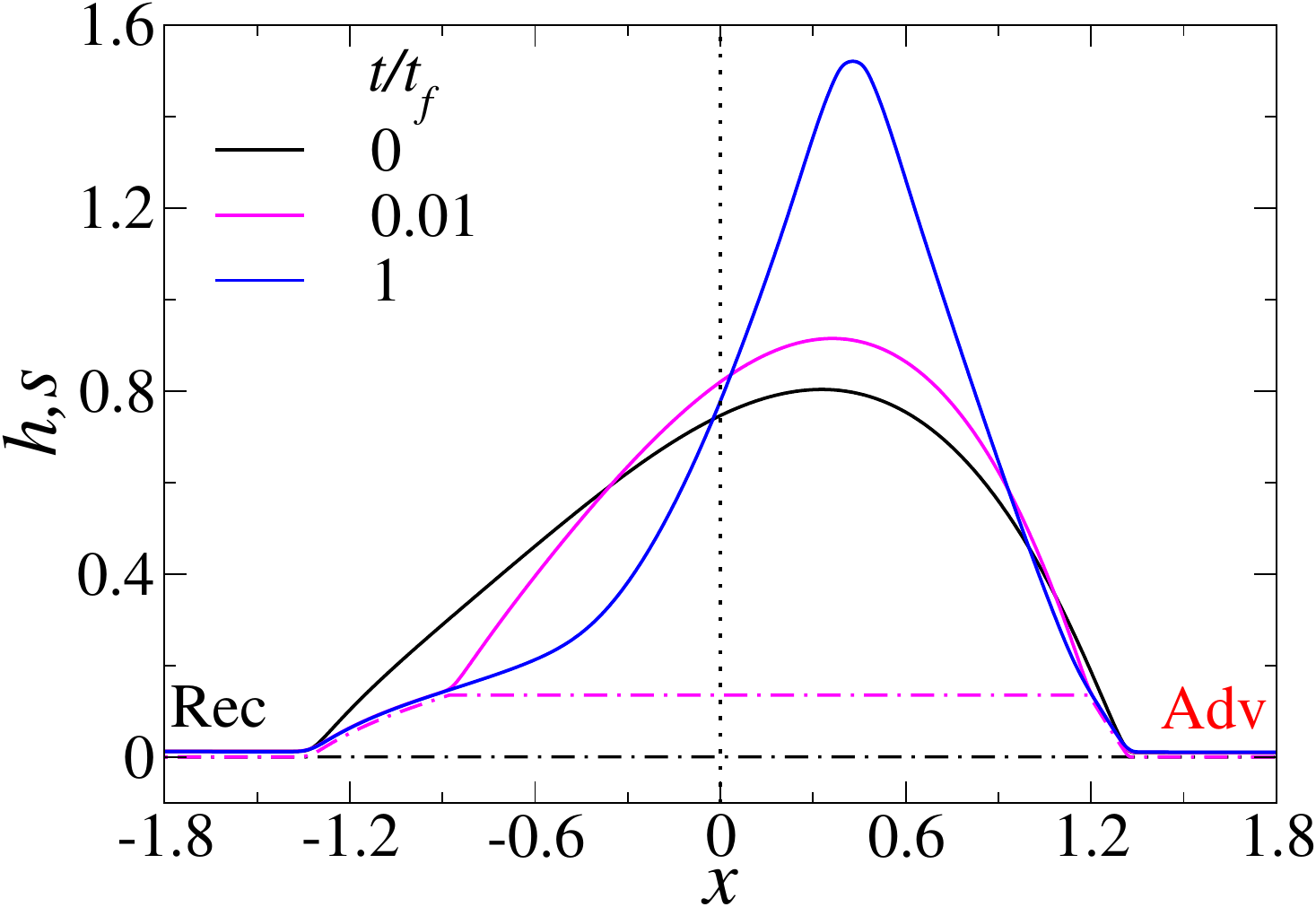}
\includegraphics[width=0.46\textwidth]{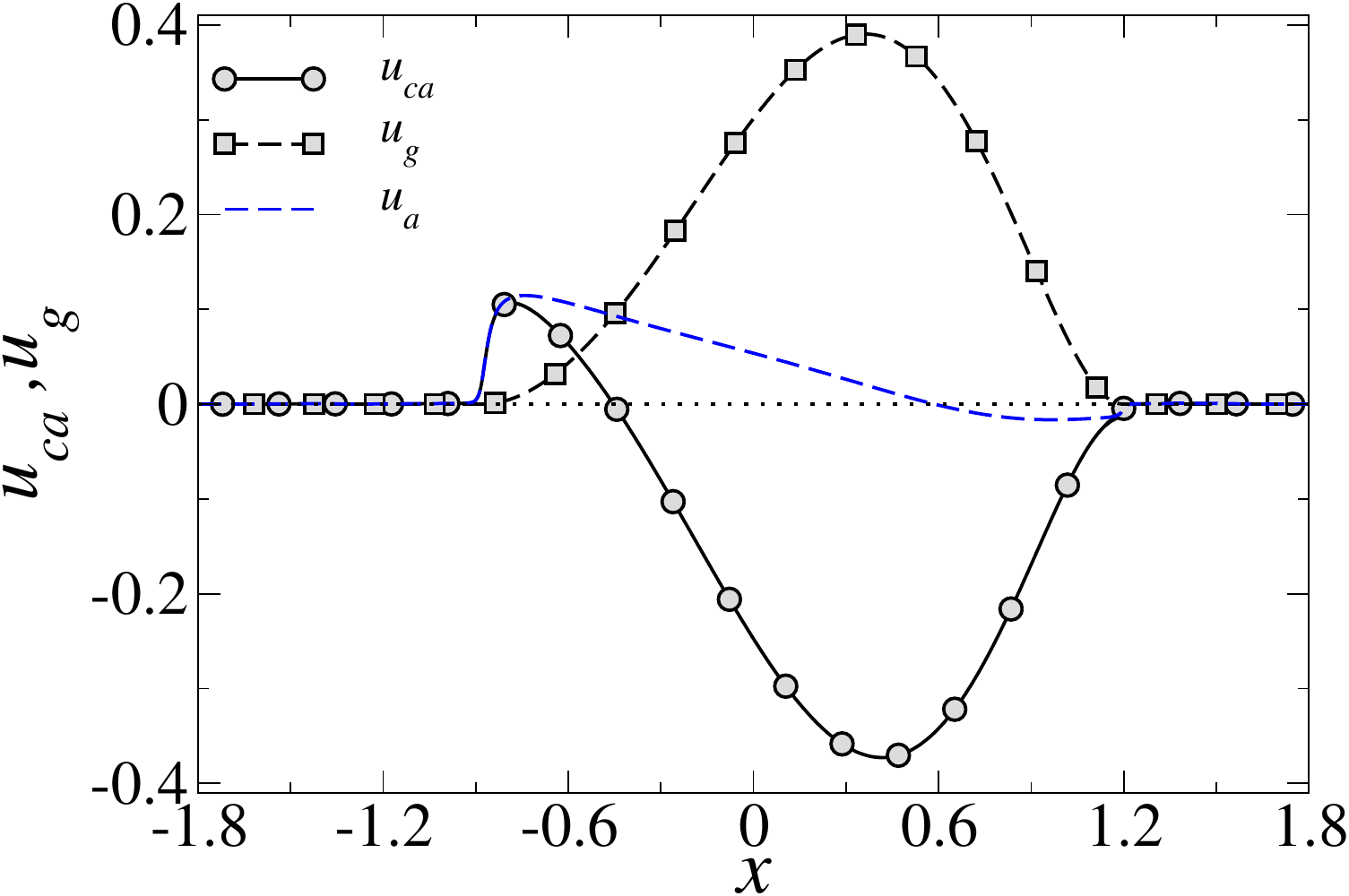}\\
\caption{Temporal evolution of (a) the advancing ($\theta_{a}$, red dashed lines) and receding ($\theta_{r}$, black solid lines) contact angles on the substrate; (b) the droplet shape, $h$ (solid lines), and the freezing front, $s$ (dot-dashed lines), for a droplet freezing on an inclined substrate with $\alpha = 75^\circ$; and (c) variations of the capillary velocity component ($u_{ca}$), gravitational velocity component ($u_{g}$), and average velocity ($u_{a}$) along the $x$-axis at a representative early freezing stage ($t/t_{f} = 0.01$) for a droplet on a substrate inclined at $\alpha = 75^\circ$, where the dotted line indicates $u = 0$. The equilibrium contact angle is $\theta_{eq} = 17.5^\circ$. The remaining dimensionless parameters and the corresponding total freezing time ($t_f$) are provided in Tables~\ref{T:dim_groups} and~\ref{tab:tf_theta}, respectively.}
\label{fig:theta_eq_17p5_alpha}
\end{figure}

Interestingly, for the substrate with the highest wettability ($\theta_{eq} = 17.5^\circ$), the pre-freezing droplet–substrate contact angle hysteresis ($(\theta_a - \theta_r)$ at $t/t_f = 0$) exceeds the post-freezing value ($(\theta_a - \theta_r)$ at $t/t_f = 1$) for $Bo_m > 0.28$, corresponding to an inclination angle $\alpha > 45^\circ$ for $Bo = 0.4$, as shown in figure~\ref{fig:Del_theta_BoSina_H}(b). To understand this behavior, in figure~\ref{fig:theta_eq_17p5_alpha}(a) we plot the temporal evolution of the advancing ($\theta_a$) and receding ($\theta_r$) contact angles for substrate inclinations of $\alpha = 15^\circ$, $45^\circ$, and $75^\circ$. It is observed that $\theta_a$ increases while $\theta_r$ decreases as the substrate inclination increases. However, the increase in $\theta_a$ from $\alpha = 45^\circ$ to $75^\circ$ is noticeably smaller than the increase observed when the inclination increases from $\alpha = 15^\circ$ to $45^\circ$. For the case of $\alpha = 75^\circ$, the evolution of the droplet shape ($h$) and the freezing front ($s$), shown in figure~\ref{fig:theta_eq_17p5_alpha}(b), indicates that the advancing contact line remains pinned throughout freezing, whereas the receding contact line progressively moves downstream. Consequently, the increase in $\theta_a$ from $\alpha = 45^\circ$ to $75^\circ$ in figure~\ref{fig:theta_eq_17p5_alpha}(a) is relatively small. To further elucidate this behaviour, we evaluate the vertically averaged velocity $(u_a(x))$ and the bulk average velocity $(\bar{u}_a)$ of the unfrozen liquid droplet using the following expressions:
\begin{equation}\label{ua_eq}
u_{a} (x) = \frac{\int_{s}^{h}udz}{\int_{s}^{h}dz} = -\frac{1}{3}\left(p_{x}-\frac{Bo ~ \sin\alpha}{\epsilon}\right)(h-s)^{2},
\end{equation}
\begin{equation}\label{ua_avg}
\bar{u}_{a} = \frac{\int_{x_{cl}}^{x_{cr}}\int_{s}^{h}udzdx}{\int_{x_{cl}}^{x_{cr}}\int_{s}^{h}dzdx} = -\frac{1}{3}\frac{\int_{x_{cl}}^{x_{cr}}\left(p_{x}-\frac{Bo ~ \sin\alpha}{\epsilon}\right)(h-s)^{3}dx}{\int_{x_{cl}}^{x_{cr}}(h-s)dx}.
\end{equation}
Here, $x_{cl}$ and $x_{cr}$ denote the left and right contact lines of the droplet (see figure~\ref{fig:geom}). To gain further insight, the velocities are analyzed by separating the contributions from capillary forces and gravitational forces. Within this framework, both the vertically averaged velocity $(u_a(x))$ and the center-of-mass velocity $(\bar{u}_a)$ of the unfrozen liquid droplet are decomposed into these two distinct components.

\begin{equation}\label{ua_eq_s}
\begin{aligned}
u_a(x) &= u_{ca}(x) + u_g(x) 
= -\frac{1}{3} p_x (h - s)^2 
+ \frac{1}{3}\left(\frac{Bo \sin\alpha}{\epsilon}\right)(h - s)^2, \\
u_{ca}(x) &= -\frac{1}{3} p_x (h - s)^2, \quad
u_g(x) = \frac{1}{3}\left(\frac{Bo \sin\alpha}{\epsilon}\right)(h - s)^2,
\end{aligned}
\end{equation}
\begin{equation}\label{ua_avg_s}
\bar{u}_a = \bar{u}_{ca} + \bar{u}_g,
\end{equation}
where
\begin{equation}
\bar{u}_{ca} = -\frac{1}{3} \frac{\int_{x_{cl}}^{x_{cr}} p_x (h-s)^3 \, dx} {\int_{x_{cl}}^{x_{cr}} (h-s) \, dx},
\quad
\bar{u}_g = \frac{1}{3} \frac{\int_{x_{cl}}^{x_{cr}} \frac{Bo \sin\alpha}{\epsilon} (h-s)^3 \, dx} {\int_{x_{cl}}^{x_{cr}} (h-s) \, dx}.
\end{equation}

\begin{figure}
\centering
\hspace{0.75cm}{\large (a)} \hspace{5.5cm}  {\large (b)} \\
\includegraphics[width=0.45\textwidth]{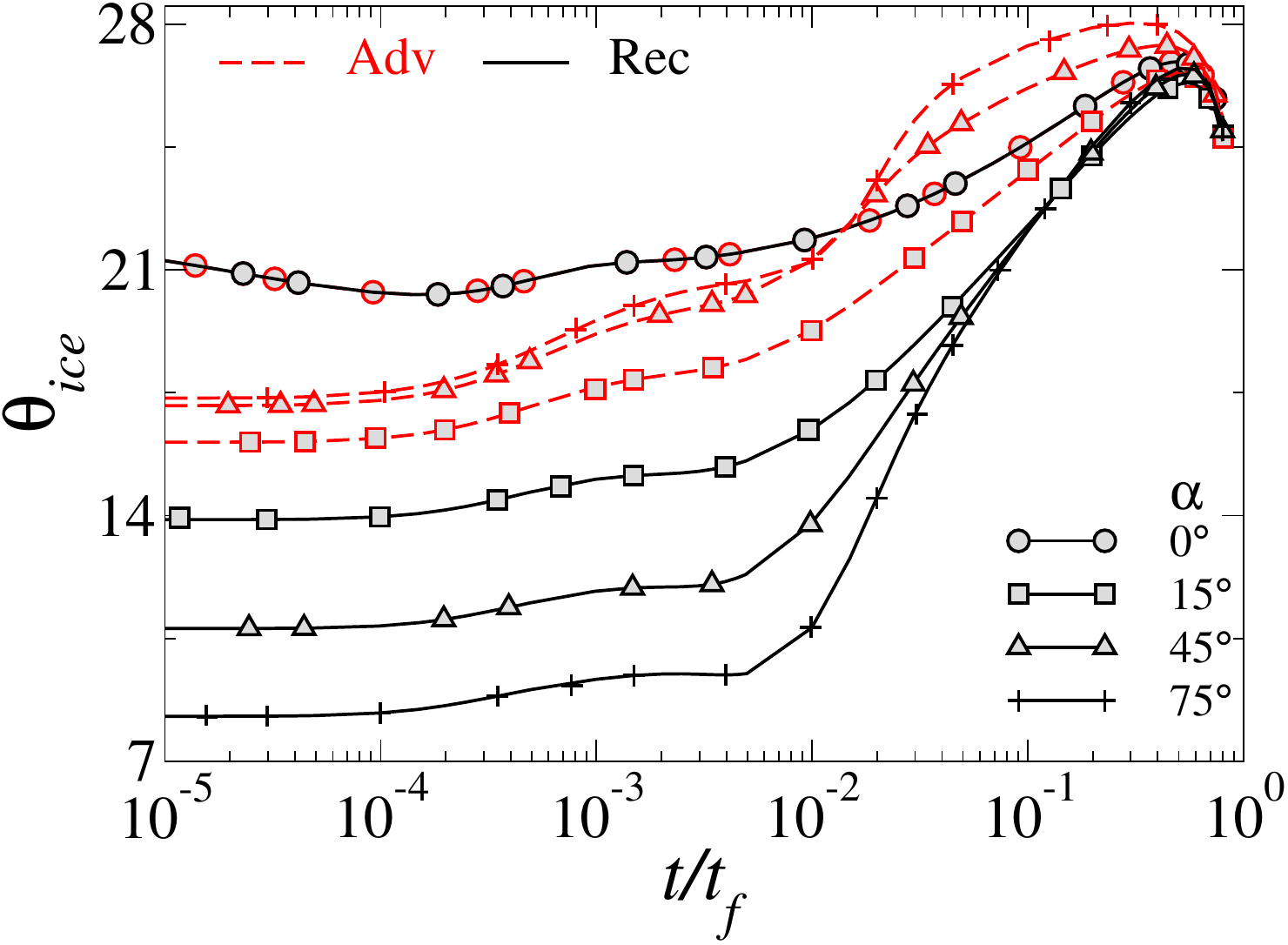} 
\hspace{0mm}
\includegraphics[width=0.45\textwidth]{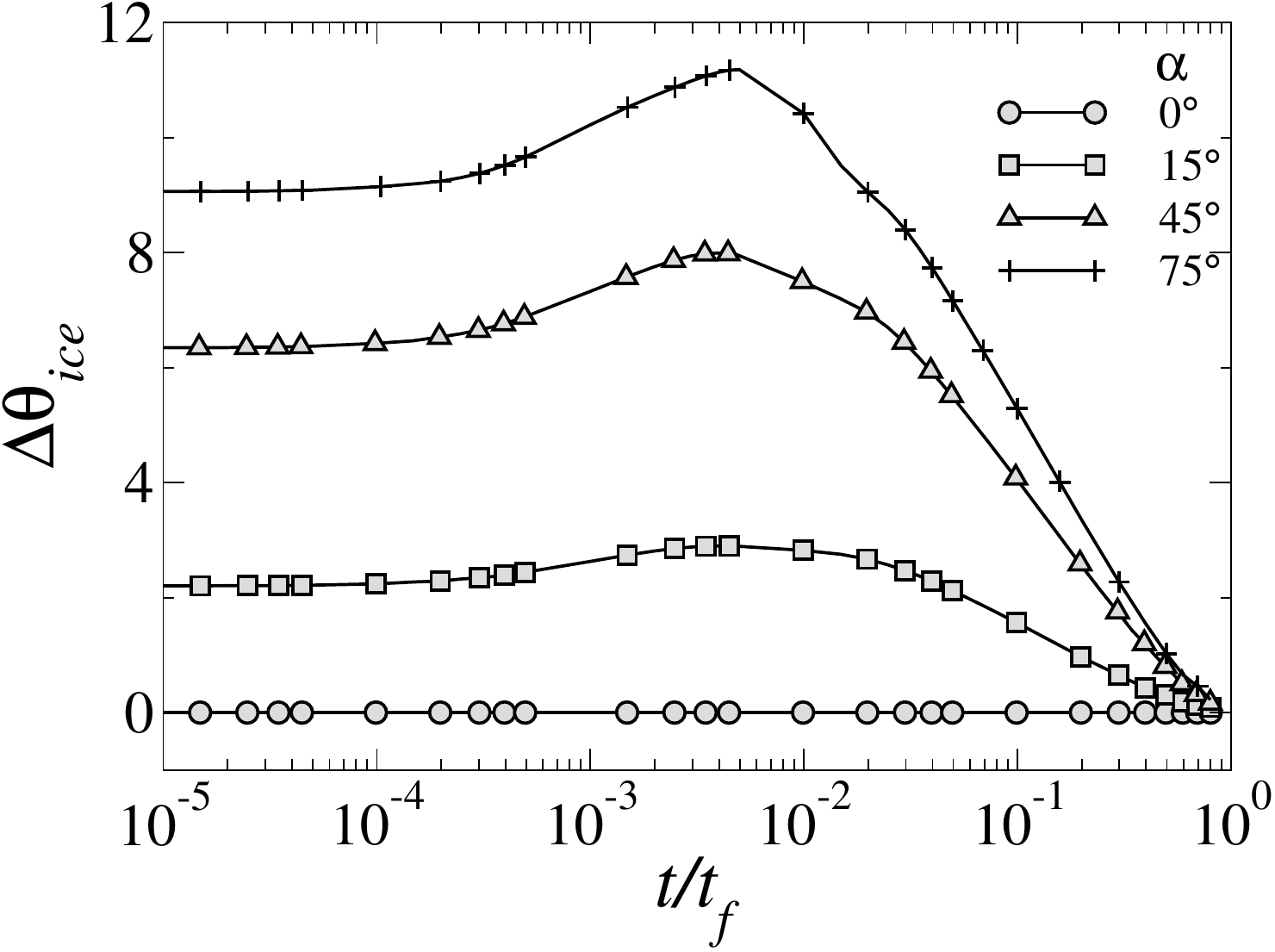}\\
\hspace{0.5cm}{\large (c)}\\
\includegraphics[width=0.45\textwidth]{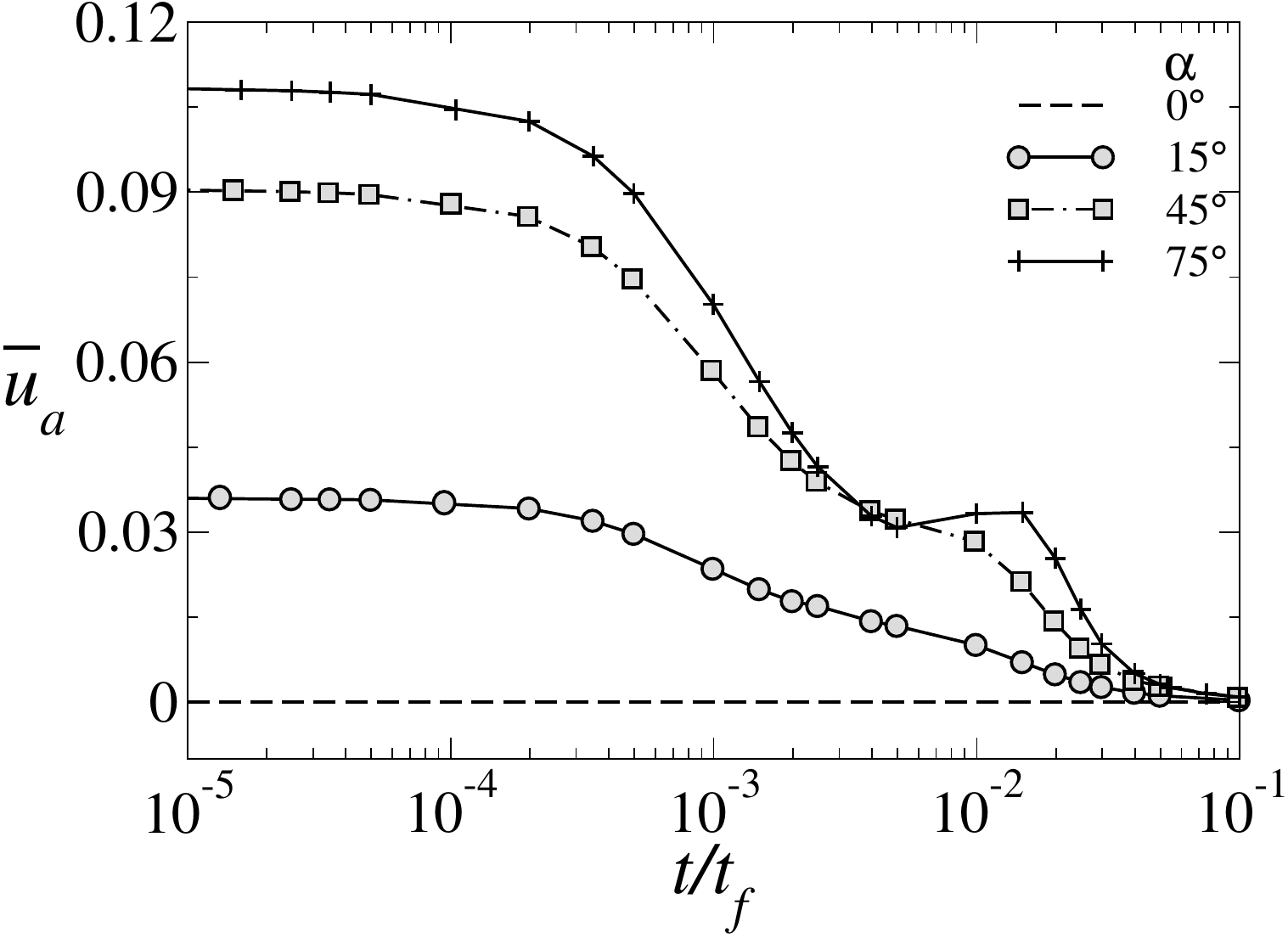}
\caption{Temporal evolution of (a) the advancing ($\theta_{ice,a}$, red dashed lines) and receding ($\theta_{ice,r}$, black solid lines) contact angles at the liquid–ice interface, (b) the difference between the advancing and receding contact angles at the liquid–ice interface ($\Delta\theta_{ice} = \theta_{ice,a} - \theta_{ice,r}$), and (c) the bulk velocity ($\bar{u}_a$) during the early stage of freezing for a droplet on substrates with different inclination angles ($\alpha$). The equilibrium contact angle is $\theta_{eq}=17.5^\circ$. The equilibrium contact angle is $\theta{eq} = 17.5^\circ$. The remaining dimensionless parameters and the values of the total freezing time ($t_f$) for different inclination angles are provided in Tables~\ref{T:dim_groups} and~\ref{tab:tf_theta}, respectively.}
\label{fig:theta_def_theta_eq}
\end{figure}

The capillary contributions $u_{ca}$ and $\bar{u}_{ca}$ capture the influence of interfacial forces that act to redistribute fluid within the medium, while the gravitational components $u_{g}$ and $\bar{u}_g$ account for the fluid motion induced by the weight of the liquid. In figure~\ref{fig:theta_eq_17p5_alpha}(c), we plot the capillary ($u_{ca}$) and gravitational ($u_{g}$) components of the average velocity ($u_{a}$) for an inclination of $\alpha = 75^\circ$. As demonstrated (and also as predicted by Eq.~\ref{ua_eq_s}), gravitational forces become significant only in regions where the droplet is relatively thick, whereas capillary forces are highly sensitive to the local curvature of the droplet interface. On the advancing side, capillary and gravitational contributions effectively counteract each other, resulting in the pinning of the contact line. Additionally, once the first layer of ice forms, the remaining liquid rests on a mesa region, further restricting the motion of the droplet on the advancing side. In contrast, on the receding side, capillary forces dominate and drive the retraction of the contact line. It is also observed that variations in the advancing and receding contact angles occur only during the early stages of freezing (up to $t/t_f = 0.01$). Beyond this point, both angles remain nearly constant (see figure~\ref{fig:theta_eq_17p5_alpha}(a)) as the liquid adjacent to the substrate solidifies.

Besides the contact line between the ice and the solid substrate, two additional contact lines form at the liquid–ice interface. Figure~\ref{fig:theta_def_theta_eq}(a) shows the temporal evolution of the corresponding contact angles on the advancing ($\theta_{ice,a}$) and receding sides ($\theta_{ice,r}$), respectively, for different substrate inclinations ($\alpha$). For a horizontal substrate ($\alpha = 0^\circ$), the liquid–ice contact angles on both sides of the droplet remain equal throughout the freezing process. In contrast, for inclinations of $\alpha = 15^\circ$, $45^\circ$, and $75^\circ$, the advancing and receding contact angles differ, with $\theta_{ice,a} > \theta_{ice,r}$, and remain unequal until $t/t_f \approx 0.65$. Figure~\ref{fig:theta_def_theta_eq}(b) illustrates that the difference $\Delta \theta_{ice} = \theta_{ice,a} - \theta_{ice,r}$ increases with $\alpha$, reflecting the larger tangential gravitational force acting on the droplet at higher inclinations. This trend is consistent with the temporal evolution of the bulk velocity $\bar{u}_a$ in figure~\ref{fig:theta_def_theta_eq}(c), which increases with $\alpha$ but vanishes after $t/t_f = 0.1$. For all nonzero inclinations, the difference between $\theta_{ice,a}$ and $\theta_{ice,r}$ increases up to $t/t_f = 0.01$, primarily due to the initial contact angle hysteresis caused by the sliding motion of the droplet prior to freezing, combined with the pinning behaviour of the contact-line on the advancing side and capillary-driven retraction on the receding side (see figures~\ref{fig:theta_eq_17p5_alpha}(b,c) and related discussion). As freezing proceeds, the loss of liquid volume from solidification reduces the effect of gravity, consequently decreasing the velocity of the center-of-mass $\bar{u}_a$. Capillary forces then become dominant, and once the remaining liquid volume is sufficiently small, gravitational effects become negligible. This leads to a nearly symmetric liquid–ice interface, with the advancing and receding contact angles equalizing, as shown in figure~\ref{fig:theta_def_theta_eq}(a).

\subsection{Interplay between wettability and inclination}\label{alpha_var}

\begin{figure}
\centering
\hspace{0.5cm}{\large (a)}   \hspace{5.5cm}  {\large (b)} \\
\includegraphics[width=0.45\textwidth]{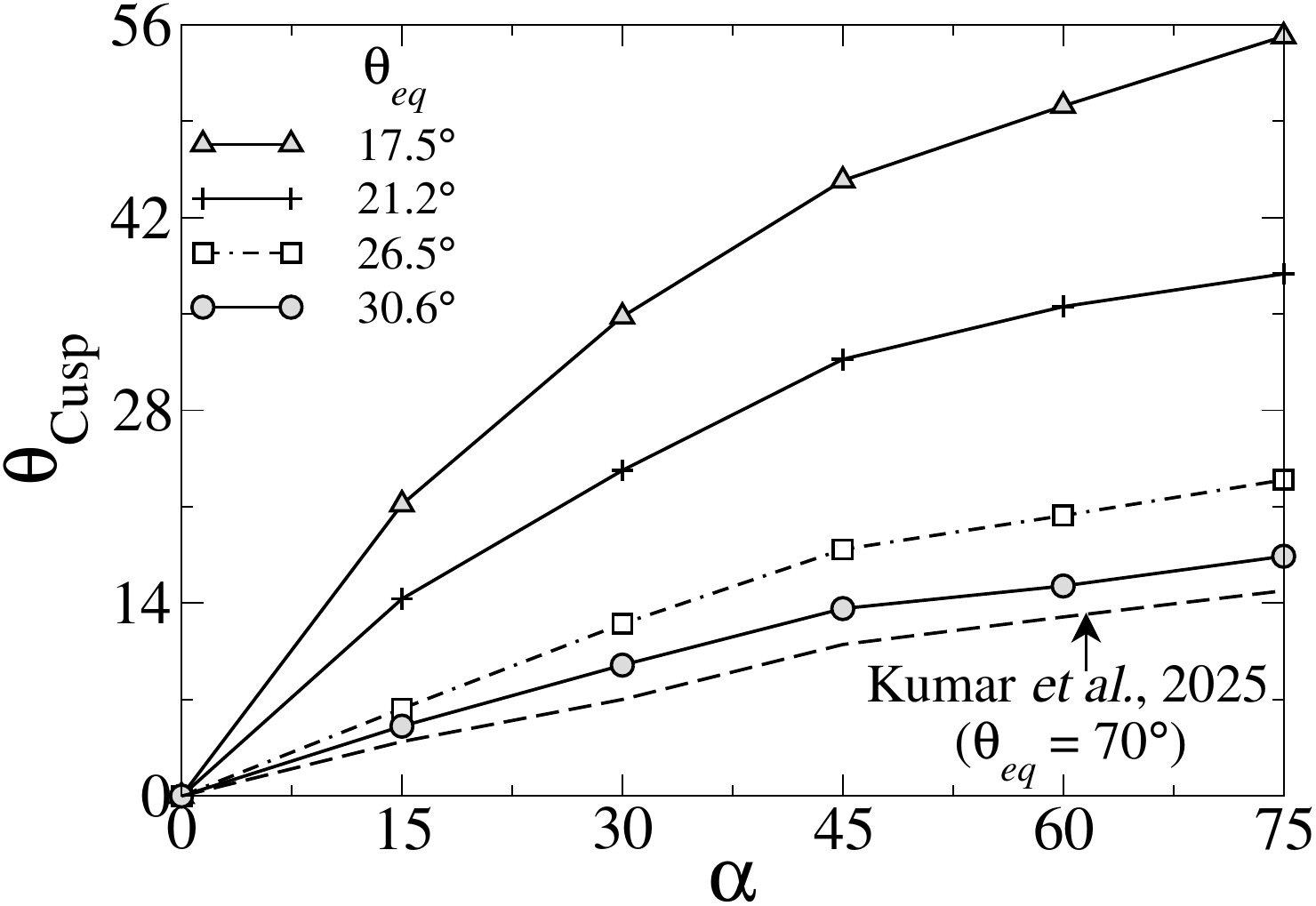}
\includegraphics[width=0.46\textwidth]{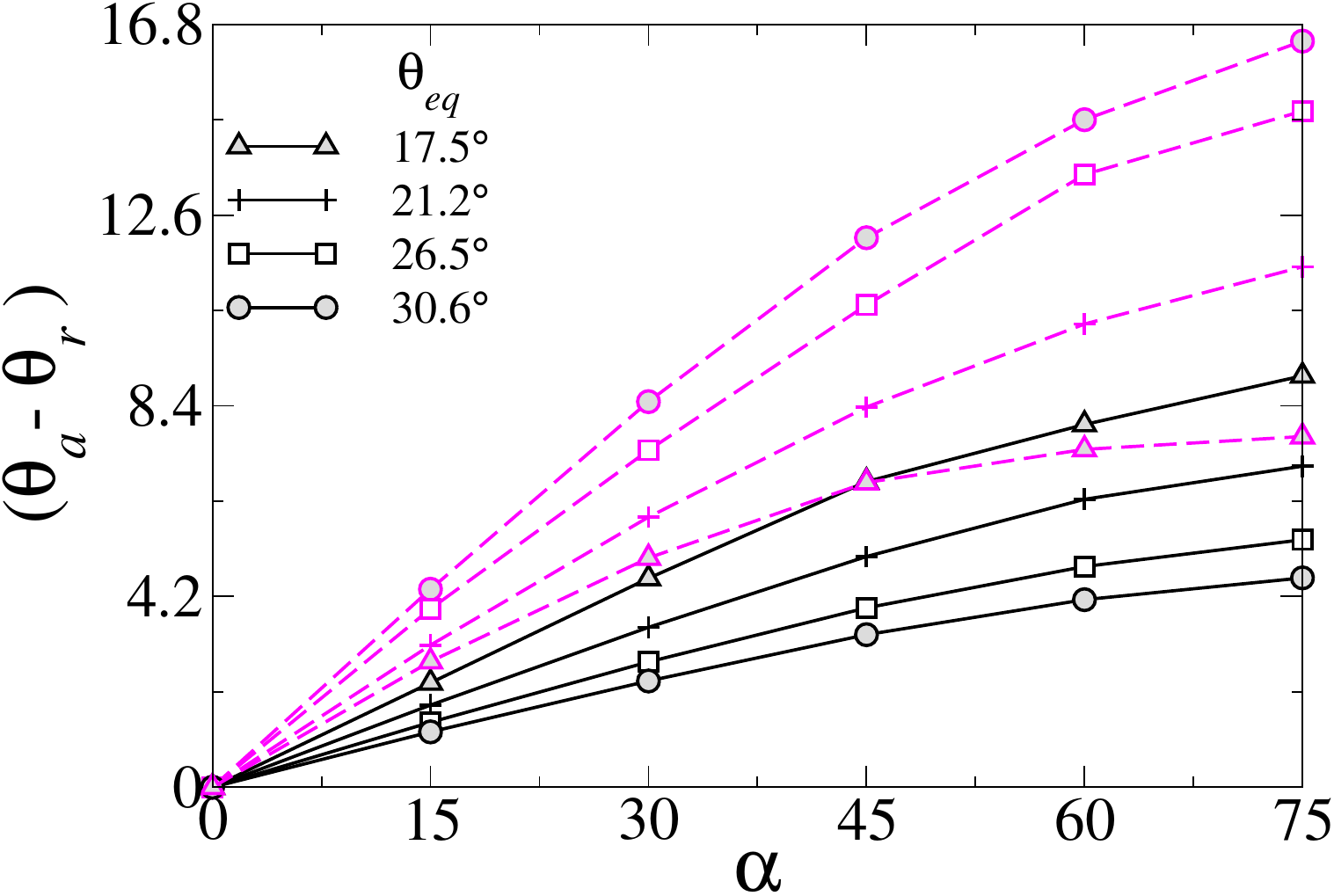}\\
\hspace{0.5cm}{\large (c)}   \hspace{5.5cm}  {\large (d)} \\
\hspace{0.4cm}\includegraphics[width=0.45\textwidth]{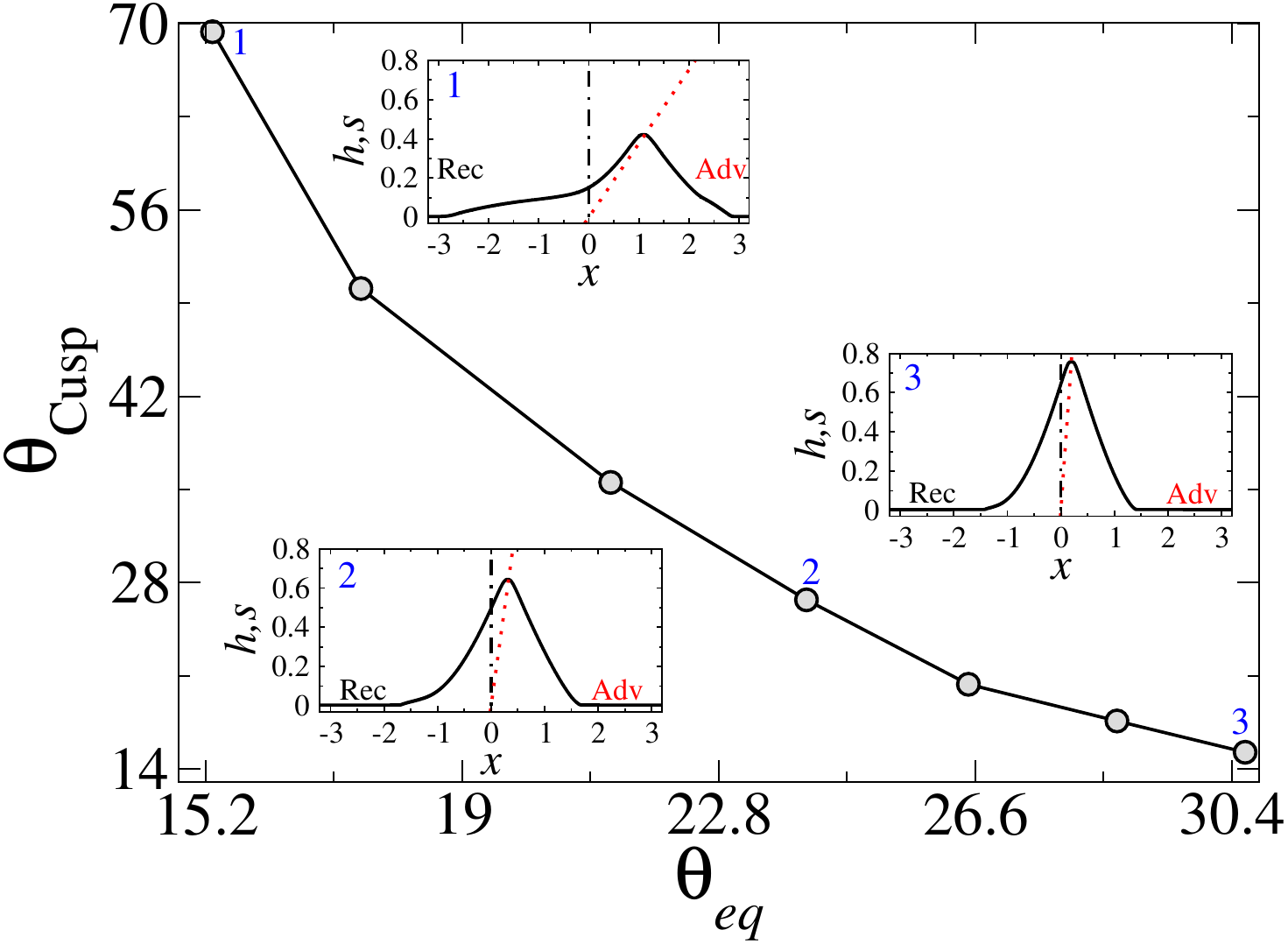}
\hspace{0.4cm}\includegraphics[width=0.48\textwidth]{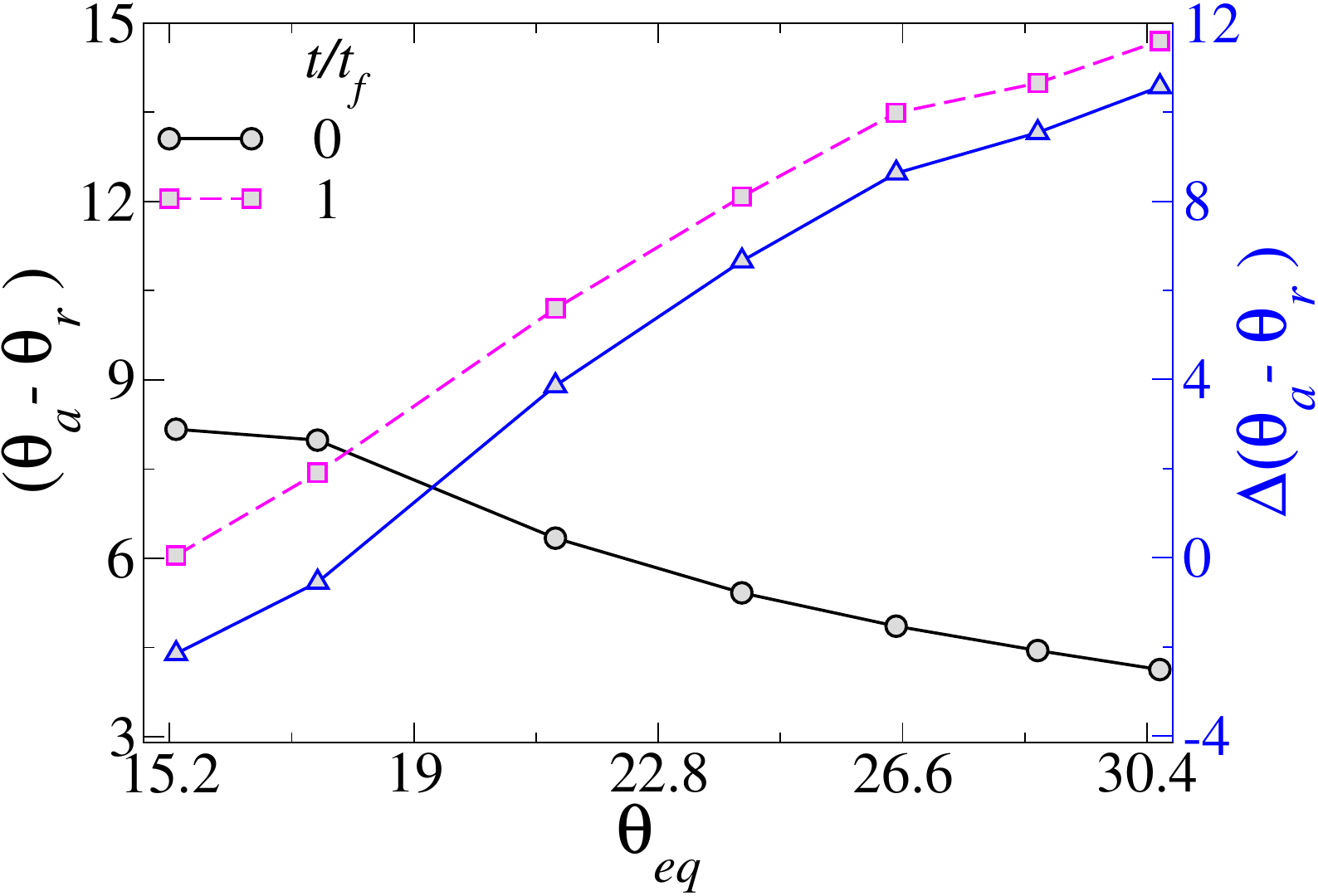}\\
\caption{Variation of (a) the cusp angle ($\theta_{\rm{Cusp}}$) at the droplet tip, measured relative to the $z$-axis, as a function of substrate inclination ($\alpha$) for different equilibrium contact angles. Here, the experimental results from \citet{kumar2025understanding} are included for comparison. (b) Variation of the difference between advancing and receding contact angles ($\theta_a - \theta_r$) with inclination angle ($\alpha$) for different equilibrium contact angles at $t/t_f = 0$ (black solid lines) and $t/t_f = 1$ (magenta dashed lines). (c) Variation of $\theta_{\rm{Cusp}}$ with $\theta_{eq}$. (d) Variation of ($\theta_a - \theta_r$) with $\theta_{eq}$ at $t/t_f = 0$ (black solid line) and $t/t_f = 1$ (magenta dashed line) on the left axis, with the change in $(\theta_a - \theta_r)$ between $t/t_f = 1$ and $t/t_f = 0$ shown on the right axis (blue solid line). In panels (c) and (d), $\alpha = 60^\circ$. The remaining dimensionless parameters are listed in Table~\ref{T:dim_groups}. The values of the total freezing time ($t_f$) for different inclination angles and equilibrium contact angles are provided in Tables~\ref{tab:tf_theta} and~\ref{tab:tf_alpha60_compact}, respectively.}
\label{fig:Del_theta_alpha_H}
\end{figure}

In the previous subsection, we focused on droplets with $\theta_{eq} = 17.5^\circ$. Here, we examine the effect of inclination ($\alpha$) and wettability ($\theta_{eq}$) on the asymmetric freezing of a sliding drop, keeping all other parameters fixed. In figure~\ref{fig:Del_theta_alpha_H}(a,b), we examine the dependence of the cusp angle ($\theta_{\rm Cusp}$) and the difference between the advancing and receding contact angles $(\theta_a-\theta_r)$ at $t/t_f=0$ (black solid lines) and $t/t_f=1$ (red dashed lines) as a function of the inclination angle ($\alpha$) for droplets with $\theta_{eq} = 17.5^\circ$, $21.2^\circ$, $26.5^\circ$, and $30.6^\circ$. As observed in figure~\ref{fig:Del_theta_alpha_H}(a) for a given inclination angle, $\theta_{\rm{Cusp}}$ decreases with decreasing substrate wettability. Interestingly, the rate of decrease progressively diminishes, and by $\theta_{eq}=30.6^\circ$, the curves exhibit a noticeable reduction in slope over the range considered, although the system remains in a transitional regime. This trend is also illustrated in figure~\ref{fig:Del_theta_alpha_H}(c), where we plot the cusp angle as a function of the equilibrium contact angle for an inclination angle $\alpha=60^\circ$. The corresponding values of $\theta_{\rm Cusp}$ are comparable in magnitude to those reported by \citet{kumar2025understanding} for a sessile water droplet on an inclined copper substrate at different inclination angles and an equilibrium contact angle of $70^\circ$; however, this comparison is qualitative, as the physical and wettability conditions differ. This behaviour can be attributed to the fact that as the wettability decreases, the droplet tends to retain its original shape due to the increased resistance to spreading. As a result, the influence of gravitational forces acting on the sliding droplet becomes less effective in deforming its shape.

\begin{figure}
\centering
\hspace{0.5cm}{\large (a)}   \hspace{5.5cm}  {\large (b)} \\
\includegraphics[width=0.45\textwidth]{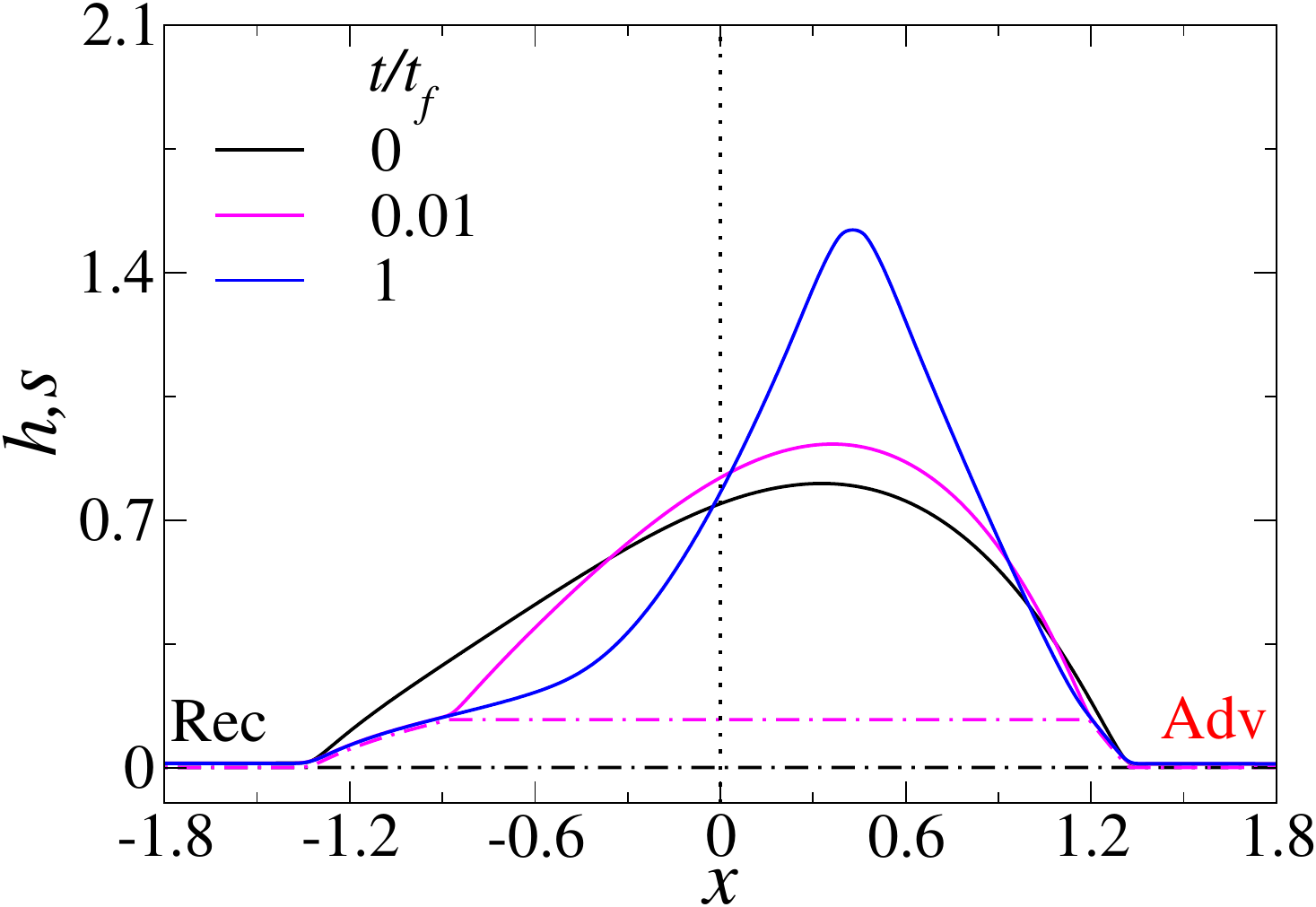}
\includegraphics[width=0.45\textwidth]{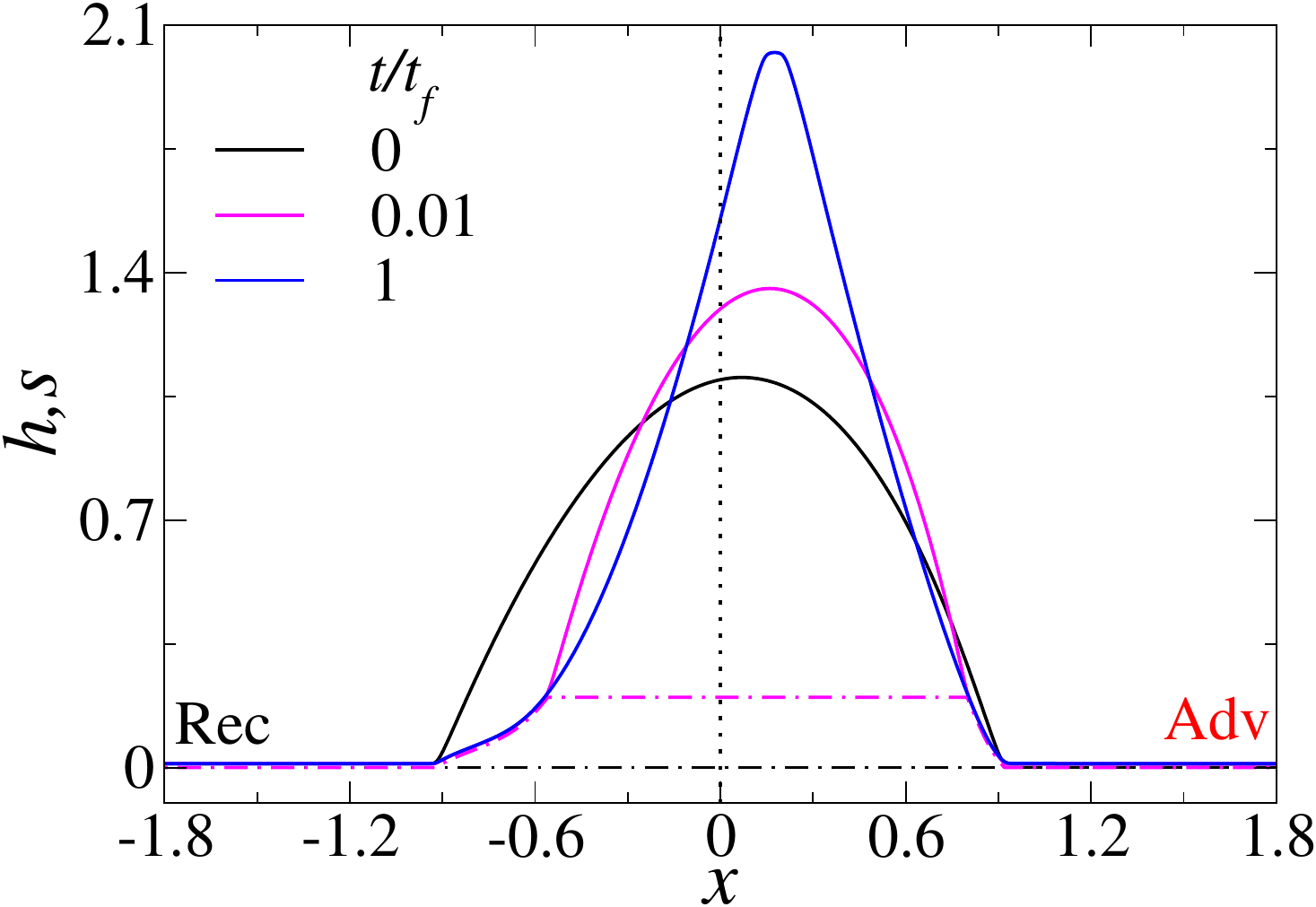}
\caption{Temporal evolution of droplet shape ($h$, solid) and freezing front ($s$, dot–dashed) on an inclined substrate ($\alpha = 75^\circ$) for (a) $\theta_{eq} = 17.5^\circ$ and (b) $\theta_{eq} = 26.5^\circ$. The remaining dimensionless parameters are listed in Table~\ref{T:dim_groups}, and the values of the total freezing time ($t_f$) for different equilibrium contact angles are provided in Table \ref{tab:tf_alpha60_compact}.}
\label{fig:h_profile}
\end{figure}

\begin{figure}
\centering
\includegraphics[width=0.95\textwidth]{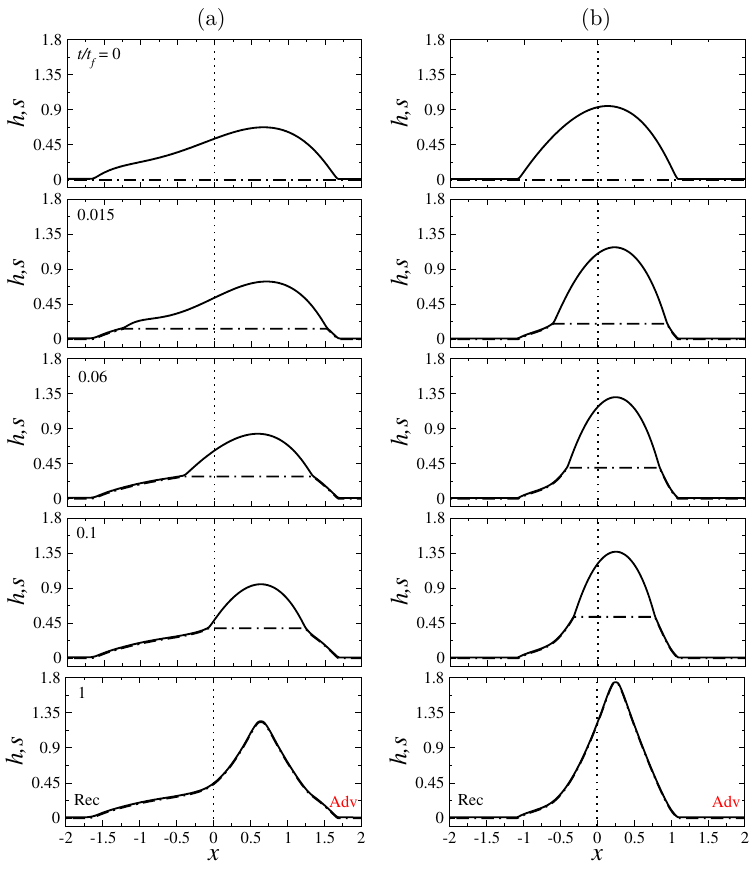} 
\caption{Evolution of the droplet shape, $h$ (solid lines), and the freezing front position, $s$ (dot-dashed lines) for (a) $\theta_{eq} = 15.3^\circ$ and (b) $\theta_{eq} = 21.2^\circ$. The normalized time, $t/t_f$, corresponding to each row is indicated in panel (a). Here, $\alpha = 60^\circ$, and the remaining dimensionless parameters and the corresponding total freezing times ($t_f$) are provided in Tables~\ref{T:dim_groups} and~\ref{tab:tf_alpha60_compact}, respectively.}
\label{fig:theta_eq_profile}
\end{figure}

In figure~\ref{fig:Del_theta_alpha_H}(b), we observe that the difference between the advancing ($\theta_{a}$) and receding ($\theta_{r}$) contact angles increases with the inclination angle ($\alpha$). For larger equilibrium contact angles ($\theta_{eq} = 21.2^\circ$, $26.5^\circ$, and $30.6^\circ$), the hysteresis ($\theta_{a}-\theta_{r}$) is larger after freezing ($t/t_{f} = 1$) than before freezing ($t/t_{f} = 0$). This behaviour is not observed for the most wettable case ($\theta_{eq} = 17.5^\circ$). To further examine the role of wettability, figure~\ref{fig:Del_theta_alpha_H}(d) shows the variation of $\theta_{a}-\theta_{r}$ with $\theta_{eq}$ for $\alpha=60^\circ$. In the unfrozen state ($t/t_f=0$), the contact angle hysteresis ($\theta_a-\theta_r$) decreases with increasing $\theta_{eq}$ (i.e., with decreasing wettability). However, the opposite trend is observed at the end of the freezing process ($t/t_f=1$). We also plot $\Delta(\theta_a-\theta_r)=(\theta_a-\theta_r)_{t/t_f=0}-(\theta_a-\theta_r)_{t/t_f=1}$ in figure~\ref{fig:Del_theta_alpha_H}(d). The value of $\Delta(\theta_a - \theta_r)$ is negative for highly wettable substrates and positive for less hydrophilic ones, and it increases monotonically with $\theta_{eq}$. To clarify this behaviour, we compare the droplet profile ($h$, solid) and the freezing front ($s$, dot-dashed) at $\alpha = 75^\circ$ for $\theta_{eq} = 17.5^\circ$ and $\theta_{eq} = 26.5^\circ$ in figure~\ref{fig:h_profile}(a) and (b). The droplet with $\theta_{eq}=17.5^\circ$ becomes more flattened at high inclination, whereas for $\theta_{eq}=26.5^\circ$ the liquid shifts more under gravity. This produces a thinner ice layer on the receding side at $t/t_f=0.01$ and leads to $(\theta_a-\theta_r)_{t/t_f=1}>(\theta_a-\theta_r)_{t/t_f=0}$ for $\theta_{eq}=26.5^\circ$, while the opposite behaviour occurs for $\theta_{eq}=17.5^\circ$. Keeping the inclination constant, the next subsection presents a detailed analysis of the influence of wettability on the asymmetric freezing of a sliding drop.

\begin{figure}
\centering
\hspace{0.50cm} {\large (a)}\\
\includegraphics[width=0.45\textwidth]{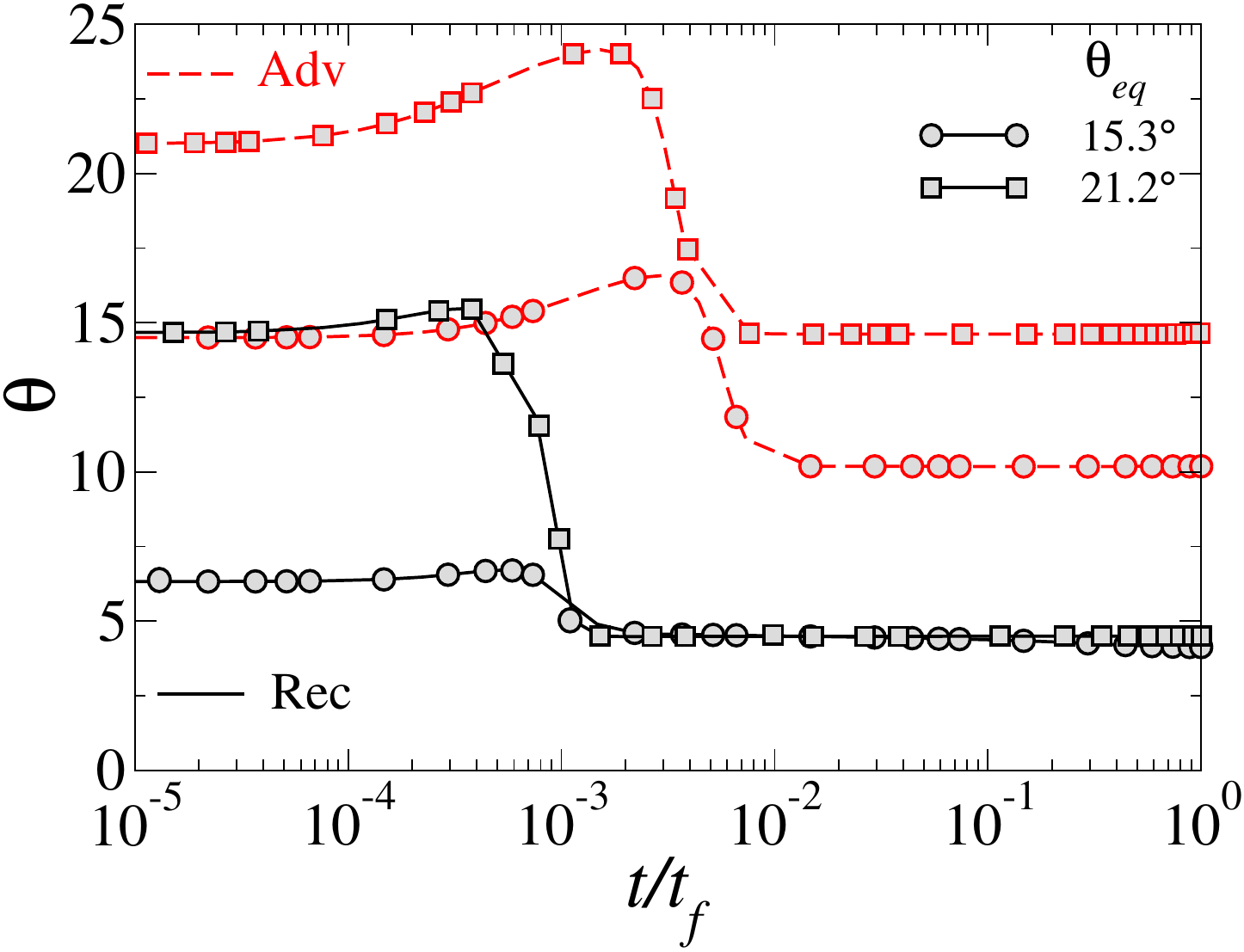}\\
\hspace{0.5cm}{\large (b)}   \hspace{5.8cm}  {\large (c)} \\
\includegraphics[width=0.45\textwidth]{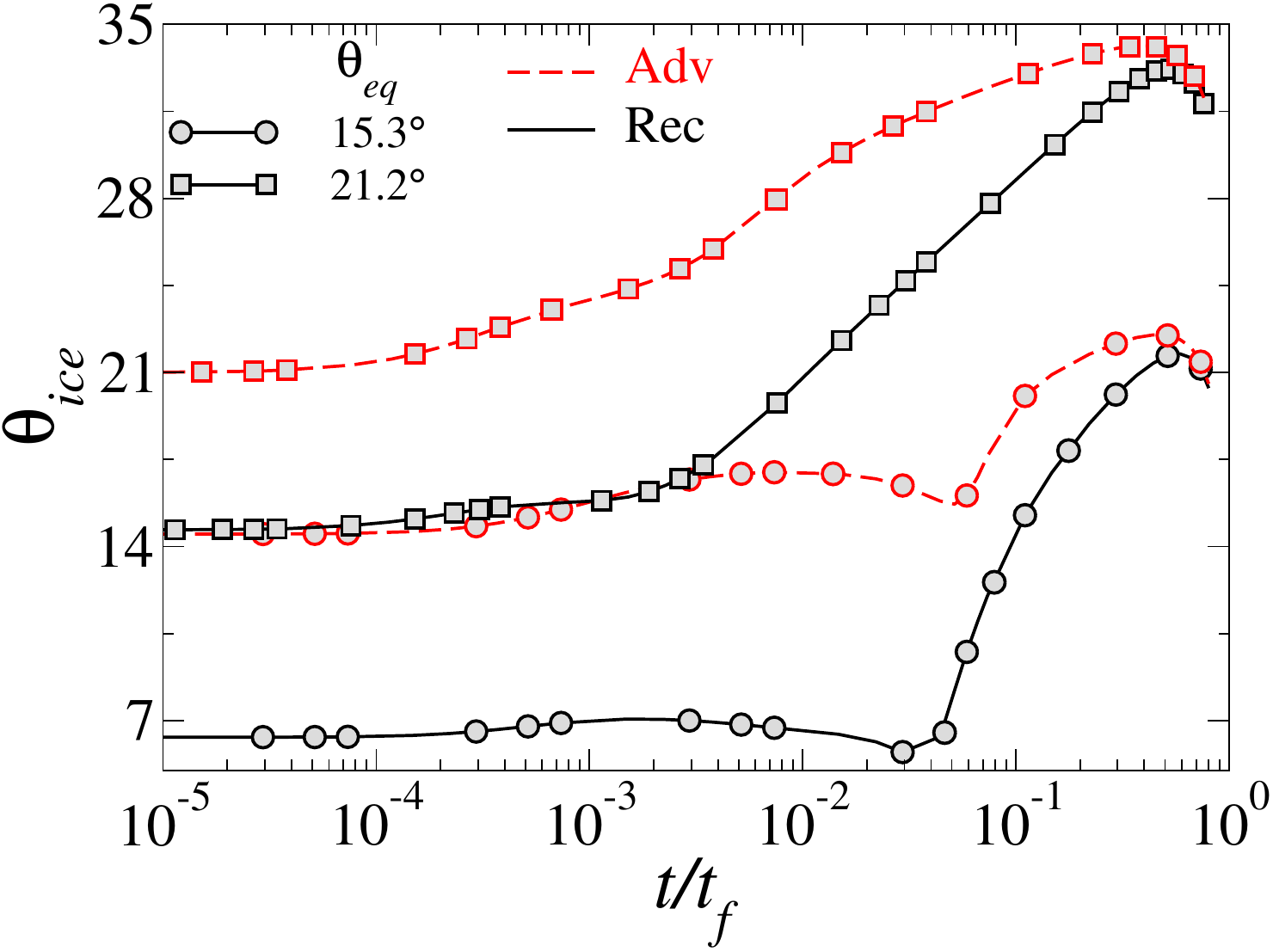}
\hspace{0mm}
\includegraphics[width=0.46\textwidth]{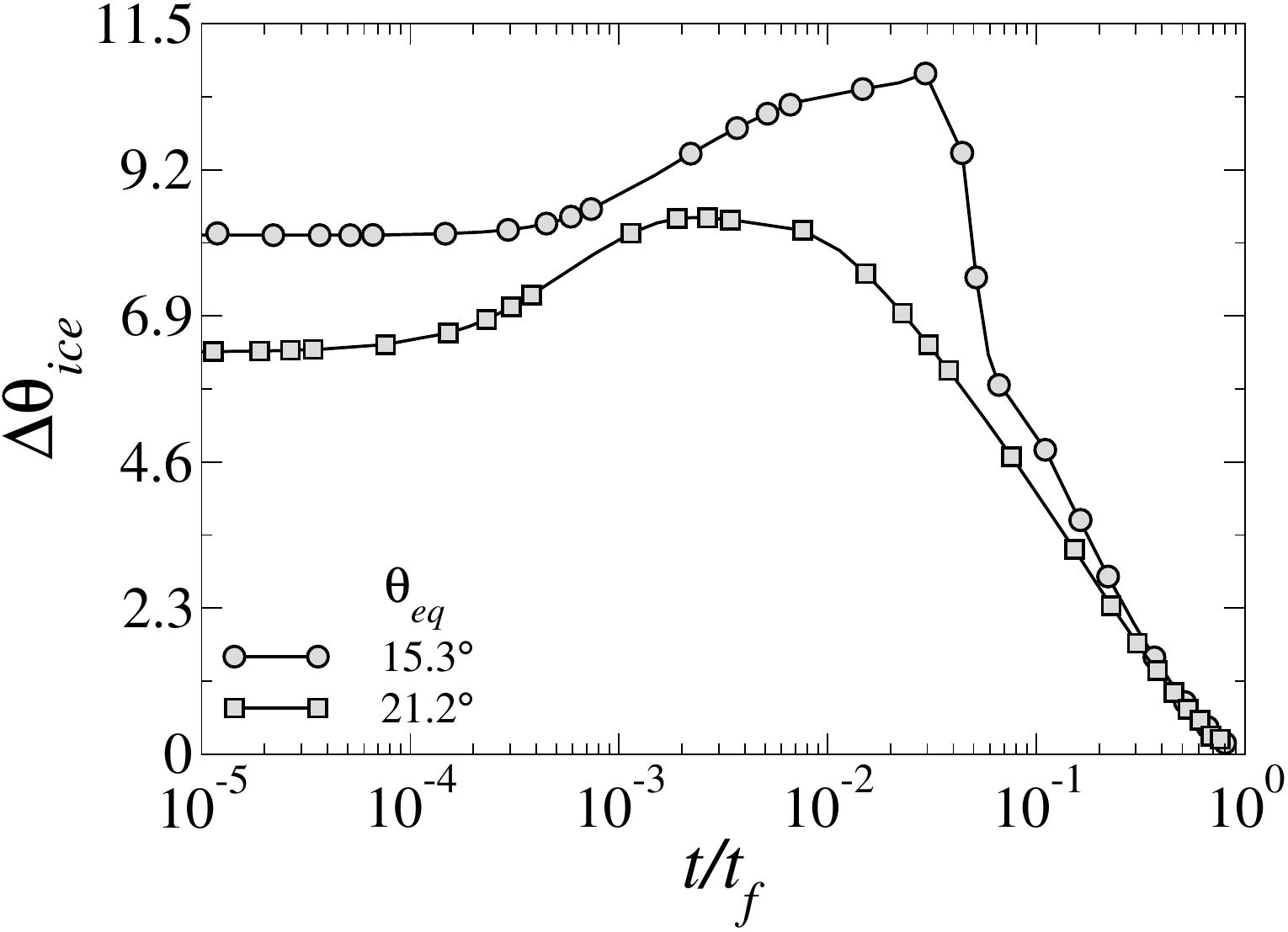}
\caption{Temporal evolution of (a) the advancing ($\theta_a$, red dashed lines) and receding ($\theta_r$, black solid lines) contact angles on the substrate, (b) the advancing ($\theta_{ice,a}$, red dashed lines) and receding ($\theta_{ice,r}$, black solid lines) contact angles at the liquid–ice interface, and (c) the difference between the advancing and receding liquid–ice contact angles ($\Delta\theta_{ice}$), for $\theta_{eq}=15.3^\circ$ and $21.2^\circ$. Here, $\alpha = 60^\circ$, and the remaining dimensionless parameters and the corresponding total freezing times ($t_f$) are given in Tables~\ref{T:dim_groups} and~\ref{tab:tf_alpha60_compact}, respectively.}
\label{fig:Del_theta_An_7p5_15}
\end{figure}

\begin{figure}
\centering
\hspace{0.5cm}{\large (a)}   \hspace{5.8cm}  {\large  (b)} \\
\includegraphics[width=0.45\textwidth]{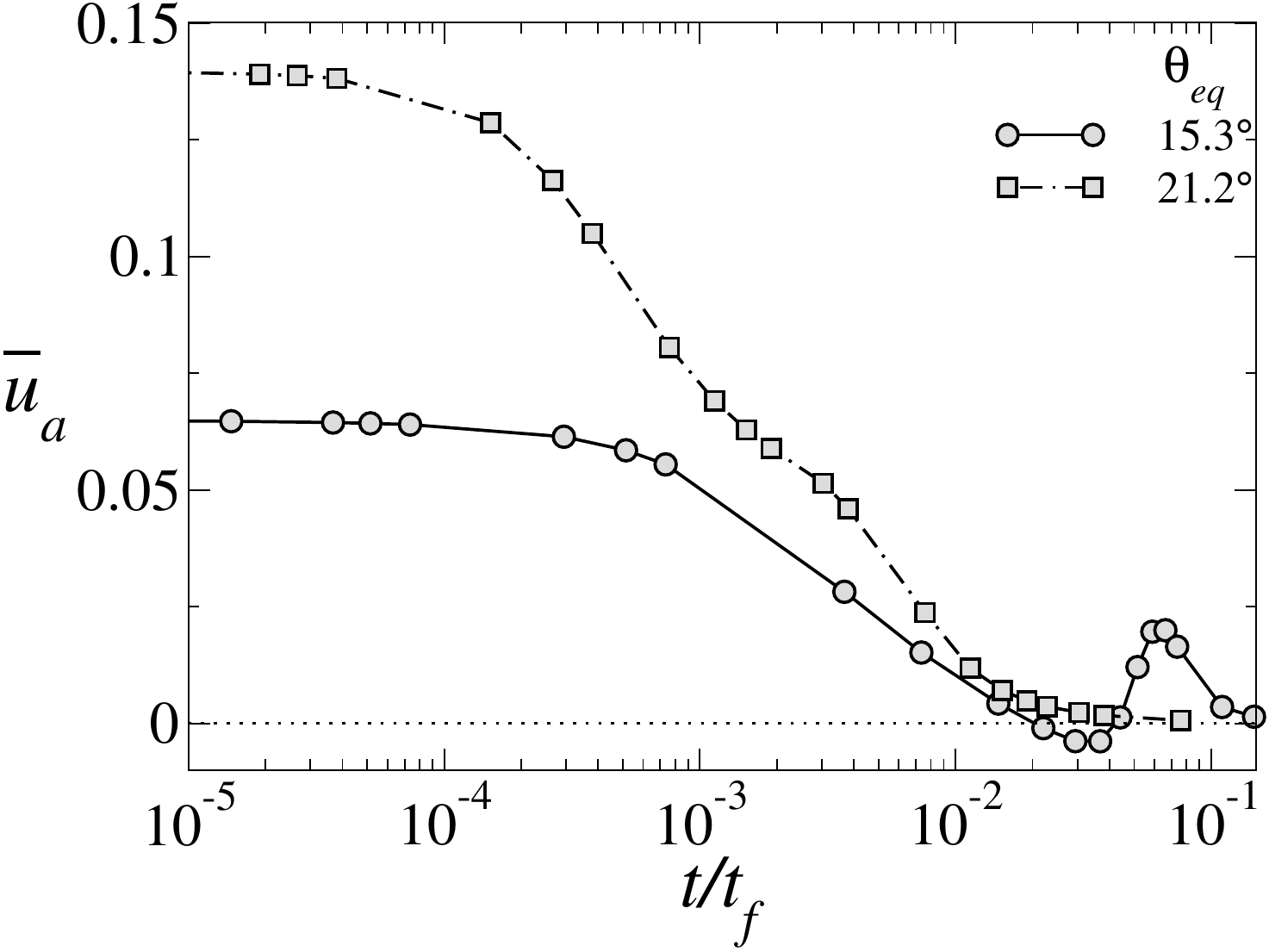} 
\hspace{0mm}
\includegraphics[width=0.465\textwidth]{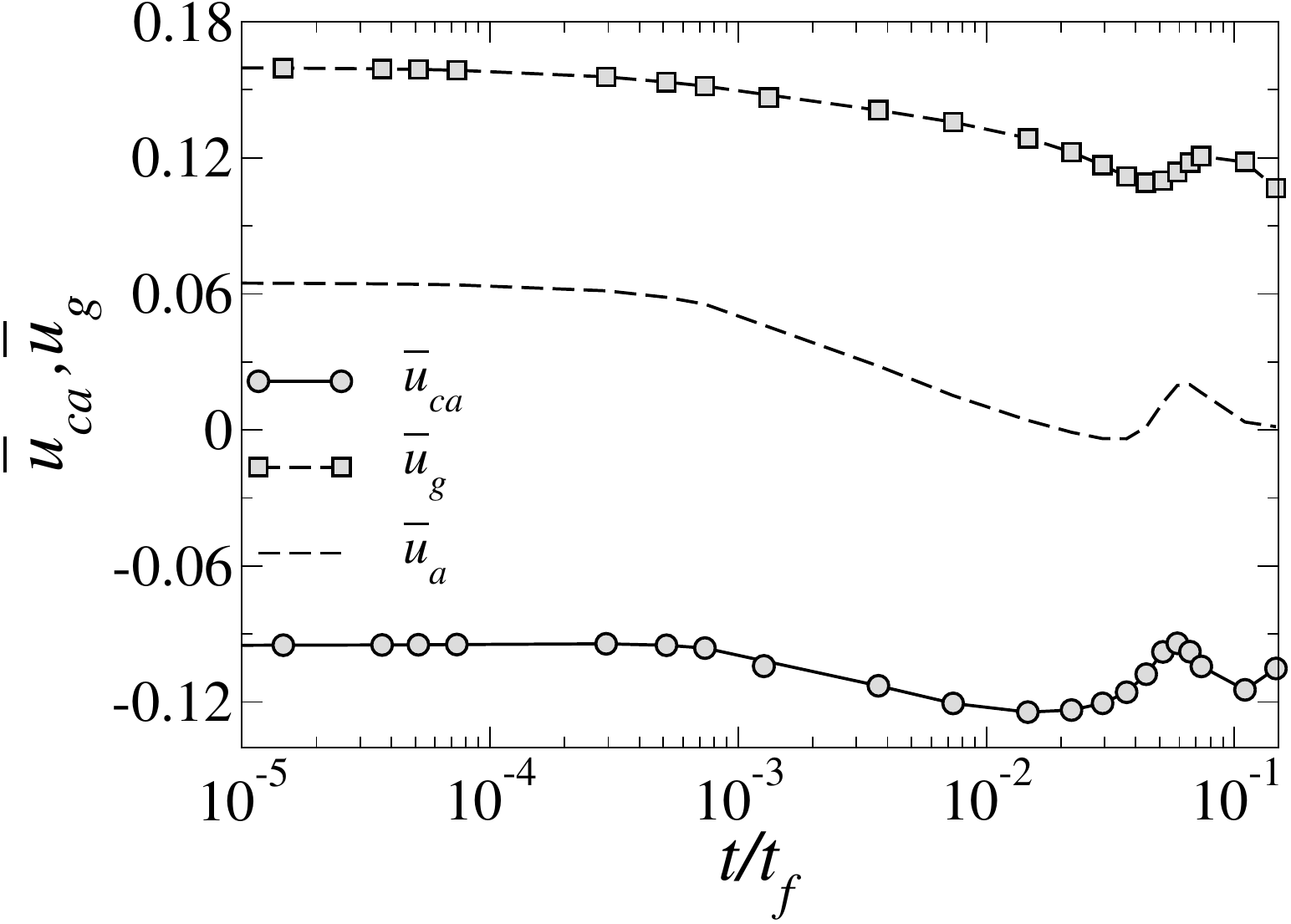}\\
\hspace{0.50cm} {\large (c)}\\
\includegraphics[width=0.47\textwidth]{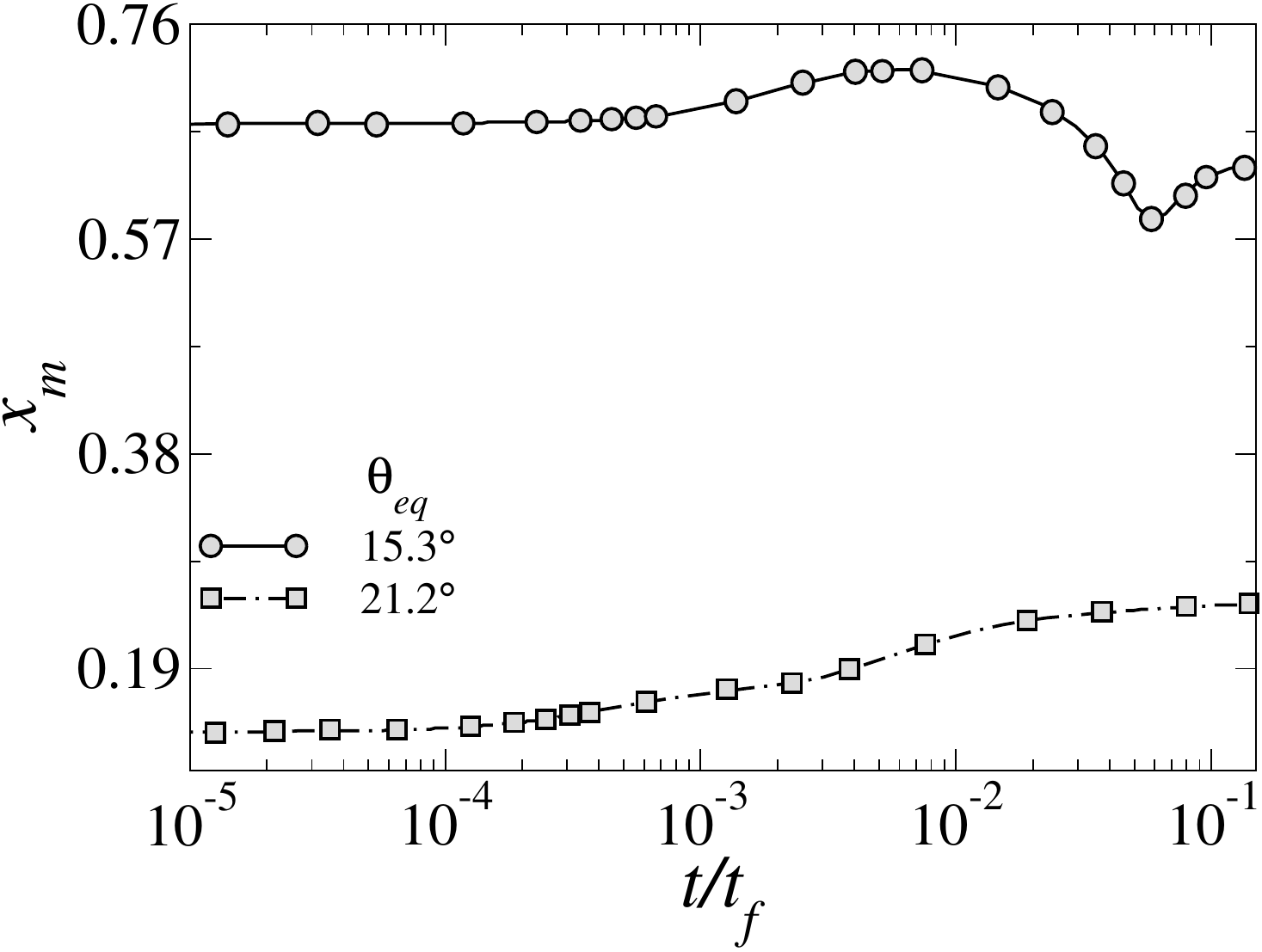}
\caption{Temporal evolution of (a) the bulk velocity ($\bar{u}_a$) during the early stages of freezing for $\theta_{eq}=15.3^\circ$ and for $\theta_{eq}=21.2^\circ$, (b) the capillary ($\bar{u}_{ca}$) and gravitational ($\bar{u}_g$) components of the bulk velocity ($\bar{u}_a$) for $\theta_{eq}=15.3^\circ$, and (c) the location of the maximum height of the drop ($x_m$) for a droplet on substrates with different equilibrium contact angles. Here, $\alpha = 60^\circ$, and the remaining dimensionless parameters and the corresponding total freezing times ($t_f$) are listed in Tables~\ref{T:dim_groups} and~\ref{tab:tf_alpha60_compact}, respectively.}
\label{fig:uavg_An_7p5_15}
\end{figure}

To further understand this behaviour, we examine two cases for $\alpha = 60^\circ$: one with a negative $\Delta(\theta_a-\theta_r)$ ($\theta_{eq}=15.3^\circ$) and another with a positive value ($\theta_{eq}=21.2^\circ$). Figure~\ref{fig:theta_eq_profile} shows the temporal evolution of the droplet shape, $h$ (solid lines), and the freezing front position, $s$ (dot-dashed lines), for these cases. For the more hydrophilic substrate ($\theta_{eq}=15.3^\circ$), the droplet shape before freezing exhibits a local concavity (negative curvature) near the receding side (see figure~\ref{fig:theta_eq_profile}(a) and (b)) at $t/t_f = 0$). This arises from the imbalance between gravity and surface tension: the tangential component of gravity stretches the droplet along the incline, while surface tension tends to restore a compact shape, leading to tail formation. This effect is stronger on hydrophilic substrates because enhanced spreading increases the wetted area and flattens the liquid–air interface, especially near the trailing edge. The resulting reduction in interfacial curvature lowers the capillary pressure resisting deformation, allowing gravity to elongate the droplet more easily. As freezing proceeds, the front advances into this concave trailing region, so the receding contact angle of the liquid–ice interface, $\theta_{ice,r}$, remains smaller than in the less hydrophilic case. In contrast, the frozen tail becomes thicker in the more hydrophilic case due to the faster freezing of the thin trailing liquid layer, producing a larger ice–substrate receding contact angle $\theta_r$ (see figure~\ref{fig:theta_eq_profile}(a) and (b) at $t/t_f = 1$). Consequently, $\Delta(\theta_a-\theta_r)<0$ for sufficiently hydrophilic substrates.

A similar behaviour can also be observed in figure~\ref{fig:Del_theta_An_7p5_15}(a), which shows the temporal evolution of the contact angles at the droplet-substrate interface. For $\theta_{eq}=15.3^\circ$, the difference between the advancing ($\theta_a$) and receding ($\theta_r$) contact angles is larger before the start of freezing ($t/t_f = 0$) than after freezing is completed ($t/t_f = 1$). In contrast, for $\theta_{eq}=21.2^\circ$, the difference between $\theta_a$ and $\theta_r$ is greater after freezing ($t/t_f = 1$) than before freezing begins ($t/t_f = 0$). We further examine the temporal evolution of the receding ($\theta_{ice,r}$) and advancing ($\theta_{ice,a}$) contact angles of the liquid-ice interface for $\theta_{eq} = 15.3^\circ$ and $\theta_{eq} = 21.2^\circ$ in figure~\ref{fig:Del_theta_An_7p5_15}(b). For both cases, $\theta_{ice,r}$ and $\theta_{ice,a}$ increase slightly at early times. For $\theta_{eq} = 15.3^\circ$, when the freezing front reaches the region of negative curvature at $t/t_f \approx 0.03$ (see figure~\ref{fig:theta_eq_profile}(a)), $\theta_{ice,r}$ decreases, followed shortly by a drop in $\theta_{ice,a}$. Both angles then recover and converge near the end of freezing. In contrast, for $\theta_{eq}=21.2^\circ$, both contact angles increase monotonically and converge only toward the end of freezing. This behaviour is reflected in figure~\ref{fig:Del_theta_An_7p5_15}(c), where $\Delta\theta_{ice}$ for $\theta_{eq}=15.3^\circ$ exhibits a pronounced maximum at $t/t_f\approx0.03$ before decaying to zero, whereas for $\theta_{eq}=21.2^\circ$ it increases smoothly to an early peak ($t/t_f\approx0.004$) and then decreases toward zero.

To rationalize this behavior, we examine the bulk velocity ($\bar{u}_a$) of the unfrozen liquid for $\theta_{eq} = 15.3^\circ$ and $21.2^\circ$ (figure~\ref{fig:uavg_An_7p5_15}a). While $\bar{u}_a$ decays to zero after $t/t_f \approx 0.1$ in both cases, it is larger for $\theta_{eq} = 21.2^\circ$, explaining the pronounced thinning of the frozen receding edge (figure~\ref{fig:theta_eq_profile}) and the larger post-freezing contact-angle contrast (figure~\ref{fig:Del_theta_alpha_H}d). At $t/t_f \approx 0.03$, $\bar{u}_a$ briefly becomes negative as the freezing front passes a concave region, reducing capillary pressure at the receding edge and driving liquid toward it. The capillary ($\bar{u}_{ca}$) and gravitational ($\bar{u}_g$) contributions (figure~\ref{fig:uavg_An_7p5_15}b) show that liquid temporarily accumulates near the receding edge, opposite gravity. This is reflected in the $x-$location of the maximum droplet height $x_m$ (figure~\ref{fig:uavg_An_7p5_15}c), which briefly decreases before resuming motion in the gravitational direction after $t/t_f \approx 0.05$.

\subsection{Effect of the Stefan number ($Ste$)}\label{Ste_var}

\begin{figure}
\centering
\hspace{0.5cm}{\large (a)}   \hspace{5.5cm}  {\large (b)} \\
\includegraphics[width=0.45\textwidth]{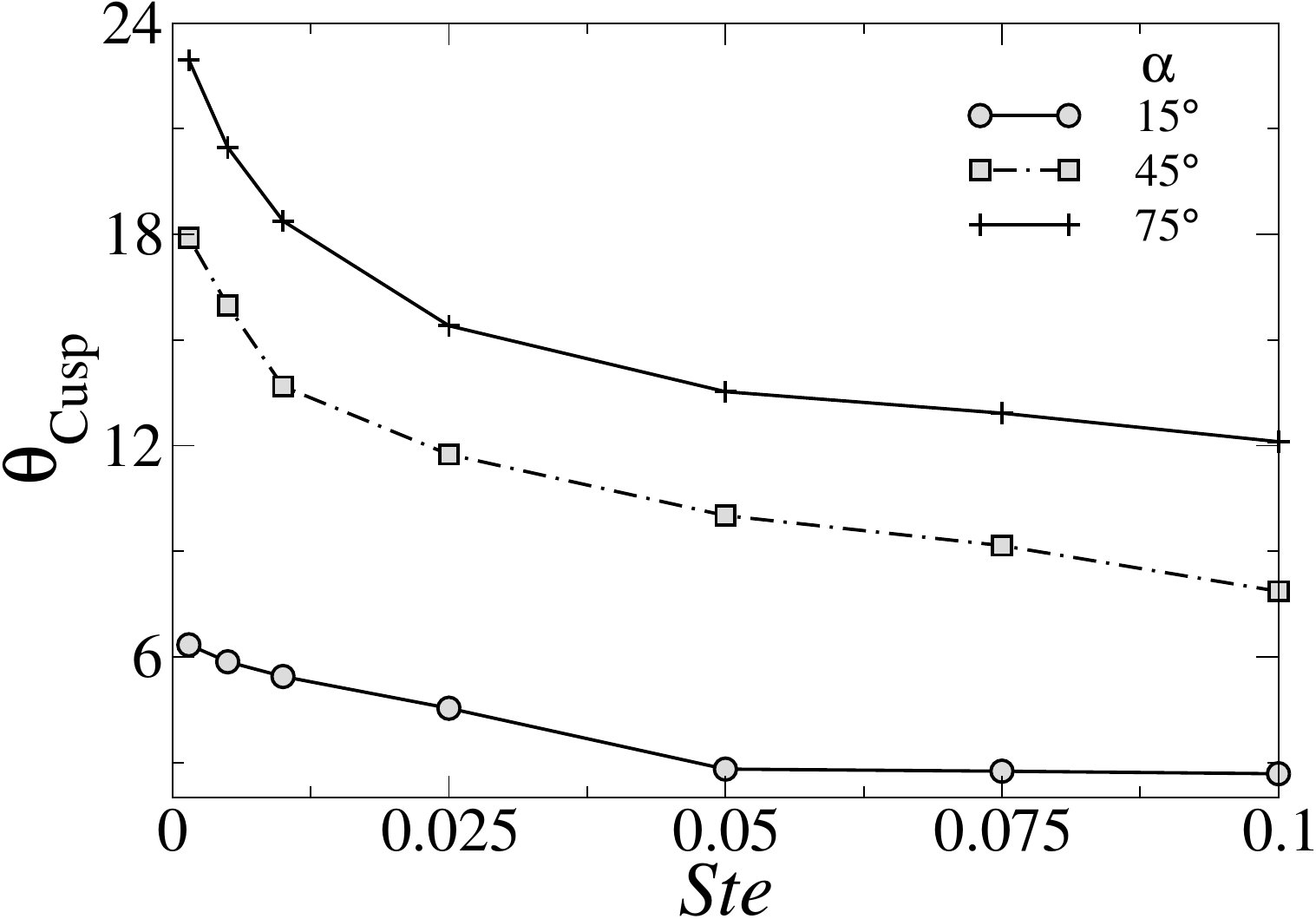}
\includegraphics[width=0.45\textwidth]{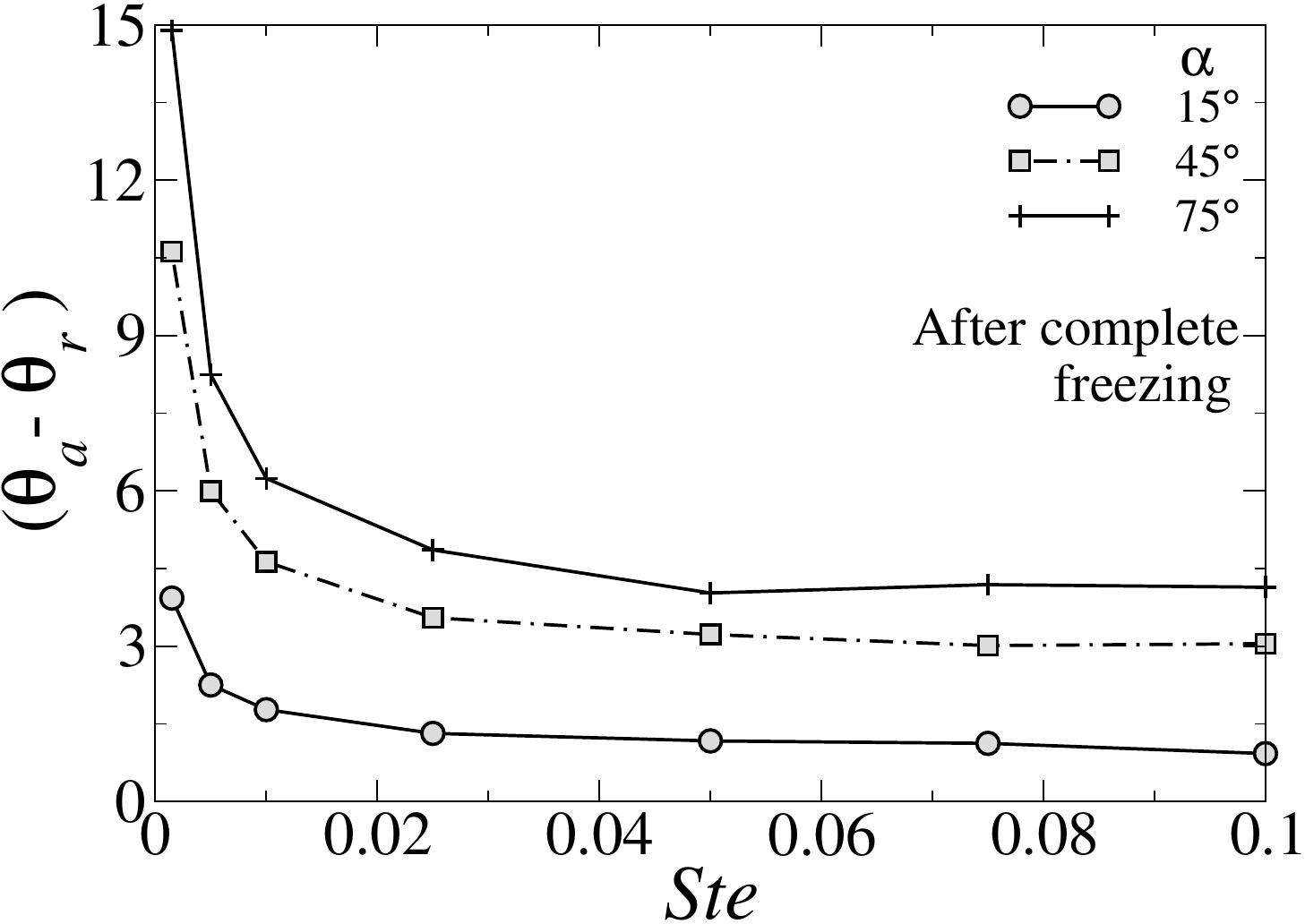}\\
\caption{Effect of the Stefan number ($Ste$) on (a) the cusp angle ($\theta_{\rm Cusp}$) and (b) the difference between the advancing and receding contact angles $(\theta_a-\theta_r)$ at $t/t_f=1$, for a droplet on substrates with different inclination angles ($\alpha$). Here, the equilibrium contact angle is $\theta_{eq}=26.5^\circ$. The remaining dimensionless parameters and the corresponding total freezing times ($t_f$) are listed in tables~\ref{T:dim_groups} and~\ref{tab:Ste_tf}, respectively.}
\label{fig:Del_theta_Ste_H}
\end{figure}

Finally, we examine the effect of the Stefan number ($Ste$) on the cusp angle ($\theta_{\rm Cusp}$) and the droplet–substrate contact-angle hysteresis $(\theta_a-\theta_r)$, evaluated at $t/t_f=0$ and $t/t_f=1$, for different substrate inclinations ($\alpha$). As shown in figure~\ref{fig:Del_theta_Ste_H}(a), the cusp angle $\theta_{\rm Cusp}$ decreases monotonically with increasing Stefan number. A larger $Ste$ implies that a greater amount of sensible heat is available relative to the latent heat of solidification, enabling more efficient removal of latent heat and thereby accelerating the freezing process. The faster advance of the freezing front leaves less time for the unfrozen liquid to slide downslope under gravity, resulting in the formation of a smaller cusp angle. As expected, prior to freezing ($t/t_f = 0$), $(\theta_a - \theta_r)$ is independent of the Stefan number. In contrast, after freezing is complete ($t/t_f=1$), $(\theta_a-\theta_r)$ decreases with increasing $Ste$, as the enhanced freezing rate at higher Stefan numbers more effectively suppresses the sliding motion of the unfrozen liquid. Consistent with this interpretation, an increase in the Stefan number leads to a substantial reduction in the dimensionless total freezing time, which decreases from $t_f=344$ for $Ste=1.49\times10^{-3}$ to $t_f=5$ for $Ste=0.1$, for $\alpha=45^\circ$ and $Bo=0.4$.

\section{Conclusions} \label{sec:Conc}

We investigate the asymmetric freezing dynamics of a liquid droplet on an inclined cold substrate through numerical simulations based on the lubrication approximation. The model has been validated against previously published numerical and experimental studies for both non-freezing and freezing droplets on cold substrates, demonstrating good quantitative agreement. In particular, we observe that the predicted tip angle is consistent with experimental observations for a horizontal substrate reported by \citet{marin2014universality}, as well as for inclined substrates in the studies of \citet{starostin2022universality,starostin2023effects}. For a fixed inclination angle, the model predicts that $\theta_{\rm Cusp}$, the angle between the cusp singularity and the vertical, decreases with decreasing wettability, with the reduction gradually approaching saturation. At $\theta_{eq} = 30.6^\circ$, the predictions converge to a lower bound that closely matches the experimental measurements of \citet{kumar2025understanding} for a sessile water droplet on an inclined copper substrate ($\theta_{eq} \approx 70^\circ$). The computed frozen droplet shapes qualitatively capture key experimental features reported in the literature \citep{anderson1996case,marin2014universality,hu2010icing,zhang2019shape}, including volume expansion, the evolution of the freezing front, and cusp formation. Furthermore, the equilibrium droplet shapes prior to freezing are in good agreement with the numerical results of \citet{park2017droplet} and are also consistent with experimental observations by \citet{podgorski2001corners}.

By systematically varying the substrate inclination, wettability, effective Bond number, and Stefan number, we elucidate the coupled roles of gravity, capillarity, and solidification in governing droplet motion, interfacial deformation, and the resulting frozen morphology. Our results demonstrate that droplet motion prior to and during freezing fundamentally alters the final structure, thereby challenging conclusions drawn from studies of sessile droplets. While inclination and wettability primarily control the degree of asymmetry, the extent of deformation can also be influenced by substrate and ambient temperatures, an aspect that warrants further investigation. The interplay between gravity and capillarity emerges as the dominant factor shaping the frozen structure. Inclination and wettability jointly dictate the degree of asymmetry, while freezing kinetics modulate the extent of deformation. Highly wetting substrates exhibit unexpected behaviour, including transient motion opposite to gravity, underscoring the complexity of early-stage dynamics.

Notably, the model predicts a transient motion opposite to gravity on highly wetting substrates during the early stages of freezing, a phenomenon that, to our knowledge, has not yet been reported experimentally and may motivate future investigations. By decomposing the liquid motion into capillary- and gravity-driven components, we provide mechanistic insight into contact-line pinning, receding-edge thinning, and the emergence of asymmetric ice–liquid contact angles.An increase in the Stefan number accelerates freezing, suppresses sliding-induced deformation, and reduces both the cusp angle and the contrast in post-freezing contact angles. These findings have practical implications for anti-icing strategies, thermal management, and surface design in aerospace and energy systems. Future work should extend this framework to three-dimensional droplets, textured or chemically heterogeneous substrates, and coupled evaporation–freezing scenarios to better capture real-world conditions. \\

\noindent{\bf Acknowledgement:} We sincerely thank the anonymous reviewers for their insightful comments and constructive suggestions, which significantly improved the quality and depth of the manuscript. \\

\noindent{\bf Funding:} This work was supported by the Suzuki Next Bharat Fellowship Program (Grant No. NBV/CHE/F011/2025-26/S401). GK would also like to acknowledge the ThermEnTrans project funded by the European Union's Horizon Europe research and innovation programme under grant agreement ID: 101236597.\\

\noindent{\bf Declaration of interests:} The authors report no conflict of interest.

\appendix
\numberwithin{equation}{section}
\makeatletter
\newcommand{\section@cntformat}{Appendix \thesection:\ }
\makeatother
\section{Validation of the numerical model} \label{sec:Val}

\subsection{Comparison with \citet{park2017droplet}} \label{A1}

\begin{figure}
\centering
\includegraphics[width=0.8\textwidth]{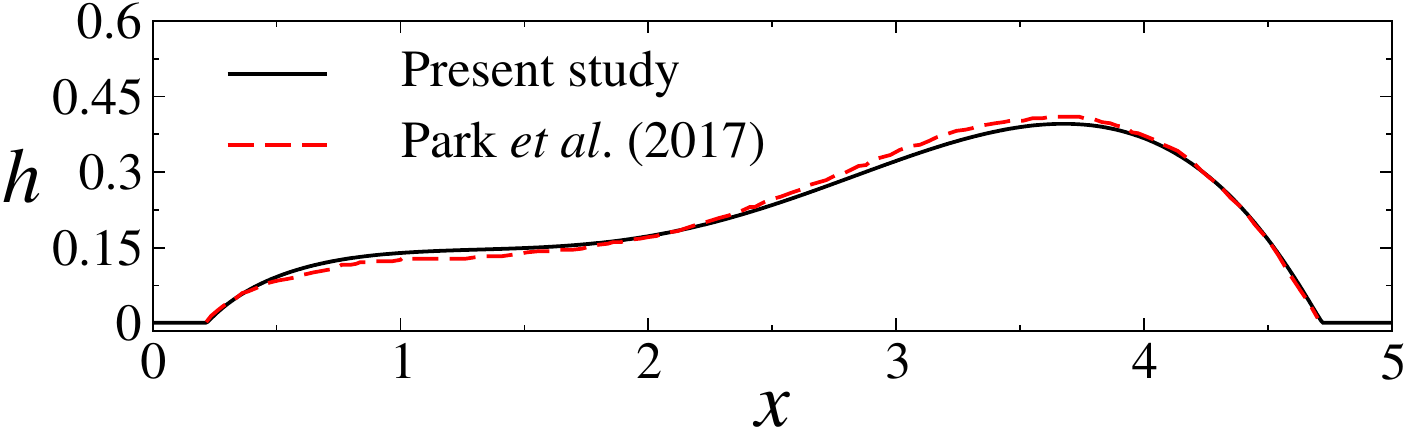}
\caption{Comparison of the equilibrium shape of a droplet sliding on a substrate inclined at $\alpha = 25^\circ$, as obtained from the present simulations, with the corresponding results reported in \cite{park2017droplet}. For comparison with \cite{park2017droplet}, the effect of freezing is neglected in the present formulation by setting $Ste = 0$. The other dimensionless parameters are $\theta_{eq} \approx 20^\circ$, $Bo = 0.9$, $A_{n} = 121.8$, and $\epsilon = 0.44$.}
\label{fig:Slid_val_A1}
\end{figure}

We validate our model by first simulating a non-freezing droplet sliding on a smooth inclined substrate, a configuration previously examined by \citet{park2017droplet}. Figure~\ref{fig:Slid_val_A1} compares the equilibrium droplet shape obtained from our simulations at an inclination angle of $\alpha = 25^\circ$ with the corresponding case shown in figure~2(c) of \citet{park2017droplet}. To ensure a consistent comparison, freezing effects are neglected by setting $Ste = 0$. The remaining dimensionless parameters are $\theta_{eq} \approx 20^\circ$, $Bo = 0.9$, $A_n = 121.8$, and $\epsilon = 0.44$. Since the exact values of the scaled Hamaker constant ($A_n$) and the aspect ratio ($\epsilon$) are not reported in \citet{park2017droplet}, this missing information may account for the minor discrepancies observed between the two results.

\subsection{Validation using the configuration of \citet{zadravzil2006droplet}} \label{A2}

\begin{figure}
\centering
\includegraphics[width=0.9\textwidth]{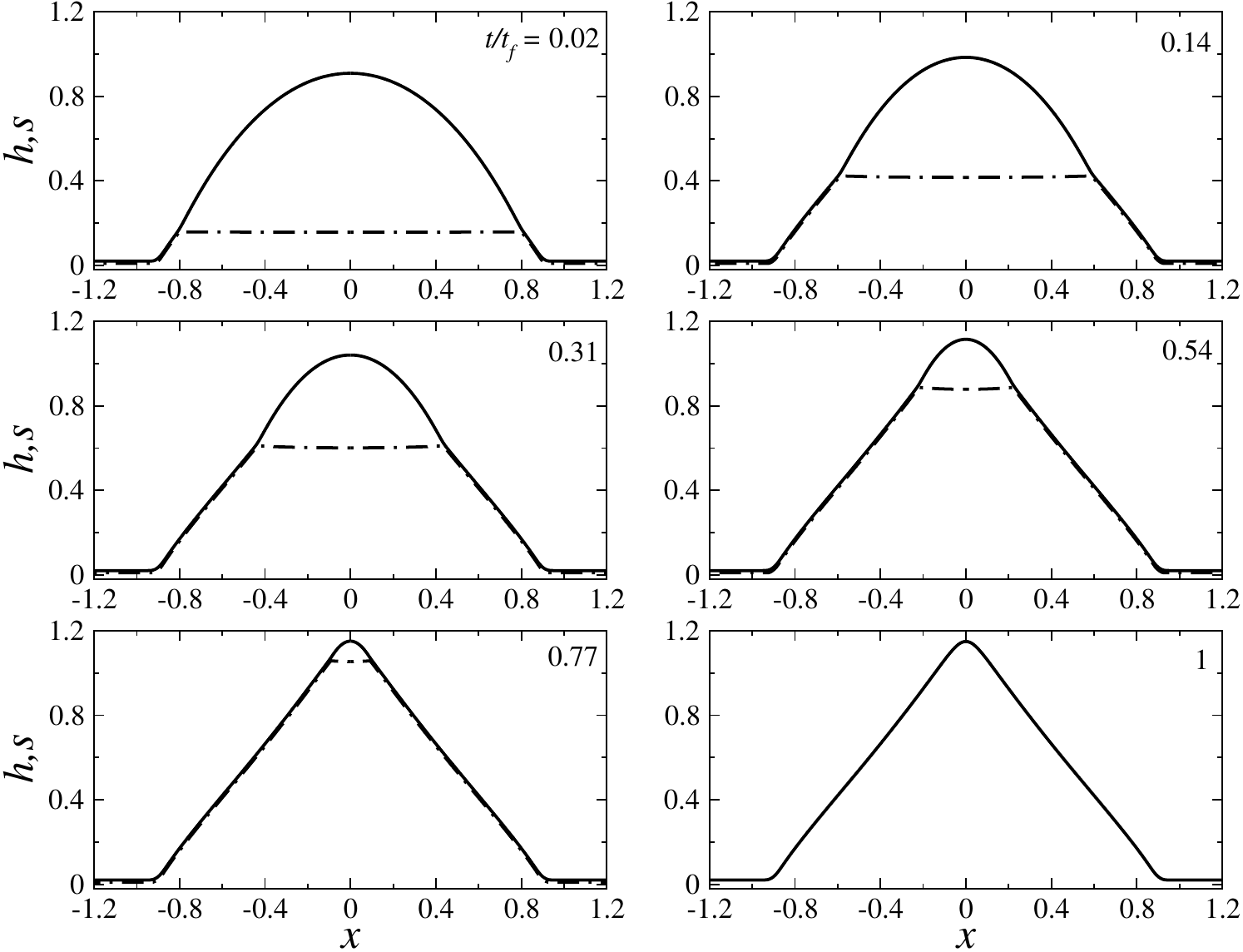}
\caption{Evolution of the freezing front, $s$ (dot-dashed lines), and the shape, $h$ (solid line) of the droplet undergoing freezing on a horizontal ($\alpha = 0^\circ$) cold substrate. The dimensionless parameters considered in the simulations are $\epsilon = 0.2$, $Ste = 0.04$, $T_{v} = 0.5$, $A_{n} = 8$, $D_{s} = \Lambda_{S} = Bi = Bo = 1$, and $D_{w} = 0$. This set of parameters corresponds to those used by \citet{zadravzil2006droplet}. Here, the dimensionless total freezing time of the droplet is found to be $t_{f} = 14$, which is in good agreement with the value reported by \citet{zadravzil2006droplet}.}
\label{fig:Zad_val_A2}
\end{figure}

We further validate our freezing model by simulating a configuration previously investigated by \citet{zadravzil2006droplet}. Figure~\ref{fig:Zad_val_A2} shows the temporal evolution of the droplet shape, $h$ (solid line), and the freezing front, $s$ (dot-dashed line), for a droplet on a cold substrate, corresponding to the configuration studied by \citet{zadravzil2006droplet}. The droplet evolution observed here is consistent with the behaviour reported in figure~15 of their manuscript. The remaining dimensionless parameters are $\epsilon = 0.2$, $Ste = 0.04$, $T_v = 0.5$, $A_n = 8$, $D_s = \Lambda_S = Bi = Bo = 1$, and $D_w = 0$. The dimensionless total freezing time obtained in figure~\ref{fig:Zad_val_A2} is $t_f = 14$. To reproduce a droplet shape consistent with the simulations of \citet{zadravzil2006droplet}, we adopt the same initial droplet profile at the onset of freezing. In their study, the droplet undergoes spreading and imbibition prior to freezing; accordingly, we use the droplet shape obtained after these processes have completed as the initial condition for our simulations. This initial profile is taken from figure~14 of \citet{zadravzil2006droplet}. The total dimensionless time reported by \citet{zadravzil2006droplet} is $t_f = 16$, of which $t = 2.08$ corresponds to spreading and imbibition, with the remaining duration associated with freezing ($t_f = 13.92$). Consequently, the freezing time alone reported in \citet{zadravzil2006droplet} agrees closely with the total freezing time $t_f = 14$ obtained in our simulations. The only differences between our simulations and those of \citet{zadravzil2006droplet} are the thickness of the precursor layer, which is $\beta = 0.01$ in the present study compared with $\beta = 0.005$ in \citet{zadravzil2006droplet}, and the form of the penalty function employed in our model (eq.~\ref{pre_model1}), which differs from that used in their work.

\subsection{Grid convergence test} \label{A3}

We perform a grid independence study for an equilibrium contact angle $\theta_{eq} = 17.5^\circ$ and an inclination $\alpha = 75^\circ$ (near the extreme) using 12,001, 25,601, and 30,001 grid points. The remaining parameters (`base' parameters) used in the simulations are $Bo = 0.4$, $Ste = 1.49 \times 10^{-3}$, $A_n = 10$, $D_s = 0.9$, $\Lambda_S = 3.89$, $\Lambda_W = 675$, $Bi = 0.16$, $D_w = 2.94$, and $\epsilon = 0.2$, as listed in Table~\ref{T:dim_groups}. The total freezing time ($t_f$) remains unchanged across all cases, with $t_f = 201$. We also note here that this solver has been extensively validated and employed in our previous studies \citep{kavuri2023freezing,kavuri2025evaporation}.

\begin{figure}
\centering
\includegraphics[width=0.9\textwidth]{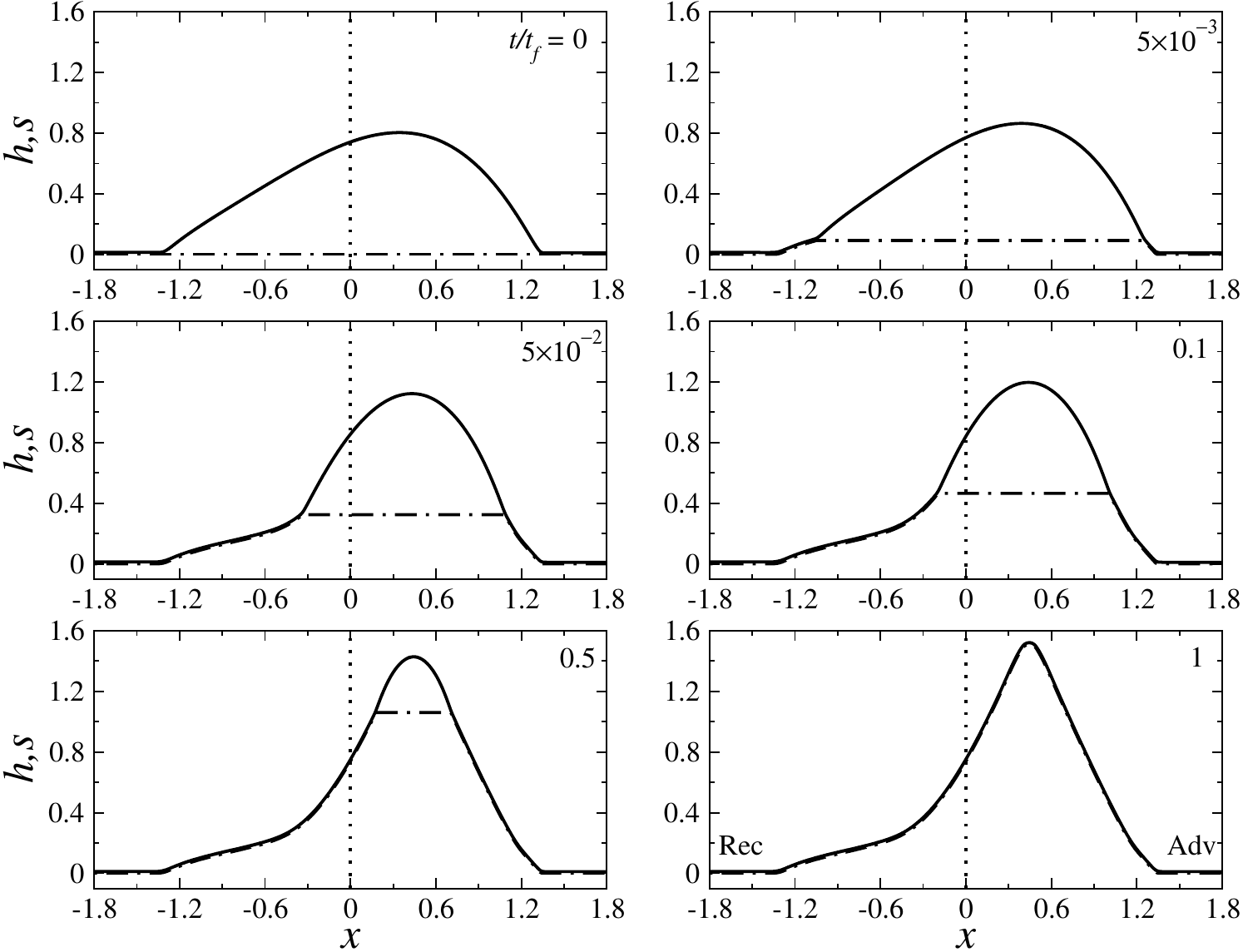}
\caption{Evolution of the freezing front, $s$ (dot-dashed lines), and the droplet shape, $h$ (solid lines), on a cold substrate inclined at $\alpha = 75^\circ$. The remaining dimensionless parameters are $Bo = 0.4$, $Ste = 1.49 \times 10^{-3}$, $A_n = 10$, $D_s = 0.9$, $\Lambda_S = 3.89$, $\Lambda_W = 675$, $Bi = 0.16$, $D_w = 2.94$, and $\epsilon = 0.2$. The total dimensionless freezing time is $t_f = 201$. The results obtained using 12001, 25601, and 30001 grid points are indistinguishable, demonstrating grid independence.}
\label{fig:Grid_study}
\end{figure}

\section{Droplet shape evolution at $\alpha = 75^\circ$ for different $\theta_{eq}$} \label{sec:the_eq_alpha_75}

\begin{figure}
\centering
\includegraphics[width=0.95\textwidth]{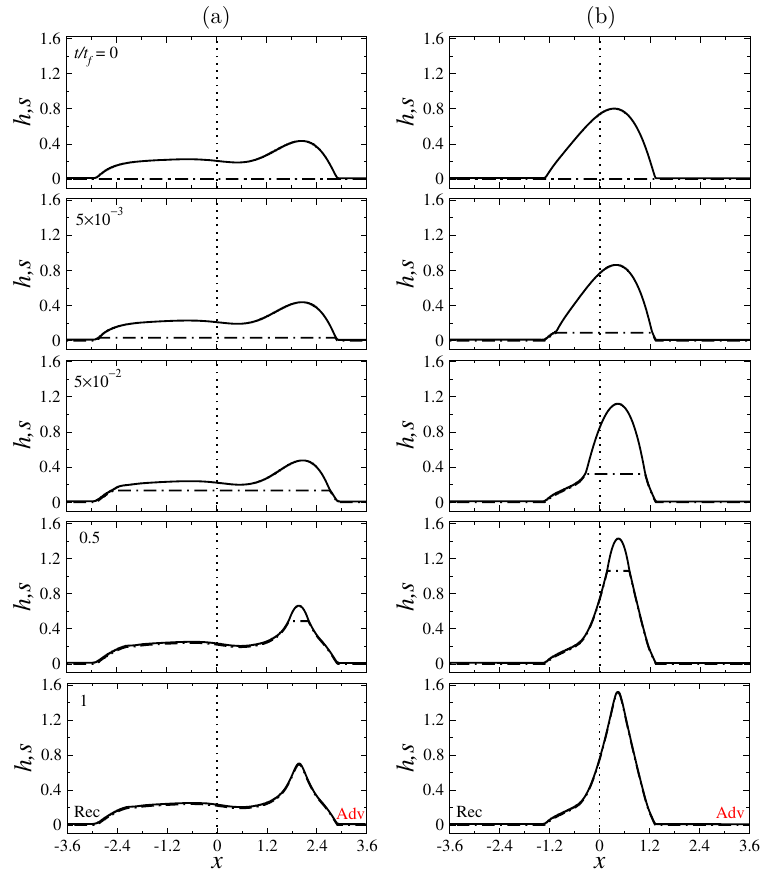} 
\caption{Evolution of the droplet shape, $h$ (solid lines), and the freezing front, $s$ (dot–dashed lines), on a substrate inclined at $\alpha = 75^\circ$ for (a) $\theta_{eq} = 12^\circ$ and (b) $\theta_{eq} = 17.5^\circ$. The normalized time, $t/t_f$, corresponding to each row is indicated in panel (a). All other parameters are provided in Table~\ref{T:dim_groups}. The corresponding total freezing times are $t_f = 44$ and $t_f = 201$ for panels (a) and (b), respectively.}
\label{fig:Theta_eq_var_profile}
\end{figure}

Here, in figure~\ref{fig:Theta_eq_var_profile}, we compare the evolution of the droplet shape $h$ (solid line) and the freezing front $s$ (dot–dashed line) for droplets with $\theta_{eq} = 12^\circ$ and $\theta_{eq} = 17.5^\circ$ on a substrate inclined at $\alpha = 75^\circ$. The equilibrium contact angle $\theta_{eq} = 12^\circ$ corresponds to experimentally reported values for water on ice near the melting temperature. It can be seen that, for $\theta_{eq} = 12^\circ$, the droplet becomes significantly more elongated along the inclined substrate, with pronounced swelling on the advancing side and tapering on the receding side. These shapes differ markedly from those obtained for $\theta_{eq} = 17.5^\circ$, highlighting the strong influence of small equilibrium contact angles on the droplet morphology during freezing.

\clearpage

\end{document}